\documentclass[prl,showpacs,amsmath,amssymb,superscriptaddress,floatfix,reprint,nofootinbib,onecolumn]{revtex4-2}
\usepackage{amssymb}
\usepackage{amsbsy}
\usepackage{amsmath}
\usepackage{mathrsfs}
\usepackage{epsfig}
\usepackage{graphicx}
\usepackage{array}
\usepackage{textcomp}
\usepackage{xcolor}
\usepackage{braket}
\usepackage{hhline}
\usepackage{dsfont}
\usepackage{comment}
\usepackage{enumerate}
\usepackage{bm,hyperref}
\usepackage[titletoc,toc,title]{appendix}
\usepackage[normalem]{ulem}
\usepackage[labelformat=simple]{subcaption}

\usepackage[justification=raggedright,font=small]{caption}
\usepackage{caption}
\usepackage{tikz}
\usepackage{adjustbox}

\graphicspath{{Artworks/}}

\newcommand{\be}{\begin{equation}}
\newcommand{\ee}{\end{equation}}
\newcommand{\iea}{\begin{equation}\begin{aligned}}
\newcommand{\fea}{\end{aligned}\end{equation}}

\begin{document}

\author{Larry Li}
\affiliation{Department of Physics and Astronomy, University of Southern California, Los Angeles, CA 90089-0484, USA}
\author{Marcin Abram}
\affiliation{Department of Physics and Astronomy, University of Southern California, Los Angeles, CA 90089-0484, USA}
\author{Abhinav Prem}
\affiliation{School of Natural Sciences, Institute for Advanced Study, Princeton, New Jersey 08540, USA}
\author{Stephan Haas}
\affiliation{Department of Physics and Astronomy, University of Southern California, Los Angeles, CA 90089-0484, USA}

\title{Hofstadter Butterflies in Topological Insulators}

\begin{abstract} In this chapter, we investigate the energy spectra as well as the bulk and surface states in a two-dimensional system composed of a coupled stack of one-dimensional dimerized chains in the presence of an external magnetic field. Specifically, we analyze the Hofstadter butterfly patterns that emerge in a 2D stack of coupled 1D Su-Schrieffer-Heeger (SSH) chains subject to an external transverse magnetic field. Depending on the parameter regime, we find that the energy spectra of this hybrid topological system can exhibit topologically non-trivial bulk bands separated by energy gaps. Upon introducing boundaries into the system, we observe topologically protected in-gap surface states, which are protected either by a non-trivial Chern number or by inversion symmetry. We examine the resilience of these surface states against perturbations, confirming their expected stability against local symmetry-preserving perturbations. 
\end{abstract}

\maketitle


\section{Introduction}
\label{sec:intro}

The classification and characterization of quantum phases of matter in terms of the topology of their many-body ground state(s) is one of the central concepts in modern condensed matter physics. The field of topological phases has witnessed remarkable synergy between theory and experiment, with the theoretical prediction and experimental discovery of electronic insulators~\cite{kanemele,bernevigzhang,fukanemele,moorebalents,roy2009,konig2007,hsieh2008} and superconductors~\cite{readgreen,ivanov,stoneroy,zhang2018} with topological band structures rapidly followed by the ten-fold way classification of gapped phases of non-interacting fermions~\cite{ryu2010,kitaev2009,chiu2016rmp}. Given the spatial dimension, this classification specifies the distinct topological phases that are protected by the the ten Altland-Zirnbauer symmetry classes~\cite{AZclass}. Even within the restricted setting of non-interacting fermionic systems, it is by now well-understood that the non-trivial topology of electronic bands manifests in remarkable universal phenomena, including quantized topological invariants and robust gapless boundary modes which are stable to local symmetric perturbations and disorder~\cite{hasankanermp,qizhangrmp,hasanmoore}.

More recently, the topological classification was extended to include systems with crystalline point-group symmetries, such as reflection or rotation. Generally speaking, gapped topological phases protected by such symmetries are referred to as topological crystalline phases and include as a subset higher-order topological insulators (HOTIs) and superconductors~\cite{fuTCI,hsiehTCI,okada2013,sessi2016,ma2017,schindlerHOTI,benalcazar2017,langbehnHOTI,songHOTI,khalaf2018}. In contrast to conventional topological states in $d$ spatial dimensions, which host protected gapless modes on their $d-1$ dimensional surfaces, an n$^{th}$ order topological phase in $d$ dimensions only hosts gapless modes on a $d-n$ dimensional surface and is gapped elsewhere. Such higher-order corner/hinge modes are correspondingly protected by higher-order bulk topological invariants, leading to a generalized notion of a bulk-boundary correspondence~\cite{khalaf2021botp}. Under the accepted framework for classifying quantum phases of matter, all of the above phases (topological insulators/superconductors, HOTIs, Chern insulators) fall under the broad umbrella of \textit{invertible} topological states, whose general classification includes the role of many-body correlations~\cite{chen2013,guwen2014,song2017,buildingblock,elsethorngren,else2021}. Invertible phases (a subset of which are so-called symmetry protected topological (SPT) phases) are characterized by a trivial bulk (i.e., no fractionalized excitations above the ground state), a unique ground state on arbitrary closed manifolds, and anomalous boundary modes~\cite{senthilSPT}.

Despite considerable progress in the classification of strongly correlated quantum matter, rich and interesting phenomena are still being uncovered in non-interacting electronic systems. In this chapter, we discuss the relatively new concept of ``hybrid" topology~\cite{otaki2019,jonah2020,zuo2021,hybrid1,hybrid2,kim2022,jonah2023} in a two-dimensional (2D) system. Specifically, we consider a 2D stack of dimerized one-dimensional (1D) chains that are coupled and are subjected to an external magnetic field, resulting in a model that inherits the topological features of both the celebrated 1D Su-Schrieffer-Heeger (SSH) model~\cite{ssh1979} and the 2D Hofstadter model~\cite{hofstadter1976}. The former permits a topologically non-trivial phase which is characterized by an integer bulk invariant and protected boundary modes; meanwhile, the latter describes charged particles hopping on a lattice in the presence of an external magnetic field and provides a lattice realization of the integer quantum Hall effect (IQHE), which exhibits a quantized Hall conductance and chiral edge modes. We show explicitly in this chapter that the combination of time-reversal symmetry breaking and dimerization leads to qualitatively different kinds of surface states, protected either by a non-trivial Chern number or by inversion symmetry. However, rather than focusing on these bulk invariants, in this chapter we will argue on general principles for the existence of edge states and demonstrate the presence of these states explicitly by exact diagonalization in the presence of a boundary. Our goal is to demonstrate the key non-trivial physics that emerges from these relatively simple models without invoking abstract mathematical concepts. In what follows, we first review the essential background for the SSH and the Hofstadter models, before considering the combination of the two, which will be the main focus of this chapter. 


\section{The Su-Schrieffer-Heeger Model}
\label{sec:SSH}

\begin{figure}[t]
    \centering
    \begin{subfigure}[t]{0.4\textwidth}
        \centering
        \raisebox{1.5cm}{ 
            \begin{tikzpicture}
                \node[anchor=north west] (image) at (0,-1.5) {\includegraphics[width=\textwidth]{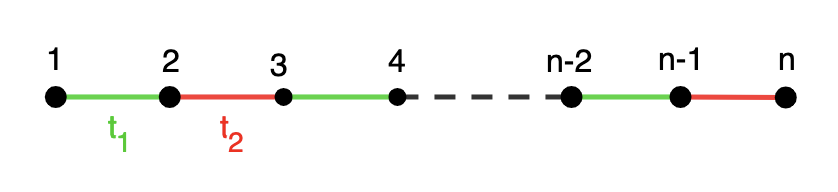}};
                \node[anchor=north west, text=black, font=\bfseries, xshift=-0.2cm, yshift=1.7cm] at (image.north west) {(a)};
            \end{tikzpicture}
        }
    \end{subfigure}
    \hspace{0.3cm}
    \begin{subfigure}[t]{0.4\textwidth}
        \centering
        \begin{tikzpicture}
            \node[anchor=north west] (image) at (0,0) {\includegraphics[width=\textwidth]{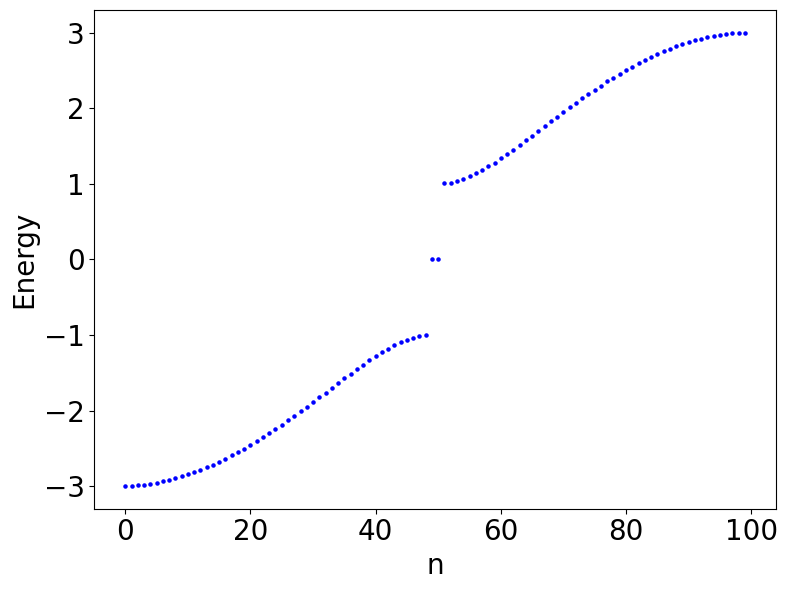}};
            \node[anchor=north west, text=black, font=\bfseries, xshift=-0.2cm, yshift=0.2cm] at (image.north west) {(b)};
        \end{tikzpicture}
    \end{subfigure}
    \vspace{-1.5\baselineskip}  
    \begin{subfigure}[t]{0.4\textwidth}
        \centering
        \begin{tikzpicture}
            \node[anchor=north west] (image) at (0,0) {\includegraphics[width=\textwidth]{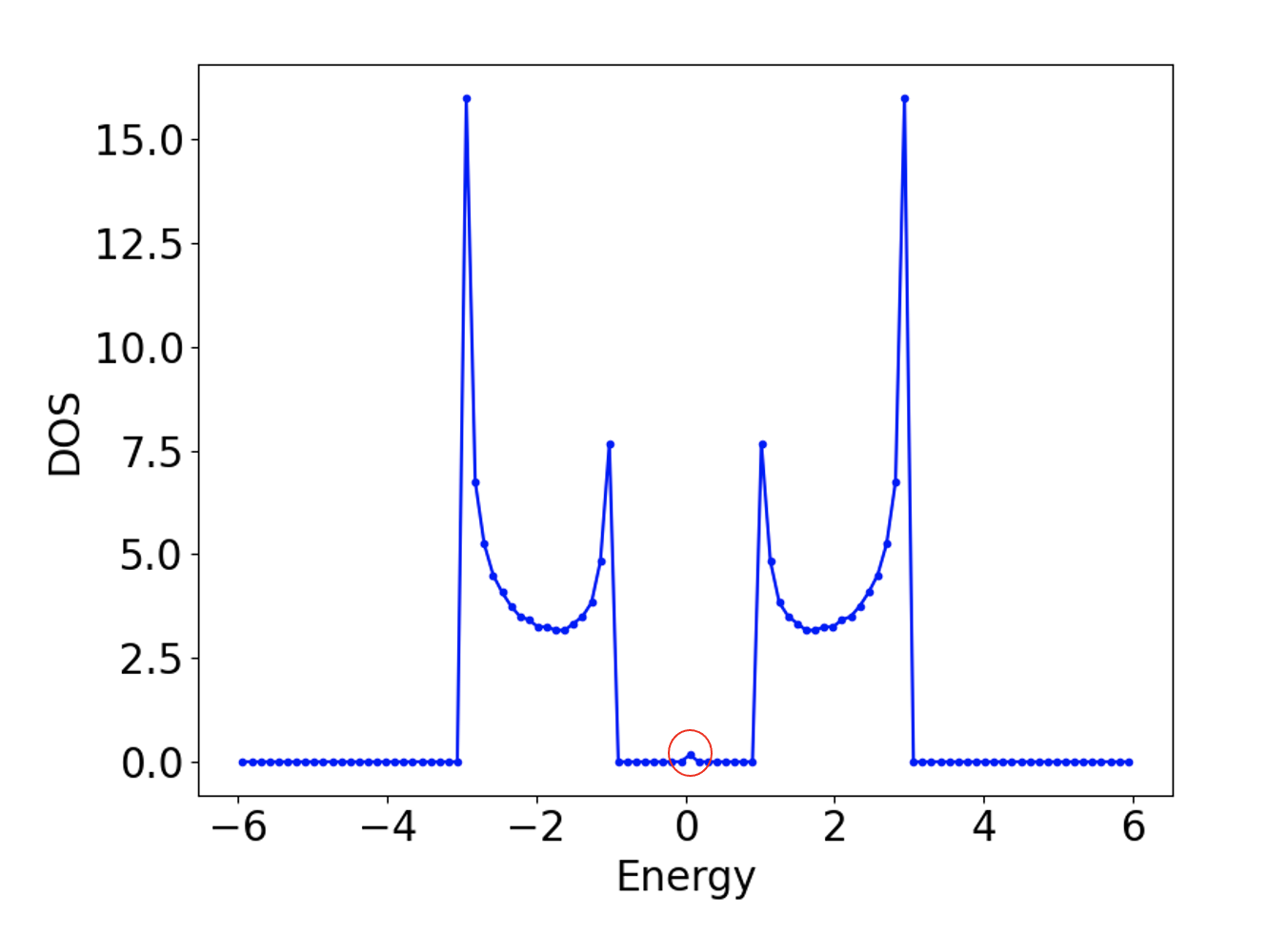}};
            \node[anchor=north west, text=black, font=\bfseries, xshift=-0.2cm, yshift=0.2cm] at (image.north west) {(c)};
        \end{tikzpicture}
    \end{subfigure}
    \hspace{0.3cm}
    \begin{subfigure}[t]{0.4\textwidth}
        \centering
        \begin{tikzpicture}
            \node[anchor=north west] (top) at (0,2.2) {\includegraphics[width=\textwidth]{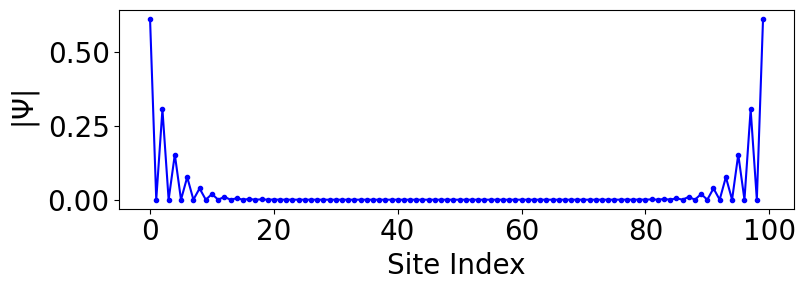}};
            \node[anchor=north west, text=black, font=\bfseries, xshift=-0.2cm, yshift=0.2cm] at (top.north west) {(d)};
            \node[anchor=north west] (bottom) at (0,0) {\includegraphics[width=\textwidth]{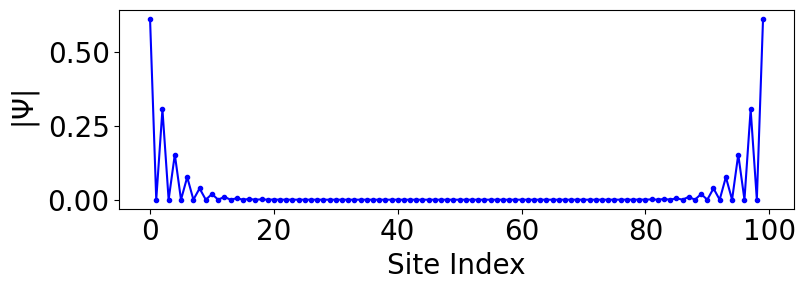}};
            \node[anchor=north west, text=black, font=\bfseries, xshift=-0.2cm, yshift=0.2cm] at (bottom.north west) {(e)};
        \end{tikzpicture}
    \end{subfigure}
    \vspace{0.5cm}
  \caption[\textbf{Electronic structure of the Su-Schrieffer-Heeger model.}]{\textbf{Electronic structure of the Su-Schrieffer-Heeger (SSH) model.} (a) Illustration of the real-space SSH model on a bipartite tight-binding chain. (b) Corresponding energy spectrum of the SSH chain for a 100-site chain with open boundary conditions (OBC). We observe two zero-energy edge states in the topological phase, whereas no zero-energy edge states are present in the trivial phase. (c) Corresponding density of states for a 2000-site chain featuring two bulk bands that are separated by a topological gap, with two degenerate surface states at E=0. (d),(e)) Magnitude of the wave functions of the two-fold degenerate zero-energy surface states, localized at the two chain edges.}  
    \label{SSH_Models}
\end{figure}
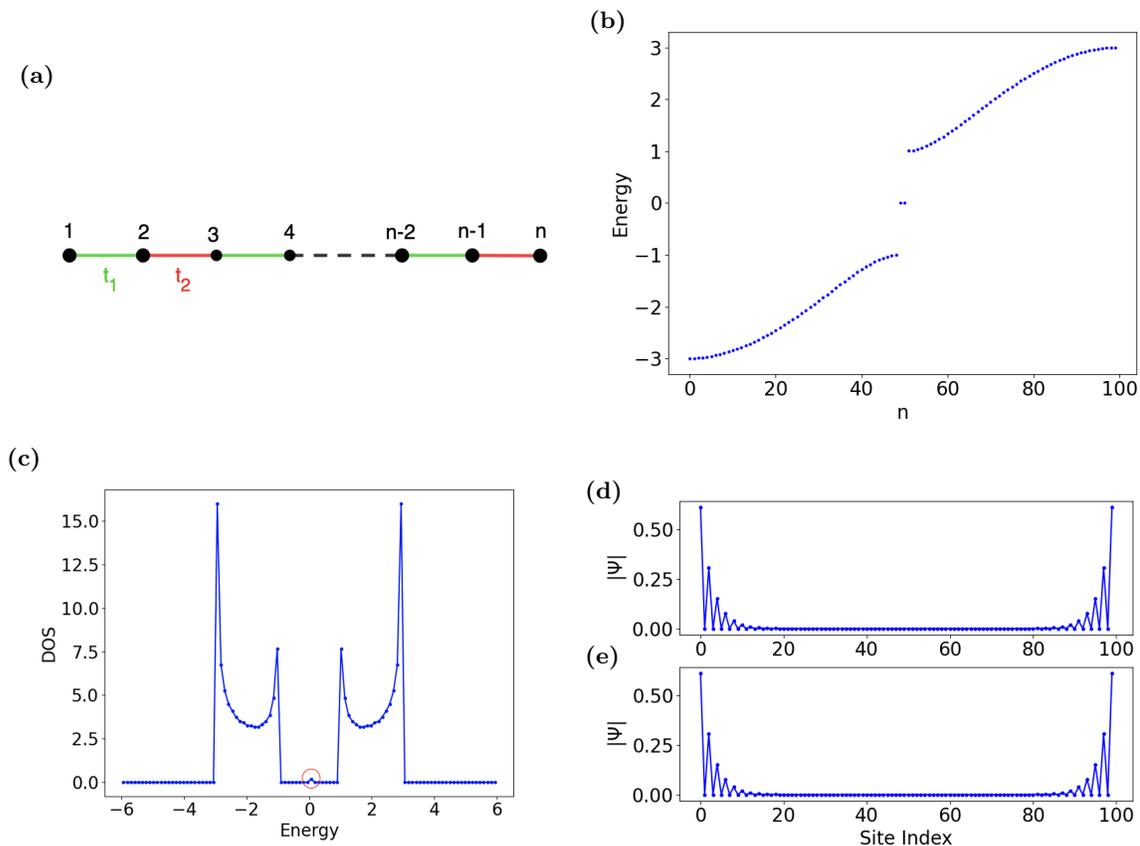

We begin by considering the one-dimensional (1D) \textbf{Su-Schrieffer-Heeger (SSH)} model~\cite{ssh1979} and briefly review its topological phase diagram. The 1D SSH model is a bipartite chain with alternating hopping parameters, as illustrated in Fig.~\ref{SSH_Models}(a). Its Hamiltonian is given by
\begin{align}
H_{SSH} =   t_1\sum_{m=1}^N (c_{m,\text{B}}^\dagger c_{m,\text{A}} + \text{H.c.})  + t_2\sum_{m=1}^{N-1} (c_{m+1,\text{A}}^\dagger c_{m,\text{B}} + \text{H.c.}),
\end{align}    
where $N$ denotes the number of unit cells, and $A$ and $B$ label the two sublattice sites within a unit cell $m$. Here, $t_1$ and $t_2$ are the intra- and inter-cell hopping parameters, respectively. In the presence of periodic boundary conditions (PBC), the Hamiltonian can be Fourier-transformed into momentum space. For each momentum $k$, we then have the single particle Hamiltonian~\cite{asboth2016short}
\begin{align}
 H_{SSH}(k) = 
 \begin{pmatrix}
  0 & h^*(k) \\
  h(k) & 0 \\
 \end{pmatrix},
\end{align}
where $h(k)=h_x(k)+ih_y(k)$, with $h_x(k) = \mathrm{Re}(t) + |t_2|\cos[ka+\arg(t_2)]$ and $h_y(k) = -\mathrm{Im}(t) + |t_2|\sin[ka+\arg(t_2)]$. Here, $a$ is the lattice spacing.

The bulk topological invariant for the SSH chain is given by the the winding number $\mathcal{W}$, which can be evaluated via the momentum space integral:
\begin{align}
\mathcal{W} = \frac{1}{2\pi i} \int_{-\pi}^{\pi}dk \frac{d}{dk} \ln [h(k)].
\end{align}
In the following, we choose real-valued hopping parameters i.e., $t_1,t_2 \in \mathbb{R}$. For $t_1>t_2$, we obtain $\mathcal{W}=0$, i.e., the system is  in a topologically trivial phase, whereas for $t_1<t_2$, the winding number evaluates to $\mathcal{W}=1$, and the system is in the topologically non-trivial phase. The phase transition occurs at $t_1=t_2$ at which point the bulk band gap closes and the winding number is ill-defined. This is consistent with the $\mathbb{Z}$ classification for 1D systems in the BDI symmetry class~\cite{kitaev2009,ryu2010}.

Due to the bulk-boundary correspondence, the non-trivial topology of the SSH model is reflected in the number of pairs of zero-energy edge states $N_\text{p-es}$ in the case of open boundary conditions (OBC). These zero-energy edge states come in pairs due to the \textit{chiral symmetry} of the SSH model, where the chiral symmetry operator is the Pauli matrix $\sigma_z$ such that
\begin{align}
    \sigma_z H_{SSH}(k) \sigma_z = - H_{SSH}(k).
\end{align}
This immediately leads to the result that for any eigenvalue $E(k)$ of $H(k)$ with eigenvector $\ket{\Psi(k)}$, there exists an orthogonal eigenvector $\sigma_z \ket{\Psi(k)}$ with eigenvalue $-E(k)$. These two states are called chiral partners; in fact, the number of zero-energy edge state pairs in the SSH model is equal to the bulk winding number i.e., $N_\text{p-es} = \mathcal{W}$.

In Fig.~\ref{SSH_Models}(b) we show the energy spectrum of the SSH chain in the topological ($t_1=0.75 < t_2 = 1.25$) phase for a system with $100$ sites. The electronic structure in the trivial phase corresponds to a gapped particle-hole symmetric insulator with $N_\text{p-es}=0$, whereas in the topologically non-trivial phase we observe $N_\text{p-es}=1$, i.e., we find a pair of degenerate zero-energy electronic states as shown in Fig.~\ref{SSH_Models}(b). Moreover, the corresponding eigenstates for these zero-energy modes are exponentially localized at the edges of the chain, as shown explicitly in Figs.~\ref{SSH_Models}(d) and (e). In this case, these two zero-energy edge states are chiral partners of each other.

\begin{figure}[b]
    \centering
    \begin{subfigure}[t]{0.3\textwidth}
        \centering
        \begin{tikzpicture}
            \node[anchor=north west] (image) at (-10,0) {\includegraphics[width=\textwidth]{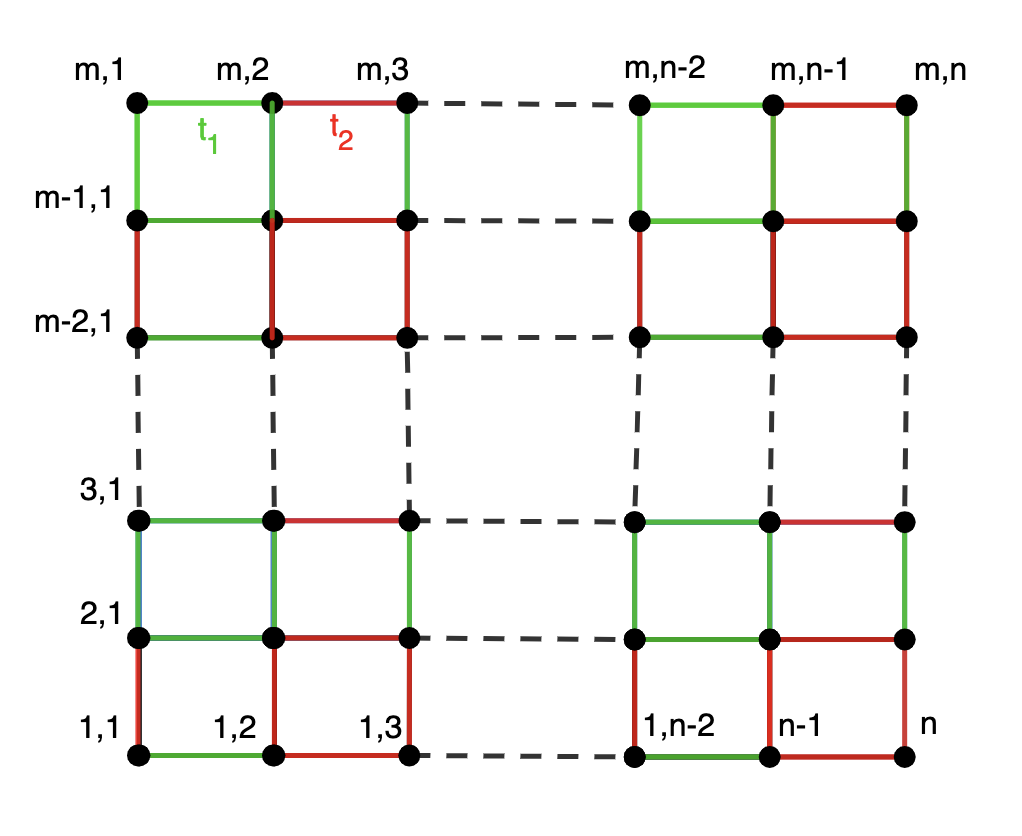}};
            \node[anchor=north west, text=black, font=\bfseries, xshift=-0.2cm, yshift=0.2cm] at (image.north west) {(a)};
        \end{tikzpicture}
    \end{subfigure}
    \hspace{0.3cm}
    \raisebox{0.3cm}{
        \begin{subfigure}[t]{0.60\textwidth}
        \centering
            \begin{tikzpicture}
                \node[anchor=north west] (image) at (0,0) {\includegraphics[width=\textwidth]{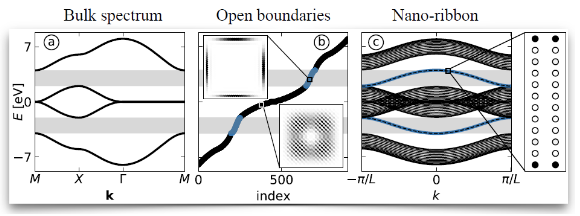}};
                \node[anchor=north west, text=black, font=\bfseries, xshift=-0.2cm, yshift=0.2cm] at (image.north west) {(b)};
            \end{tikzpicture}
        \end{subfigure}
    }
    \caption[\textbf{Two-dimensional Su-Schrieffer-Heeger model.}]{\textbf{Two-dimensional Su-Schrieffer-Heeger model.} (a) Real-space illustration of the 2D SSH model on a square lattice. There are four basis atoms in each unit cell. The intra- and inter-cell hopping parameters are $t_1$ and $t_2$ respectively in both the $x$- and $y$-directions. (b) Energy spectra for this system for different boundary conditions. For PBC in both directions, we only observe bulk states, and the momenta $k_x,k_y$ remain good quantum numbers. For OBC in both directions, we observe additional 0D and 1D topological surface states. For OBC in only one direction and PBC in the other direction, we encounter a nano-ribbon structure, with only 0D topological edge states.}
    \label{fig_2d_ssh}
\end{figure}

We now briefly review the two-dimensional (2D) extension of the SSH model (see Ref.~\cite{liu2017novel} for more details). The 2D SSH model is defined on a 2D square-lattice with alternating hopping parameters in both the $x$- and the $y$-directions, as illustrated in Fig.~\ref{fig_2d_ssh}(a). Consequently, there are four basis atoms: A, B, C, and D within a unit cell, and the intra- and inter-cell hopping parameters are again denoted by $t_1$ and $t_2$, respectively.

In contrast with conventional 2D topological insulators, the 2D SSH model has zero Berry curvature everywhere in the Brillouin zone due to the coexisting time-reversal and inversion symmetries. Hence, it is neither a Chern insulator nor is it a quantum spin Hall insulator; instead, the topological nature of this model can be characterized by the 2D Zak phase~\cite{liu2017novel}, which is given by
\begin{align}
   \bm{\gamma} = \int_\text{BZ} d\vec k \, \mathrm{Tr} \left[\braket{\psi|i\nabla_{\vec{k}}|\psi}\right] = 
	\begin{cases}
	(0,0) & \quad \text{if } t > t_2, \\
	(\pi,\pi) & \quad \text{if } t < t_2.
	\end{cases} \label{2D_Zak_phase} 
\end{align}
The topologically non-trivial phase of the 2D SSH model nevertheless displays in-gap edge and corner localized modes. In particular, 0D corner states and 1D edge states exist along the open boundaries of a square lattice, as well as on the 0D end points and 1D edges of a nano-ribbon structure. The corresponding energy spectra for these two cases are shown in Fig.~\ref{fig_2d_ssh}(b).


\section{The Hofstadter Model}
\label{sec:hofstad}

In this Section, we turn to the problem of an electron hopping on a 2D square lattice in the presence of a transverse magnetic field, which provides a non-interacting lattice model for a Chern insulator. The presence of a lattice introduces competition between the lattice spacing $a$ and the magnetic length $\ell_0$ (set by the magnetic field), with the incommensuration between these two length-scales leading to the paradigmatic Hofstadter butterfly~\cite{hofstadter1976}.

\subsection{Electron in a Magnetic Field}
\label{sec:2.1}

First, we briefly review the canonical example of a charged particle moving in a 2D plane subject to a uniform transverse magnetic field $\vec{B} = B \hat{z}$, with dynamics governed by the Hamiltonian:
    \begin{equation}
    \hat{H} = \frac{1}{2 M} \left(- i \hbar \mathbf{\nabla} + \frac{e}{c}\textbf{A} \right)^2 \, ,
    \end{equation}
where $M$ is the mass of the electron and $\textbf{A}$ is the vector potential. The net flux through the system is $\Phi = B L_1 L_2 = N_{\phi} \phi_0$, where $L_1$ and $L_2$ are the dimensions along $\hat{x}$ and $\hat{y}$ respectively, $\phi_0 = hc/e$ is the flux quantum, and $N_{\phi}$ denotes the number of flux quanta through the plane. Defining the magnetic length $\ell_0 = \sqrt{\frac{\hbar c}{e B}} = \sqrt{\frac{\phi_0}{2\pi B}}$ and working in units where $\hbar = c = e = 1$, we have
    \begin{equation}
    \label{eq:3}
   \phi_0 = 2\pi \text{ ; } N_{\phi} = \frac{L_1 L_2}{2 \pi \ell_0^2} \text{ ; } \ell_0 = \sqrt{\frac{1}{B}} \, .
    \end{equation}
It is straightforward to diagonalize this Hamiltonian (see e.g. Ref.~\cite{macdonaldrev}), whose energy levels are the degenerate Landau levels $E_n = \omega_c \left(n + \frac{1}{2} \right)$, where $n \in \mathbb{Z}_{\geq 0}$ is the Landau level (LL) index and $\omega_c = 1/(M \ell_0^2)$ is the cyclotron frequency. 

The macroscopic degeneracy of the LLs can be understood in terms of the magnetic translation algebra, as first noted by Zak~\cite{zak}. Due to the vector potential, the na\"ive translation operator $e^{i \textbf{p}\cdot\textbf{R}}$ does not commute with $\hat{H}$. We can instead introduce the generator of infinitesimal magnetic translations $\textbf{k} = \textbf{p} - \textbf{A}$, with the magnetic translation operator given by $\hat{T}(\textbf{R}) = e^{i \textbf{k}\cdot\textbf{R}}$. These operators form a projective representation of the group of magnetic translations
    \begin{equation}
    \hat{T}(\textbf{R}_1) \hat{T}(\textbf{R}_2) = \text{exp} \left(\frac{i}{2 \ell_0^2}(\textbf{R}_1\times\textbf{R}_2)\cdot\hat{z} \right)\hat{T} (\textbf{R}_1 + \textbf{R}_2)
    \end{equation}
where $\hat{z}$ is normal to the plane. The Hamiltonian in terms of the conjugate momentum operator $\mathbf{\pi} = \textbf{p} + \textbf{A}$ then takes the simple form: $\hat{H} = \mathbf{\pi}^2/(2 M)$. Working in the symmetric gauge, $A = \frac{1}{2}\textbf{B}\times\textbf{r}$, it is clear that the magnetic translation operators commute with the Hamiltonian $[k_i, \hat{H}] = 0$ but not amongst themselves since $[k_i,k_j] = -\frac{i}{\ell_0^2}\epsilon_{ij} \, (i,j = 1,2)$. Thus, we can use either $k_1$ or $k_2$ to label the degenerate LL states. To explicitly see this degeneracy, define
    \begin{equation}
    \hat{T}_x = \hat{T} \left(\frac{L_1}{N_{\phi}}\hat{x} \right) \text{ ; }
    \hat{T}_y = \hat{T} \left(\frac{L_2}{N_{\phi}}\hat{y} \right) \, ,
    \end{equation}
which satisfy the magnetic algebra
    \begin{equation}
    \hat{T}_x \hat{T}_y = e^{i \frac{2 \pi}{N_{\phi}}} \hat{T}_y \hat{T}_x \, .
    \end{equation}
(Note that this encodes the fact that the full symmetry group in the presence of a magnetic field is a central extension of translations $\mathbb{Z}^2$ by U(1) charge conservation). Now suppose $\psi$ is a simultaneous eigenstate of $\hat{H}$ and $\hat{T}_x$ such that $\hat{H} \psi = E \psi, \, \hat{T}_x \psi = t \psi$. Then, the state $\psi_m = \hat{T}_y^m \psi$ is an orthogonal eigenstate, with 
    \begin{equation}
    \hat{H} \psi_m = E \psi_m \text{\,\,;\,\,} \hat{T}_x \psi_m = t e^{i\frac{2\pi m}{N_{\phi}}} \psi_m \, ,
    \end{equation}
with the degeneracy of each LL lower bounded by the number of flux quanta $N_{\phi}$.

\subsection{Square Lattice in a Magnetic Field}
\label{sec:2.2}

We now consider the same problem as before but now regularized on a square lattice. Setting the lattice spacing $a = 1$, the Hofstadter model is defined as
    \begin{equation}
    \hat{H}_{H} = - t \sum_{<i,j>} \left[c^{\dagger}_i\,c_j\, \text{exp}\left( i\int_{\textbf{r}_i}^{\textbf{r}_j} \textbf{A}\cdot d\textbf{l} \right) + \text{h.c} \right] \, ,
    \end{equation}
where $c^{\dagger}\,,c$ are fermionic creation/annihilation operators which satisfy canonical anti-commutation relations, $i,j$ label lattice sites ($i = (m,n)$), and $\bf{A}$ encodes the uniform transverse magnetic field $B$. The lattice translation operators are 
    \begin{align}
    \hat{t}_x = \sum_{m,n} c_{m+1,n}^{\dagger}\,c_{m,n}\,\text{exp}\left(i \int_{m}^{m+1}\textbf{A}\cdot d\textbf{x} \right) \, , \, 
    \hat{t}_y = \sum_{m,n} c_{m,n+1}^{\dagger}\,c_{m,n}\,\text{exp}\left(i \int_{n}^{n+1}\textbf{A}\cdot d\textbf{y} \right) \, ,
    \end{align}
which satisfy the algebra $\hat{t}_y \hat{t}_x = e^{i \phi} \hat{t}_x \hat{t}_y$, where $\phi = B a^2$ is the uniform flux per plaquette. In terms of these operators, $\hat{H} = -t(\hat{t}_x + \hat{t}_y + \text{h.c})$, which does not commute with the na\"ive translation operators. The incommensurability between $a$ and $\ell_0$ is apparent in the translation algebra, since $\frac{\phi}{\phi_0} = \frac{a^2}{2 \pi \ell^2}$.

In analogy with the continuum, we now look for operators that commute with the Hamiltonian. Working in the Landau gauge $\textbf{A} = B (0,x,0)$, the magnetic translation operators are
    \begin{equation}
    \hat{T}_x = \sum_{m,n} c^{\dagger}_{m+1,n} c_{m,n} e^{i n \phi} \text{ ; } \hat{T}_y = \sum_{m,n} c^{\dagger}_{m,n+1} c_{m,n} \, .
    \end{equation}
One can verify that these commute with $\hat{H}$ and satisfy the magnetic translation algebra $\hat{T}_x^q \hat{T}_y = e^{i \phi q}\,\hat{T}_y \hat{T}_x^q$. Note that for rational flux per plaquette $\phi = \frac{p}{q}\phi_0$, where $p$ and $q$ are co-prime integers, the two operators $\hat{T}_x^q$ and $\hat{T}_y$ commute. As these are still translation operators--$\hat{T}_x^q$ translates a single particle state by $q$ in the x-direction and $\hat{T}_y$ by 1 in the y-direction--we can label their eigenstates with momentum quantum numbers
    \begin{align}
    \hat{T}_x^q \ket{k_x,k_y} = e^{i q k_x} \ket{k_x, k_y} \, , \, \hat{T}_y \ket{k_x,k_y} = e^{i k_y} \ket{k_x,k_y} \, .
    \end{align}
Thus, in the presence of a magnetic field and for rational $\phi/\phi_0 = p/q$, the unit cell is enlarged. Here, the magnetic unit cell is $q \times 1$ lattice plaquettes; equivalently, the wave vectors belong to the magnetic Brillouin Zone (mBZ) $k_x \in \left[-\frac{\pi}{q},\frac{\pi}{q}\right] \text{ ; } k_y \in [-\pi, \pi]$. Finally, to impose PBC on this system, we require $L_1 = r q$ where $r\in\mathcal{Z}$. The total flux through the system, $\phi_\text{total} = \phi L_1 L_2 = \phi_0 \left(r\,p\,L_1 L_2 \right)$ is thus an integer multiple of the flux quanta.

In analogy with the continuum case, we expect degenerate energy levels. Consider an eigenstate $\ket{k_x,k_y}$ of $\hat{H}$. Since $[\hat{H},\hat{T}_x] = 0$, $\hat{T}_x\ket{k_x,k_y}$ is a degenerate eigenstate; however,
    \begin{equation}
    \label{eq:21}
    \hat{T}_y \hat{T}_x \ket{k_x,k_y} = e^{-i \phi} \hat{T}_x \hat{T}_y \ket{k_x,k_y} = e^{ i(k_y - \phi)} \hat{T}_x \ket{k_x,k_y} \, .
    \end{equation}
Since $\phi = 2 \pi p/q$ (with $p,\,q$ co-prime) and as $T_x\ket{k_x,k_y} = \ket{k_x,k_y - \phi}$ has the same energy, the spectrum is at least $q$ fold degenerate.

Moving away from the square lattice, we now introduce spatial anisotropy into the system, with the Hamiltonian (in the Landau gauge) given by
    \begin{equation}
    \hat{H}_{H} = \sum_{m,n} \left[-t_a c^{\dagger}_{m+1,n}\,c_{m,n} -t_b c^\dagger_{m,n+1} c_{m,n} e^{2 \pi i \Phi m} + \text{h.c}\right]
    \end{equation}
where $\alpha = p/q$ (with $p,q$ co-prime integers) is the finite fraction of the flux quanta $\phi_0 = 2 \pi$ through each plaquette. Since we have an enlarged $q \times 1$ magnetic unit cell, we expect $q$ energy bands. To see this, we move to Fourier-space via
    \begin{equation}
c_{m,n} = \frac{1}{(2\pi)^2} \int_{-\pi}^{\pi} d\mathrm{k_x} \int_{-\pi}^{\pi} d\mathrm{k_y} e^{i (k_x m + k_y n)} \, ,
    \end{equation}
and use the magnetic translation symmetry of the system, which results in a mBZ that is $q$-times smaller in the $x$-direction:
    \begin{align}
    \hat{H}_{H} = & \frac{1}{(2\pi)^2} \int_{-\frac{\pi}{q}}^{\frac{\pi}{q}} d\mathrm{k_x} \int_{-\pi}^{\pi} d\mathrm{k_y} \sum_{m = 0}^{q - 1} \left[ -2 t_a \cos(k_x + 2 \pi \Phi m) c^{\dagger}_{m, k_y} c_{m, k_y}  \right. \nonumber \\
    & \left. - t_b \left( e^{i k_y} c^{\dagger}_{m-1,k_y} c_{m,k_y} + e^{-i k_y} c^{\dagger}_{m+1,k_y} c_{m,k_y} \right)\right] \, .
    \end{align}
To simply notations, we denote $c_{m,k_y} \equiv c_{k_x + 2\pi \Phi m, k_y}$. This is a 1D tight binding model with an enlarged unit cell and a $q \times q$ Bloch Hamiltonian with $q$ energy levels. The energy levels are obtained by considering the action of $\hat{H}_H$ on the single particle state $\ket{\psi} = \sum_{m=0}^{q-1} \psi_m c^{\dagger}_{m,k_y}$ which leads to the Harper equation~\cite{harper}
    \begin{equation}
    \label{eq:26}
    v_m \psi_{m,k_y} - t_b \left(e^{i k_y} \psi_{m+1,k_y} + e^{-ik_y}\psi_{m-1,k_y}\right) = E \psi_{m,k_y} \, ; \quad v_m = -2 t_a \cos(k_x + 2 \pi \Phi m) \, ,
    \end{equation}
subject to $\psi_{m+q,k_y} = \psi_{m,k_y}$. In the Landau gauge, all eigenstates have trivial y-dependence and we can write $\psi_{m,k_y} = e^{-i k_y m}\psi_m$, leading to
    \begin{equation}
    v_m \psi_m - t_b \left(\psi_{m+1} + \psi_{m-1} \right) = E \psi_m \, ,
    \end{equation}
subject to the constraint that $\psi_{m+q} = e^{i k_y q} \psi_m$. From this, we recover the $q$-fold degeneracy of the spectrum
    \begin{equation}
    E(k_x,k_y) = E\left(k_x,k_y + \frac{2 \pi m}{q} \right) \text{ ; } m\in \mathbb{Z} \, , \, m\in[0,q-1] \, .
    \end{equation}

\begin{figure}[t]
    \centering
    \begin{subfigure}[t]{0.4\textwidth}
        \centering
        \includegraphics[width=\textwidth]{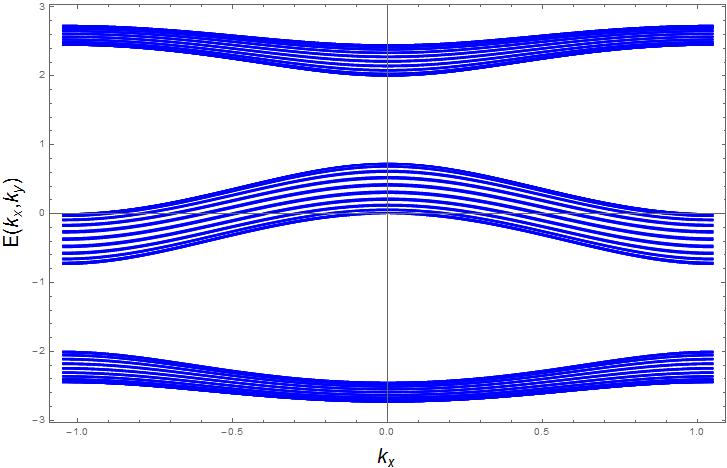}
    \end{subfigure}
    \begin{subfigure}[t]{0.4\textwidth}
        \centering
        \includegraphics[width=\textwidth]{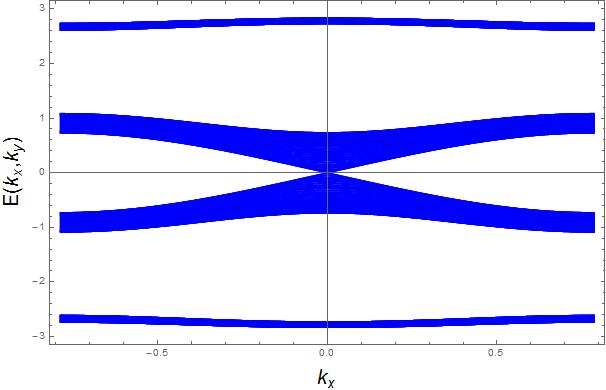}
    \end{subfigure}
    \caption{Energy spectrum at fixed $k_y$ momenta ($k_y= 0$) for the isotropic Hofstadter model ($t_a = t_b =1$) with rational flux per plaquette $\phi = p/q \phi_0$: (a) $p=2,q=3$ and (b) $p=3,q=4$. The former has a gapped spectrum while the latter exhibits degenerate Dirac nodes at $E = 0$.}
    \label{fig:hofstadbulk}
\end{figure}

In the fully anisotropic case $t_a = 1, t_b = 0$, the magnetic field does not play a role as it only enters through the hopping in the $y$-direction (in the Landau gauge). The spectrum in this case is simply given by $- 2 t \cos(k_x)$ and can be understood as the dispersion in the full BZ folded onto the magnetic BZ. For the fully isotropic case $t_a = t_b = 1$, we show the spectrum $E(k_x,k_y = 0)$ at a fixed $k_y$ in Fig.~\ref{fig:hofstadbulk} for two cases: $p=2,q=3$ and $p=3,q=4$. In general, when $q$ is even the spectrum is symmetric about $E = 0$ while for $q$ odd, the $E = 0$ level lies at the centre of the middle band. Importantly, the system has $q - 1$ gaps when $q$ is odd. Another interesting feature~\cite{wenzee} is that for $q$ even, there exist zero energy solutions when
    \begin{equation}
    2 t_a^q \cos(q k_x) + 2 t_b^q \cos(q k_y) = (-1)^\frac{q}{2} \left( t_a^q + t_b^q \right) \, ,
    \end{equation}
such the momenta at which the energy vanishes are given by $k_x = 0, k_y = 2\pi m/q$ (when $q \in 4 \mathbb{Z}$) and $k_x = \pi/q, k_y = (2m+1)\pi/q$ (when $q \in 4 \mathbb{Z} + 2$), for $m \in \mathbb{Z}, m \in [0,q-1]$. The system thus exhibits $q$ Dirac nodes which form when the bands touch around $E = 0$. In the vicinity of these nodes, the dispersion takes the relativistic form $E \propto \pm q \sqrt{t_a^q k_x^2 + t_b^q k_y^2}$, which can be seen explicitly in Fig.~\ref{fig:hofstadbulk}(b), where $q = 4$.

\begin{figure}[t]
    \centering
    \begin{subfigure}[t]{0.4\textwidth}
        \centering
        \begin{tikzpicture}
            \node[anchor=north west] (image) at (0,0) {\includegraphics[width=\textwidth]{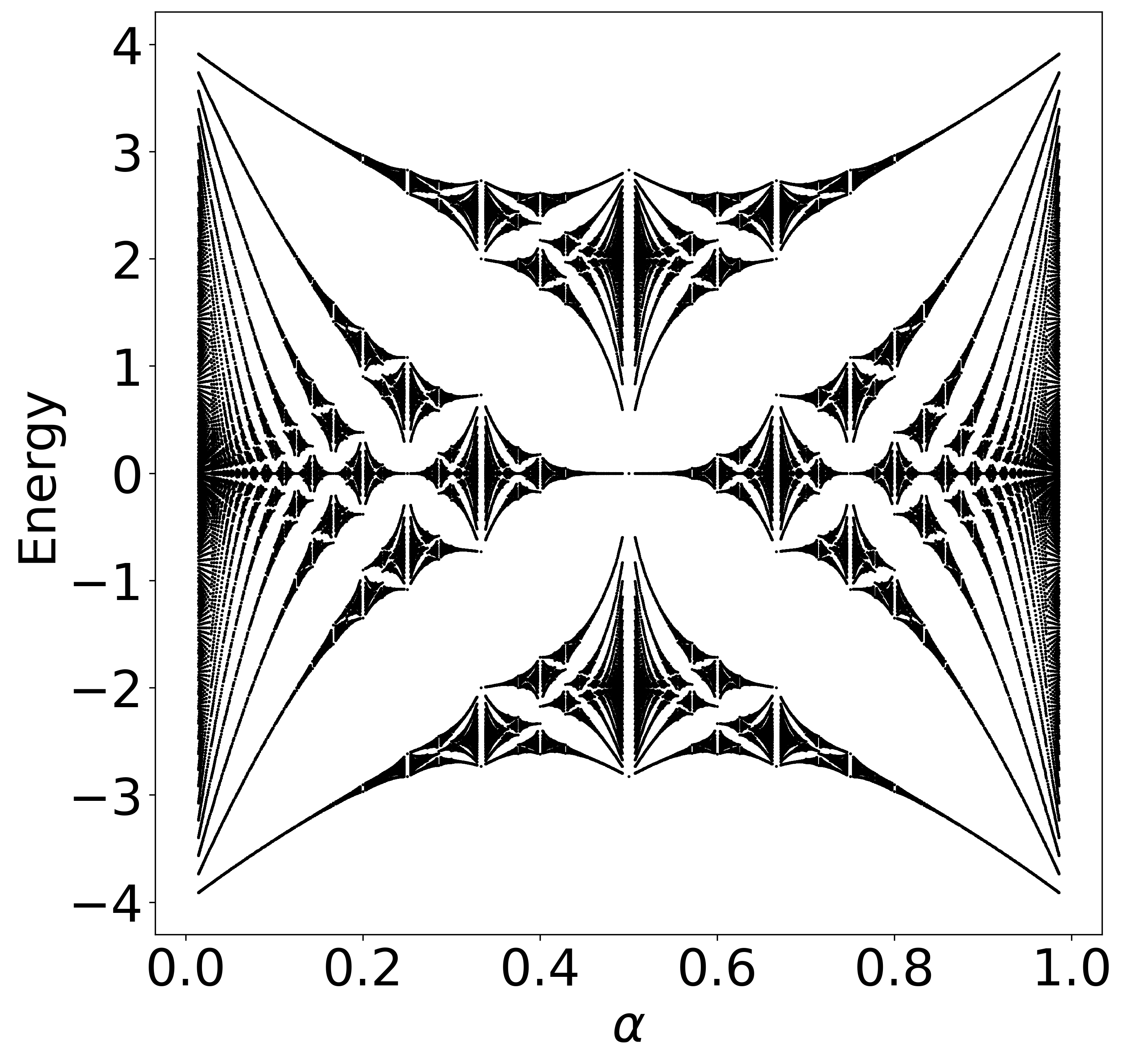}};
            \node[anchor=north west, text=black, font=\bfseries, xshift=-0.2cm, yshift=0.2cm] at (image.north west) {(a)};
        \end{tikzpicture}
    \end{subfigure}
    \hspace{0.3cm}
    \begin{subfigure}[t]{0.4\textwidth}
        \centering
        \begin{tikzpicture}
            \node[anchor=north west] (image) at (0,0) {\includegraphics[width=\textwidth]{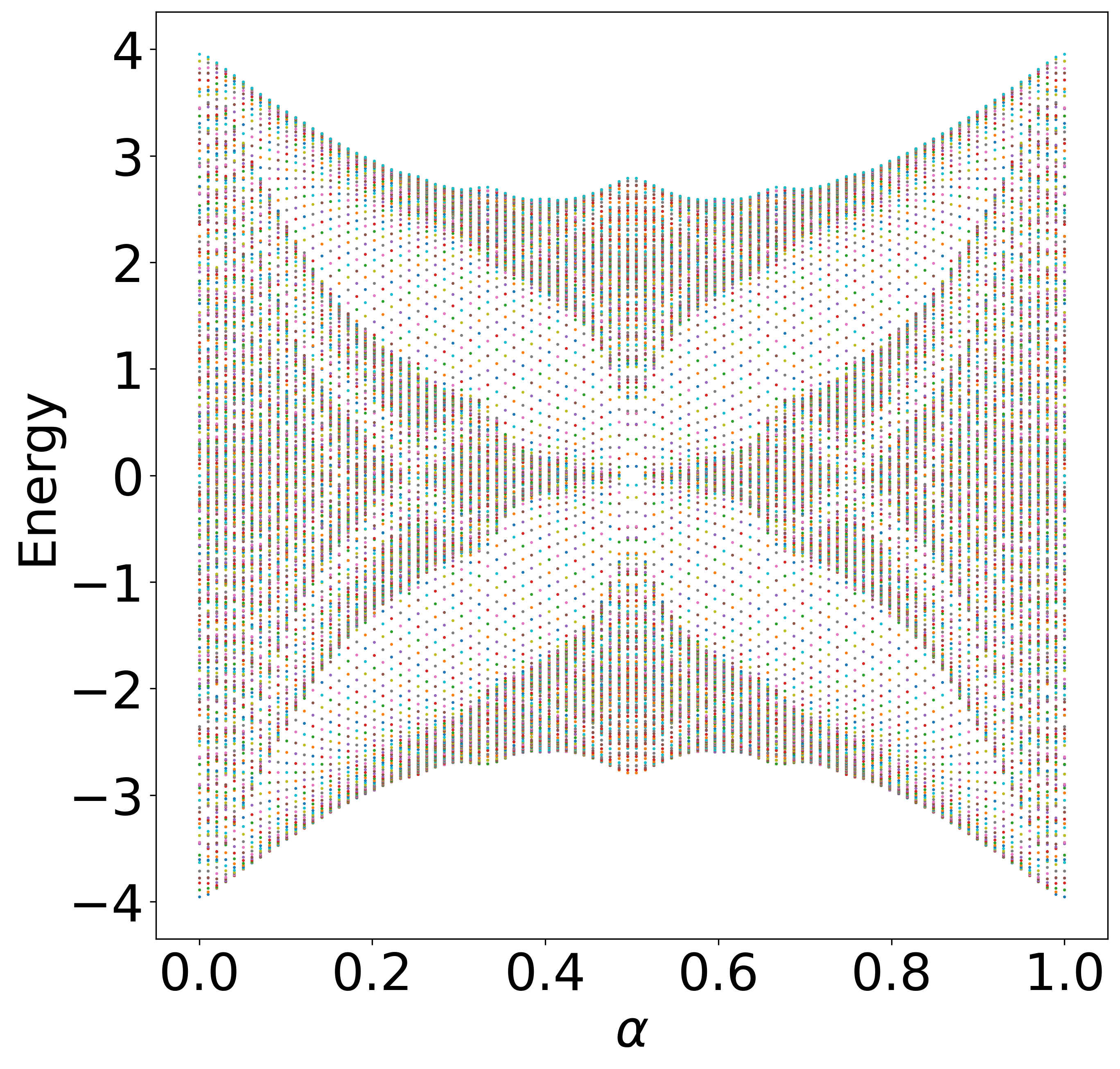}};
            \node[anchor=north west, text=black, font=\bfseries, xshift=-0.2cm, yshift=0.2cm] at (image.north west) {(b)};
        \end{tikzpicture}
    \end{subfigure}
    \caption{Hofstadter Butterfly spectrum on a $20\times20$ tight-binding square lattice: (a) using PBC and rational momenta, corresponding to magnetic flux per plaquette in integer multiples of the magnetic flux quantum; (b) using OBC in real space, leading to surface states in the otherwise "forbidden" regions. }
\label{fig:butterfly}
\end{figure}

Finally, we note that the full solution to Harper's equation, originally found by Hofstadter~\cite{hofstadter1976}, results in the famed Hofstadter butterfly which is found by plotting the energy spectrum $E$ as function of the flux through each plaquette $\alpha$. We reproduce this spectrum for both PBC and OBC in Fig.~\ref{fig:butterfly}, where the gapped regions observed in the former setting are seen to host surface states upon introducing boundaries into the system.

\subsection{Edge States on the Square Lattice}   

The Hall conductance for the Hofstadter model can be computed by evaluating the TKNN invariant~\cite{TKNN} within the gapped regions of the butterfly; as expected for a Chern insulator, the Hall conductance $\sigma_{xy} = n e^2/h$ ($n \in \mathbb{Z}$) is quantized, with different values of $n$ corresponding to different topological phases that are separated by gap closing transitions. Moreover, as expected by the bulk-boundary correspondence, a non-zero $n$ implies the existence of gapless states localized at physical boundaries of the system. This is easily understood as follows: consider a non-trivial insulator $(n \neq 0)$ which shares a boundary with a trivial insulator ($n = 0$). Since the hall conductance is a topological invariant, it can only change if the the bulk gap closes and so as long as the edge of the system respects U(1) charge conservation (required to define the Hall conductance), the only way for $\sigma_{xy}$ to change across the boundary is if there exists a state localised on the edge that crosses the Fermi level.

The original argument for the existence of edge states was given by Laughlin~\cite{Laughlin, Halperin}. Consider placing a non-trivial Chern insulator on a cylindrical geometry, through which $\Phi$ flux quanta are threaded (in the longitudinal direction). Now, suppose a voltage $V_x$ is applied transverse to $k_y$. Since the adiabatic insertion of a flux is the same as adiabatically switching on a vector potential, the flux shifts the momentum $k_y \to k_y + 2 \pi \frac{\Phi}{L}$, such that $I_y = \Delta E/\Delta \Phi$ (via the Byers-Yang formula), where $\Delta E$ is the change in energy during the process of flux insertion. Suppose that the Fermi level crosses some bands and that precisely one flux quanta is adiabatically threaded through the cylinder. The momentum of all states then changes by precisely $\frac{2\pi}{L_y}$, which (due to PBC) is the spacing between momentum states. Thus, for all of the bands that the Fermi level crosses, there will be one state occupied above the Fermi level close to one edge and one state empty below the Fermi level close to the opposite edge. If the Fermi level crosses $n$ bands, this implies that $\Delta E = n V_x$, since the two edges have a potential difference of $V_x$. Thus, $\sigma_{xy} = \frac{I_y}{V_x} = n$. Since this argument depends only on gauge invariance, the result is robust even in the presence of disorder.

\begin{figure}[t]
    \centering
    \begin{subfigure}[t]{0.4\textwidth}
        \centering
        \includegraphics[width=\textwidth]{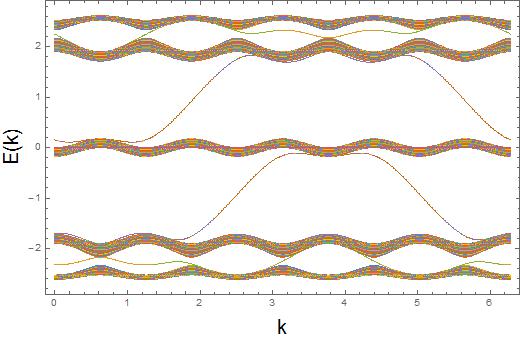}
    \end{subfigure}
    \hspace{0.3cm}
    \begin{subfigure}[t]{0.4\textwidth}
        \centering
        \includegraphics[width=\textwidth]{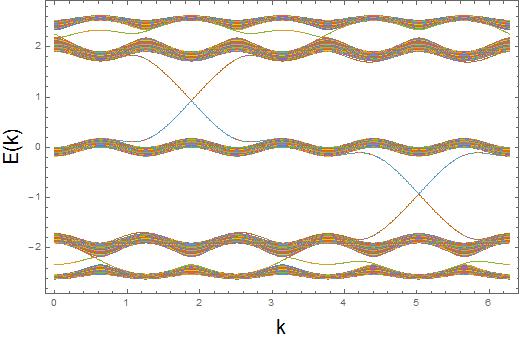}
    \end{subfigure}
    \caption{Edge states in the Hofstadter model on a cylindrical geometry. Parameters here are $t_a = t_b = 1$, $p = 2$, and $q = 5$. On the left, $L_x = 120$ (commensurate with $q$), while on the right $L_x = 121$ (incommensurate with $q$).}
    \label{fig:hofstadedge}
\end{figure}
For concreteness, consider again the Hofstadter model. We consider a cylindrical geometry, where we impose PBC in the $y$-direction (with length $L_y$) but OBC in the $x$-direction. The two edges are fixed at $x = 0$ and $x = L_x$, such that the system has $L_x - 1$ sites along the $x$-direction. We then assume that $\Phi$ flux quanta are threaded through this cylinder. In a slight abuse of notation, we now call the fraction of flux per plaquette $\phi$ ($\phi = \frac{B a^2}{\phi_0}$ where B is the magnetic field). The Hamiltonian, again in the Landau gauge ($A_y = B \hat{x}$) is
    \begin{equation}
    \hat{H}_{H} = -\sum_{m,n} \left[t_a  c^{\dagger}_{m+1,n} c_{m,n} + t_b e^{2 \pi i \frac{\Phi}{L_y}} c^{\dagger}_{m,n+1} c_{m,n}e^{2 \pi i \phi m} + \text{h.c} \right] \, .
    \end{equation}
Due to the finite geometry, we no longer have translation invariance in the $x$-direction and hence $k_x$ is no longer a good quantum number. Fourier transforming in the $y$-direction, $c_{m,n} = \frac{1}{\sqrt{L_y}}\sum_{k}e^{i k_y n} c_m(k)$, with $k = \frac{2 \pi m}{L_y}$ ($m = 1, 2, \dots , L_y$), we find
    \begin{equation}
    \hat{H}_{H} = -\sum_{m,k}\left[ t_a c^{\dagger}_{m+1,k} c_{m,k} + t_b e^{2 \pi i \frac{\Phi}{L_y}} e^{i(2 \pi \phi m - k)} c^{\dagger}_{m,k} c_{m,k} \text{h.c} \right] \, .
    \end{equation}
The action of this Hamiltonian on a single particle state, $\ket{\Psi(k,\Phi)} = \sum_m \psi_m(k,\Phi)$ results in
    \begin{equation}
    -t_a(\psi_{m+1} + \psi_{m-1}) - 2 t_b \cos\left(k - 2 \pi \frac{\Phi}{L_y} - 2 \pi \phi m \right) \psi_m = E \psi_m \, .
    \end{equation}
This equation can equivalently be written in matrix form and studied within the transfer matrix formalism developed by Hatsugai~\cite{Hatsugai}. While analytic results, such as the existence of edge modes, can be obtained in the case where $L_x$ is commensurate with q (i.e., $L_x = r q$ with $r \in \mathbb{Z}$), in general this problem can only be studied numerically. Here, we exactly diagonalize the preceding 1D Hamiltonian and present the numerical results in Fig.~\ref{fig:hofstadedge} for the isotropic case with $p=2,q=5$. In each case, the edge states connect the different bulk bands and are localized on the $x = 0$ and $x = L_x$ edges. Note that while the bulk states are indifferent to commensurability and the Hall conductance is the same in either case, the nature of the edge states is different~\cite{Hatsugai}.


\section{Hofstadter Butterflies in Coupled SSH Chains}
\label{sec:hybrid}

\begin{figure}[t]
    \centering
    \includegraphics[width=0.5\linewidth]{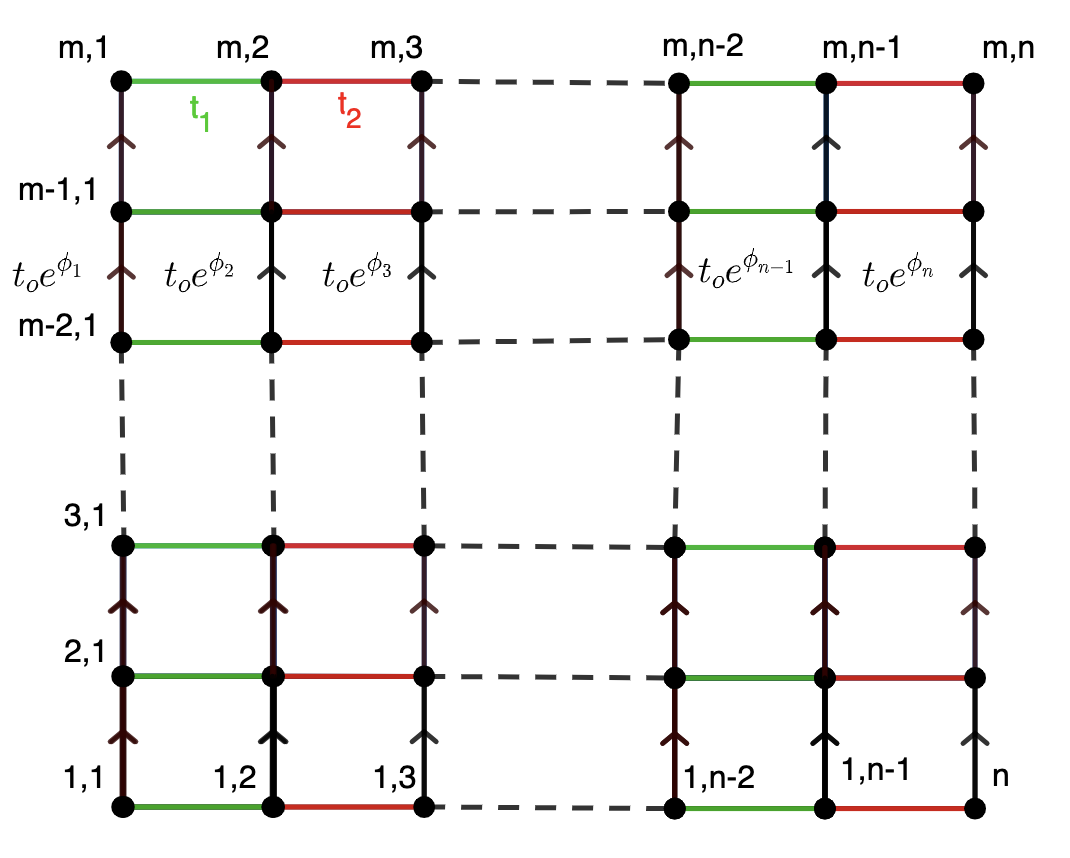}
    \caption{Illustration of an m-leg SSH ladder of chains with n sites. The horizontal intra-chain couplings, $t_1$ and $t_2$, have alternating strength. The vertical inter-chain couplings, $t_0\exp{(i2\pi \alpha x)}$, where $x$ is the coordinate in the horizontal direction, are modulated by the external magnetic field, parameterized by $\alpha = B a^2 e/h$.}
    \label{fig:mlegladder}
\end{figure}

In this Section, we investigate a hybrid model in which features of the 1D SSH model coexist with those of the Hofstadter model. We note that analogous models have previously been considered in Refs.~\cite{lau2015,otaki2019,zuo2021,he2022}. Specifically, we consider a stack of 1D SSH chains which are coupled together to form a 2D lattice that is further subjected to a uniform transverse magnetic field. The underlying Hamiltonian for this system is given by 
\begin{equation}
H = 
-t_0\sum_{i,j} \left( e^{i\phi_i}c^{\dagger}_{i,j}c_{i,j+1}+h.c.\right)
-t_1\sum_{i,j}\left(c^{\dagger}_{2i-1,j}c_{2i,j}+h.c.\right)
-t_2\sum_{i,j}\left(c^{\dagger}_{2i,j}c_{2i+1,j}+h.c.\right),
\end{equation}
where the site positions $(i,j)$ are defined on a 2D lattice of 
$m$ coupled chains with $n$ sites each, i.e., we are considering $m$-leg ladders. The horizontal intra-chain coupling is staggered in accordance with the SSH Hamiltonian, whereas the vertical inter-chain coupling is modulated by the flux to incorporate the external magnetic field, in accordance with the Hofstadter Hamiltonian (see Fig.~\ref{fig:mlegladder} for an illustration). Since our focus will be on topological surface states, we will primarily consider systems with open boundary conditions (OBC).

To get an intuition for this model, consider the case where no magnetic field is switched on. Then, we do not expect a non-trivial Chern number in this system, which we have also verified explicitly. Moreover, if we place the system on a cylinder, with PBC along the $y$-direction and OBC along the $x$-direction, we expect topological surface states that are inherited from the underlying SSH chains when $t_2>t_1$ and no such states when $t_2<t_2$. The coupling between the chains then delocalizes these 0D states along the edge, leading to topologically protected modes that disperse along the 1D surface. In fact, it is straightforward to analyze the model on this geometry and show that the edge states have a dispersion $\sim t_0 \cos(k_y)$ and that they are protected by an effectively 1D winding invariant $\mathcal{W}(k_y)$ (evaluated at fixed $k_y)$ that is non-trivial when $t_2 > t_1$. Once we turn on the external magnetic field, the system can now additionally host a non-trivial Chern number (stricitly speaking, this will only be quantized in the 2D limit i.e., as we take the number of legs $m$ to be large) which can be calculated via the TKNN invariant~\cite{TKNN}. Thus, we expect the distinct phases of this system to be characterized by two independent topological invariants: the Chern number and an effectively 1D winding invariant. To evaluate the latter, the system is placed on a cylinder geometry with PBC along $y$, resulting in an effectively 1D Hamiltonian parameterized by $k_y$. Depending on the magnetic flux, this effective 1D system may preserve inversion symmetry ($\mathcal{I} H(k) \mathcal{I}^{-1}= H(-k)$) for various values of $k_y$~\cite{lau2015}, and can hence lead to an integer invariant and corresponding inversion symmetry protected edge modes.

Let us first consider the minimal system in which we may observe the interplay between dimerization and the magnetic field i.e., the 2-leg ladder, since a single chain does not offer any plaquettes that can accommodate non-trivial magnetic flux. In the absence of a magnetic field ($\phi = 0$), the phase diagram of this hybrid model on a two-leg ladder consists of a gapped, topologically trivial phase (when $t_1>t_2$, $t_1+t_2 < t_0$ and $t_0 < t_1-t_2$), a gapless critical region (when $|t_1+t_2| > t_0 > |t_1-t_2|$), and a gapped topologically non-trivial dimerized phase (when $t_2>t_1$, $t_2+t_1 < t_0$ and $t_0 < t_2-t_1$). The latter phase inherits its non-trivial topology from the underlying SSH chain and is characterized by localized topological surface states whose eigenenergies lie within the bulk gap. Since this system is still quasi-1D, we do not expect a sharply quantized Chern number. 

\begin{figure}[h!]
    \centering
    \begin{subfigure}[t]{0.4\textwidth}
        \centering
        \begin{tikzpicture}
            \node[anchor=north west] (image) at (0,0) {\includegraphics[width=\textwidth]{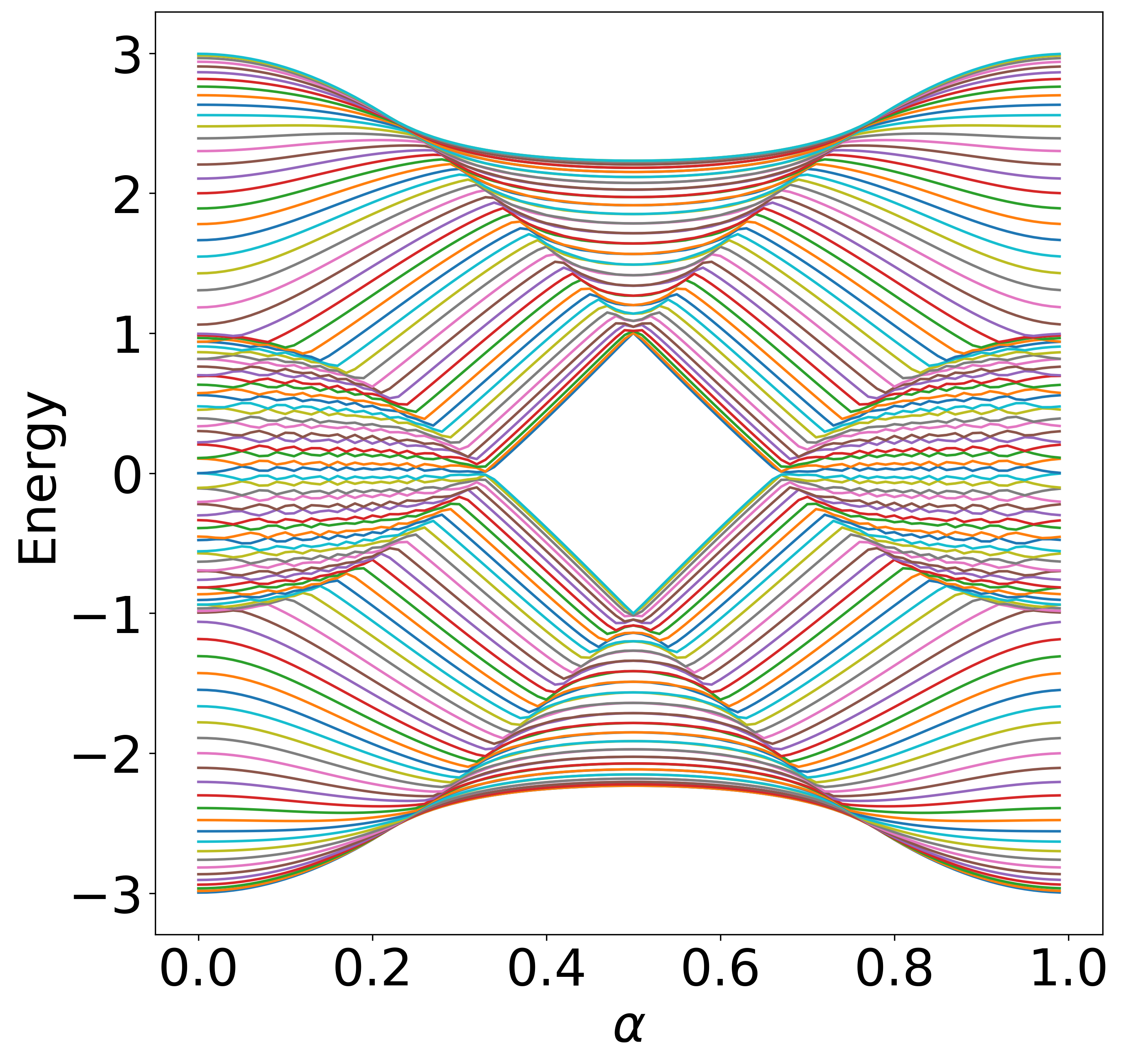}};
            \node[anchor=north west, text=black, font=\bfseries, xshift=-0.2cm, yshift=0.2cm] at (image.north west) {(a)};
        \end{tikzpicture}
    \end{subfigure}
    \hspace{0.3cm}
    \begin{subfigure}[t]{0.4\textwidth}
        \centering
        \begin{tikzpicture}
            \node[anchor=north west] (image) at (0,0) {\includegraphics[width=\textwidth]{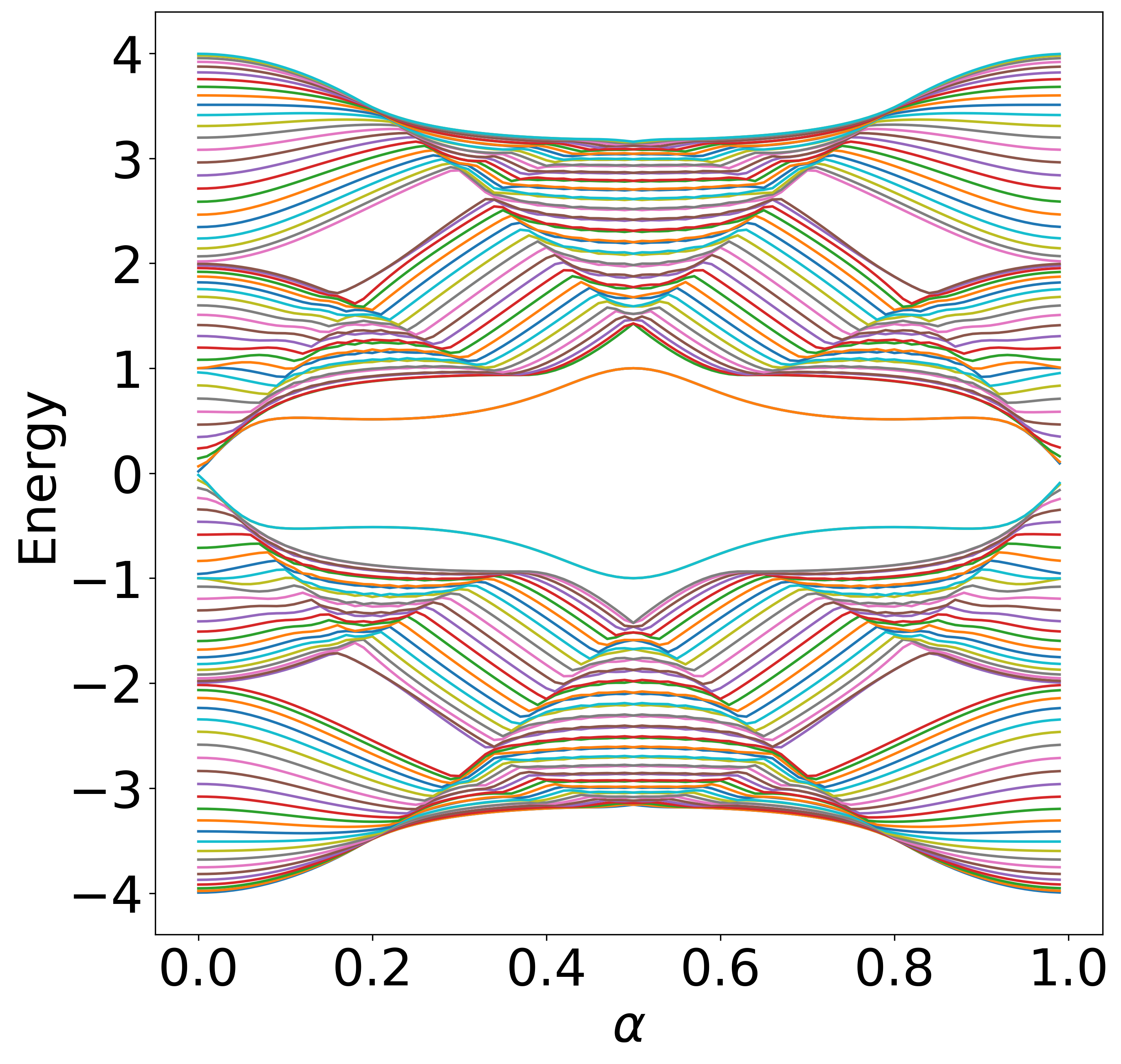}};
            \node[anchor=north west, text=black, font=\bfseries, xshift=-0.2cm, yshift=0.2cm] at (image.north west) {(b)};
        \end{tikzpicture}
    \end{subfigure}
    \vspace{-1.5\baselineskip}  
    \begin{subfigure}[t]{0.4\textwidth}
        \centering
        \begin{tikzpicture}
            \node[anchor=north west] (image) at (0,0) {\includegraphics[width=\textwidth]{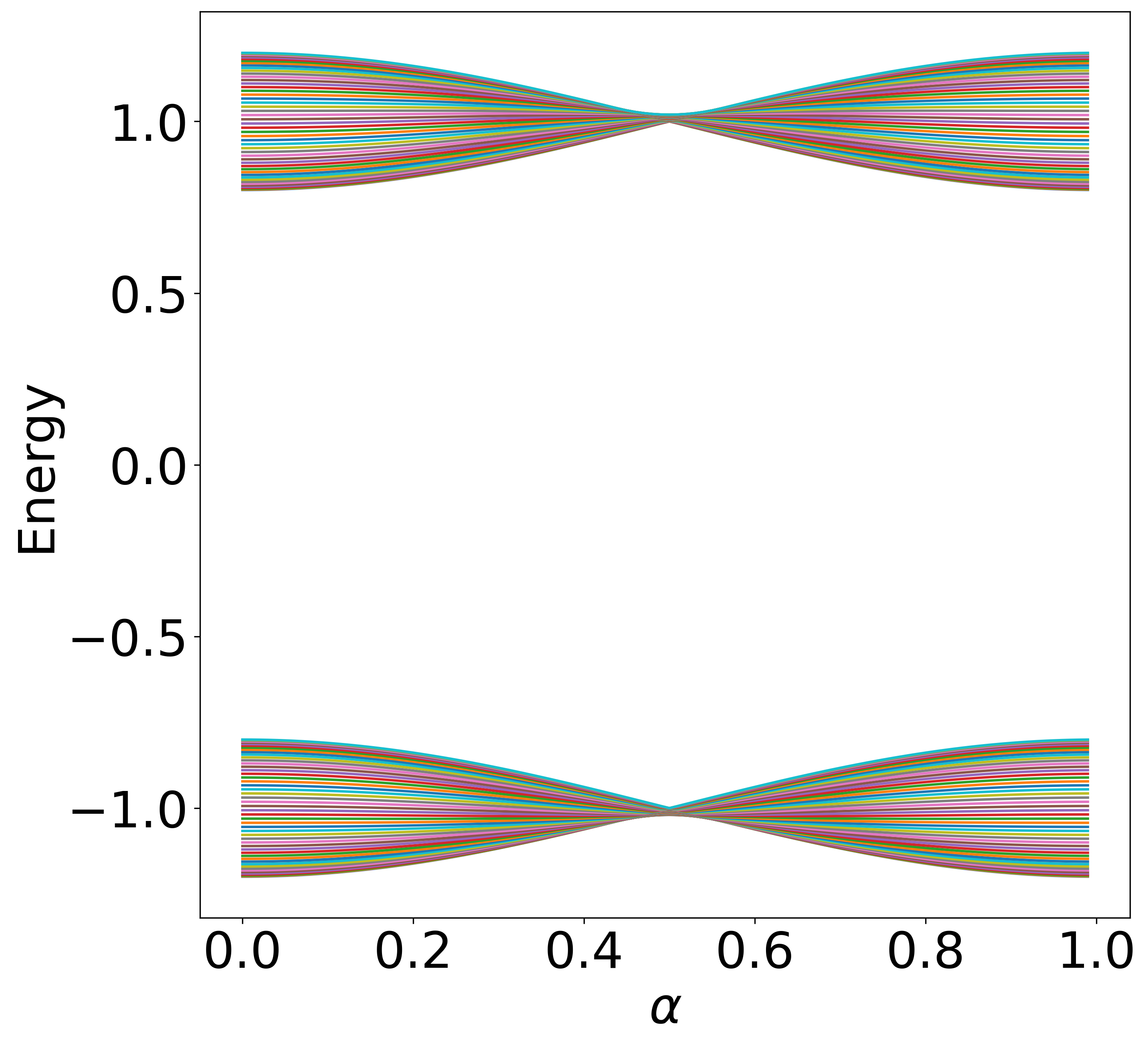}};
            \node[anchor=north west, text=black, font=\bfseries, xshift=-0.2cm, yshift=0.2cm] at (image.north west) {(c)};
        \end{tikzpicture}
    \end{subfigure}
    \hspace{0.3cm}
    \begin{subfigure}[t]{0.4\textwidth}
        \centering
        \begin{tikzpicture}
            \node[anchor=north west] (image) at (0,0) {\includegraphics[width=\textwidth]{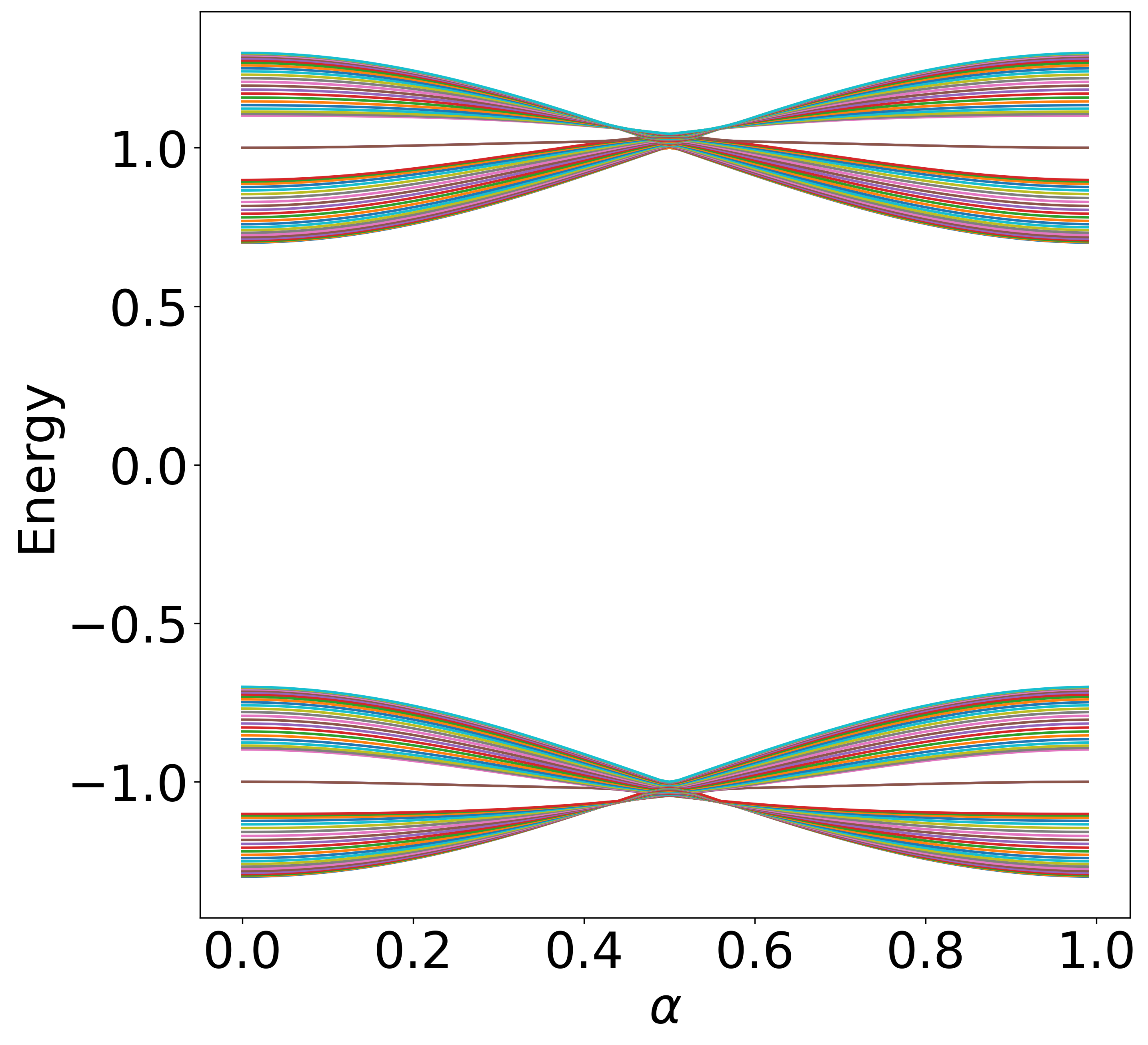}};
            \node[anchor=north west, text=black, font=\bfseries, xshift=-0.2cm, yshift=0.2cm] at (image.north west) {(d)};
        \end{tikzpicture}
    \end{subfigure}
    
    \vspace{1.5\baselineskip}  
    \begin{subfigure}[t]{0.4\textwidth}
        \centering
        \begin{tikzpicture}
            \node[anchor=north west] (image) at (0,0) {\includegraphics[width=\textwidth]{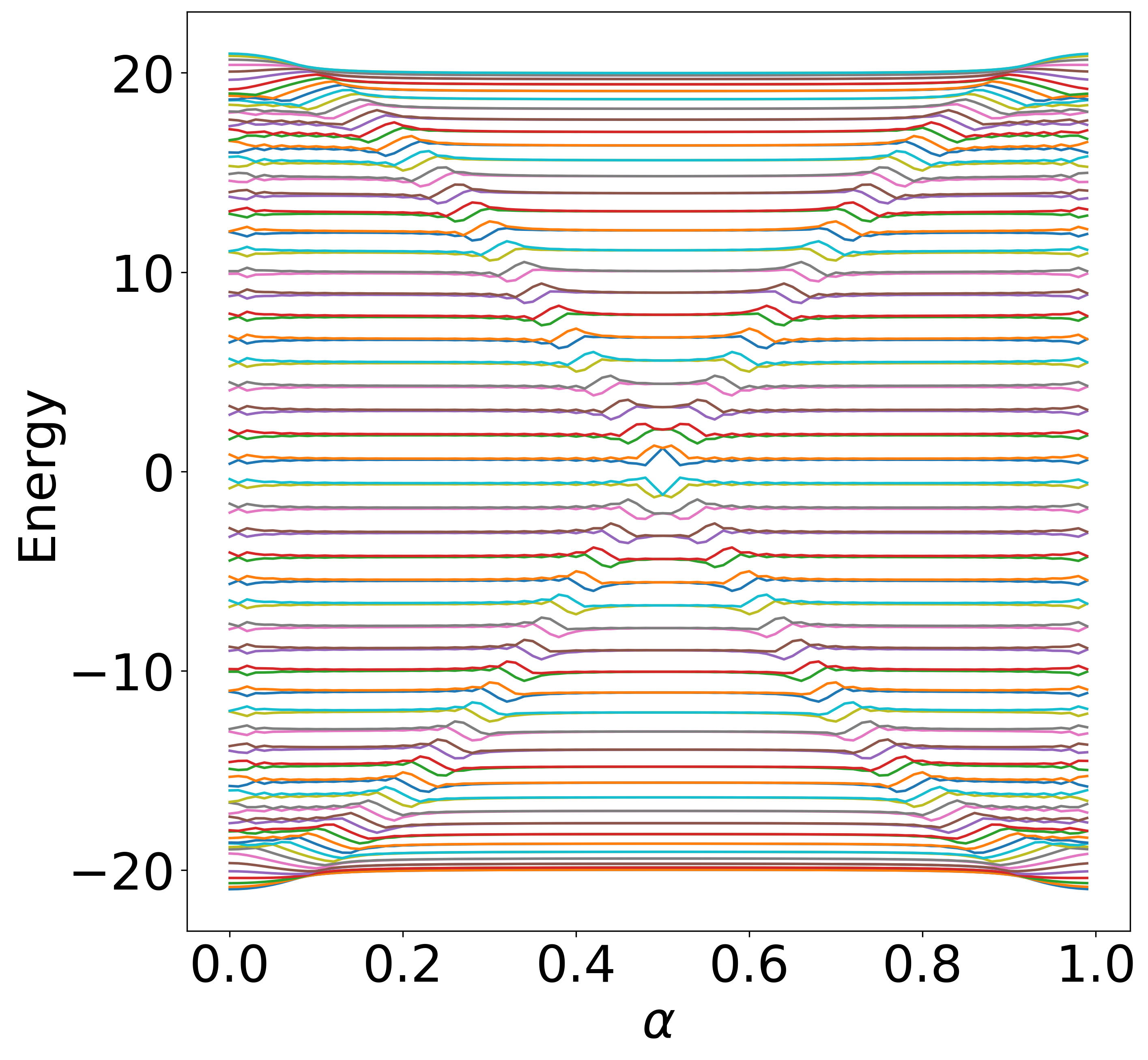}};
            \node[anchor=north west, text=black, font=\bfseries, xshift=-0.2cm, yshift=0.2cm] at (image.north west) {(e)};
        \end{tikzpicture}
    \end{subfigure}
    \hspace{0.3cm}
    \begin{subfigure}[t]{0.4\textwidth}
        \centering
        \begin{tikzpicture}
            \node[anchor=north west] (image) at (0,0) {\includegraphics[width=\textwidth]{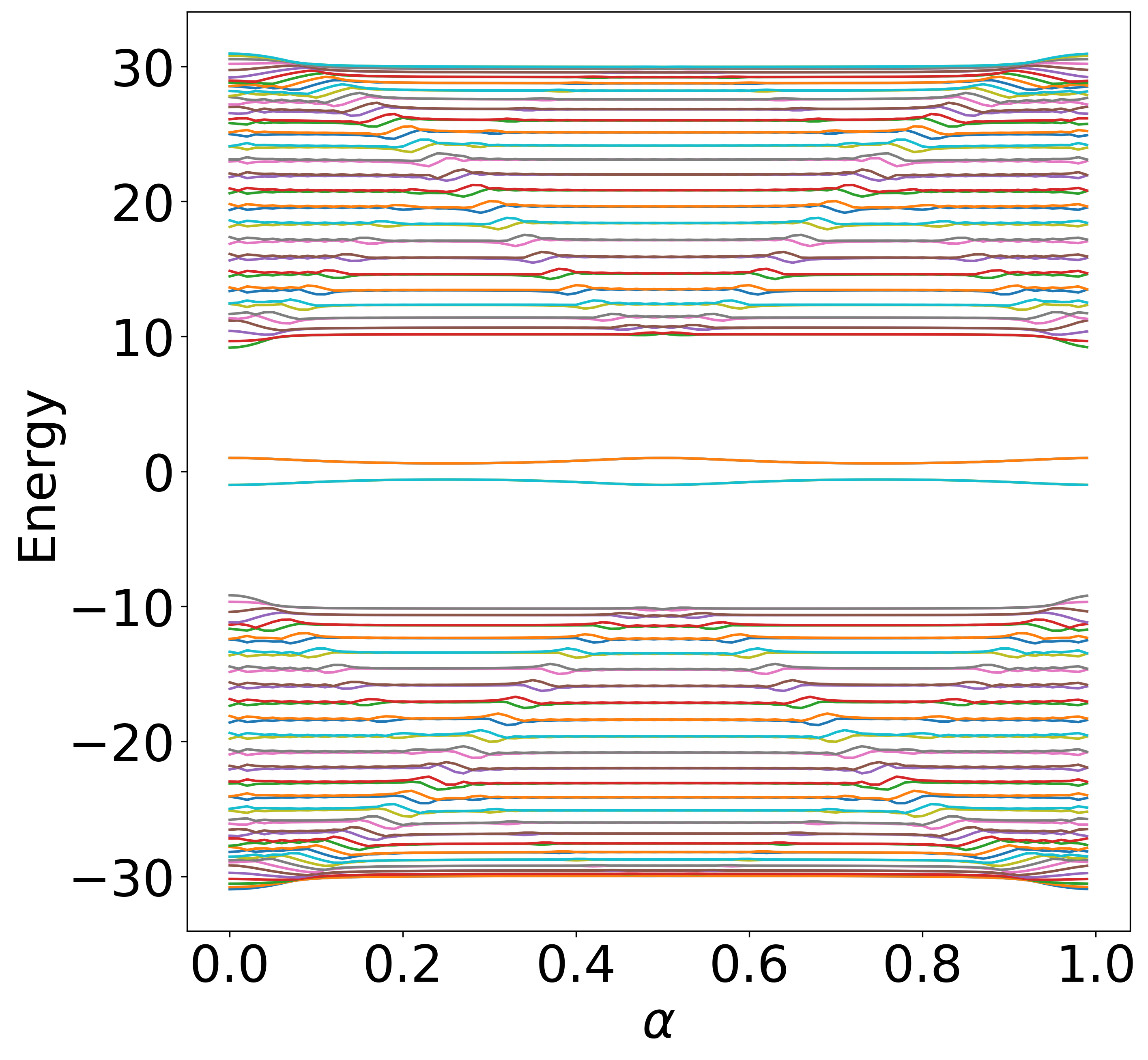}};
            \node[anchor=north west, text=black, font=\bfseries, xshift=-0.2cm, yshift=0.2cm] at (image.north west) {(f)};
        \end{tikzpicture}
    \end{subfigure}
    \caption{Hofstadter butterfly spectra of the two-leg SSH ladder. (a) Critical point - isotropic limit, $t_0=t_1=t_2=1$; (b) Topologically non-trivial dimerized phase with four two-fold degenerate surface states, $t_2=2$, $t_1=t_0=1$; (c) Trivially dimerized phase without surface states, dominant inter-chain coupling, $t_1=0.2$, $t_2=0.1$, $t_0=1$; (d) Topologically non-trivial dimerized phase, dominant inter-chain coupling, $t_1=0.1$, $t_2=0.2$, $t_0=1$;
    (e) Trivially dimerized phase without surface states, dominant intra-chain couplings, $t_1=20$, $t_2=10$, $t_0=1$; (f) Topologically non-trivial dimerized phase, dominant intra-chain couplings, $t_1=10$, $t_2=20$, $t_0=1$.}
        \label{fig:2legladder}
\end{figure}
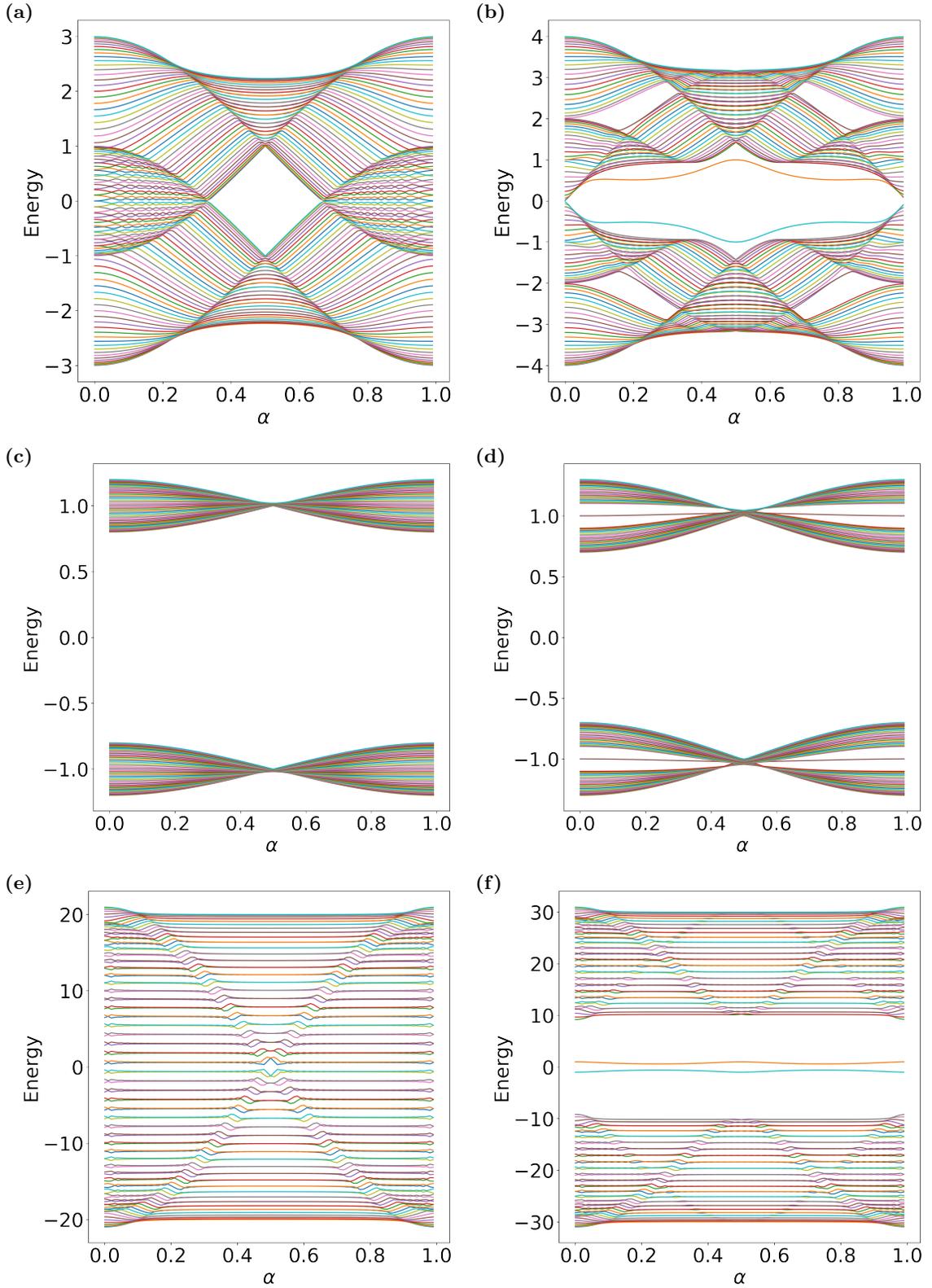

We now numerically investigate the Hofstadter butterfly spectrum in the two-leg SSH ladder, shown in Fig.~\ref{fig:2legladder} for system dimensions $m=2$ and $n=50$ with OBC. In the upper panel, we compare the energy spectrum of the critical, isotropic system ($t_0=t_1=t_2=1$) in Fig.~\ref{fig:2legladder}(a) with that of the topologically non-trivial dimerized phase ($t_0=t_1=1, t_2=2$) in Fig.~\ref{fig:2legladder}(b). In both cases, the overall bandwidth of the spectrum is given by $W=t_0+2(t_1+t_2)$, and the latter clearly displays bulk energy bands separated by an energy gap. In addition, in the topologically non-trivial dimerized phase we observe a pair of two-fold degenerate states that lie within the bulk gap and which are spatially localized at the two open ends of the ladder system. 

This phenomenology persists into the regime where the inter-chain coupling dominates, as shown in Figs.~\ref{fig:2legladder}(c) and (d). In the trivially dimerized phase ($t_0=1,t_1=0.2,t_2=0.1$) there are two bulk bands, but no surface states within the energy gap separating them (see Fig.~\ref{fig:2legladder}(c)). In contrast in the topologically non-trivial dimerized phase ($t_0=1,t_1=0.1,t_2=0.2$), additional gaps open up within the bulk bands, with two-fold degenerate non-dispersive topological surface states appearing in the center of these gaps at energies ~$\pm t_0$. 

The situation is different in the regime of dominant intra-chain interactions. As seen in  Figs.~\ref{fig:2legladder}(e) and (f), the spectrum is quasi-continuous in the trivially dimerized phase, whereas in the non-trivial phase a central gap emerges, again with a pair of two-fold degenerate surface states at energies ~$\pm t_0$. Also, note that for the 2-leg SSH ladder, all of the surface states are zero-dimensional and are located at the corners of the system. 

\begin{figure}[h!]
    \centering
    \begin{subfigure}[t]{0.4\textwidth}
        \centering
        \begin{tikzpicture}
            \node[anchor=north west] (image) at (0,0) {\includegraphics[width=\textwidth]{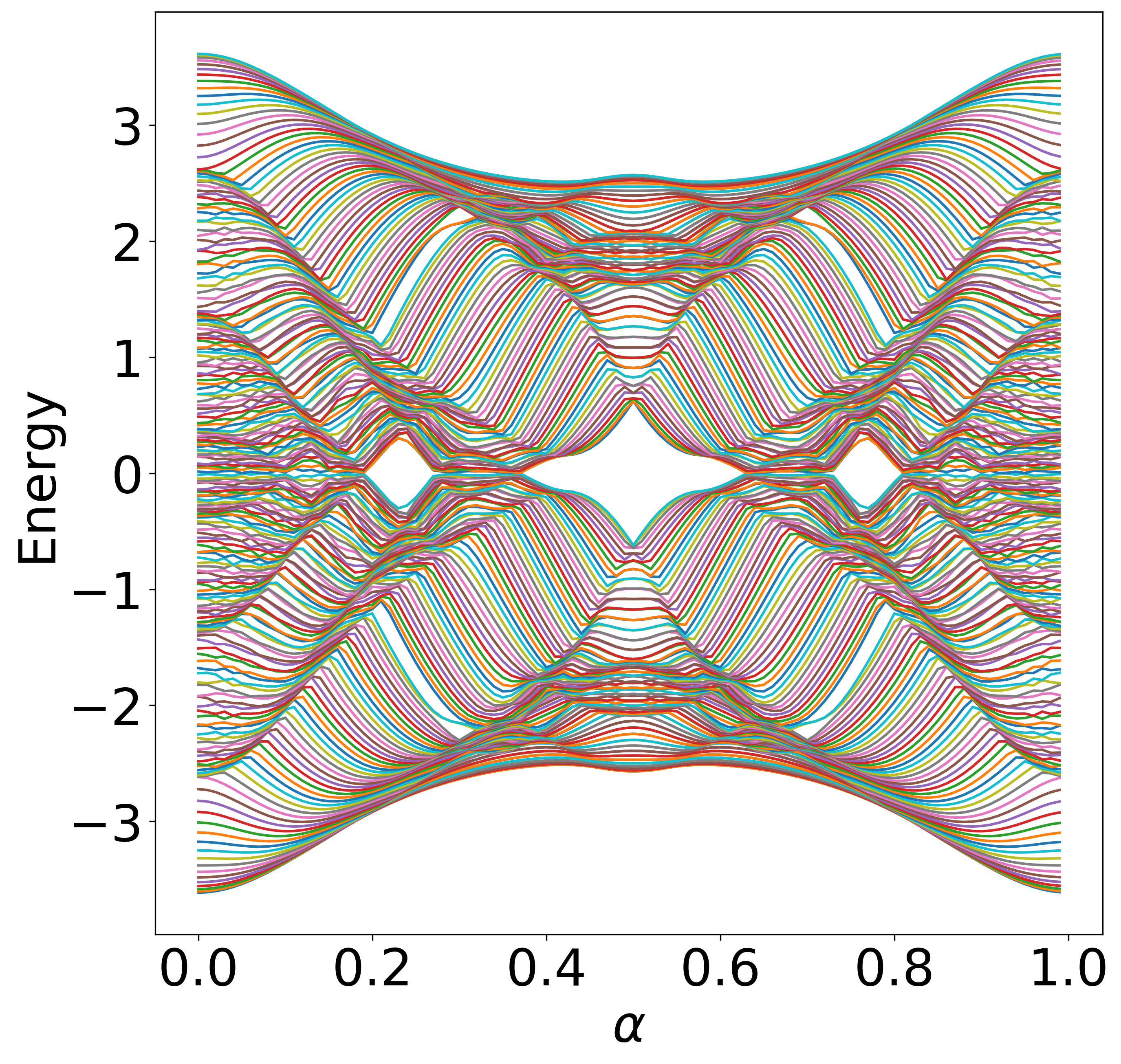}};
            \node[anchor=north west, text=black, font=\bfseries, xshift=-0.2cm, yshift=0.2cm] at (image.north west) {(a)};
        \end{tikzpicture}
    \end{subfigure}
    \hspace{0.3cm}
    \begin{subfigure}[t]{0.4\textwidth}
        \centering
        \begin{tikzpicture}
            \node[anchor=north west] (image) at (0,0) {\includegraphics[width=\textwidth]{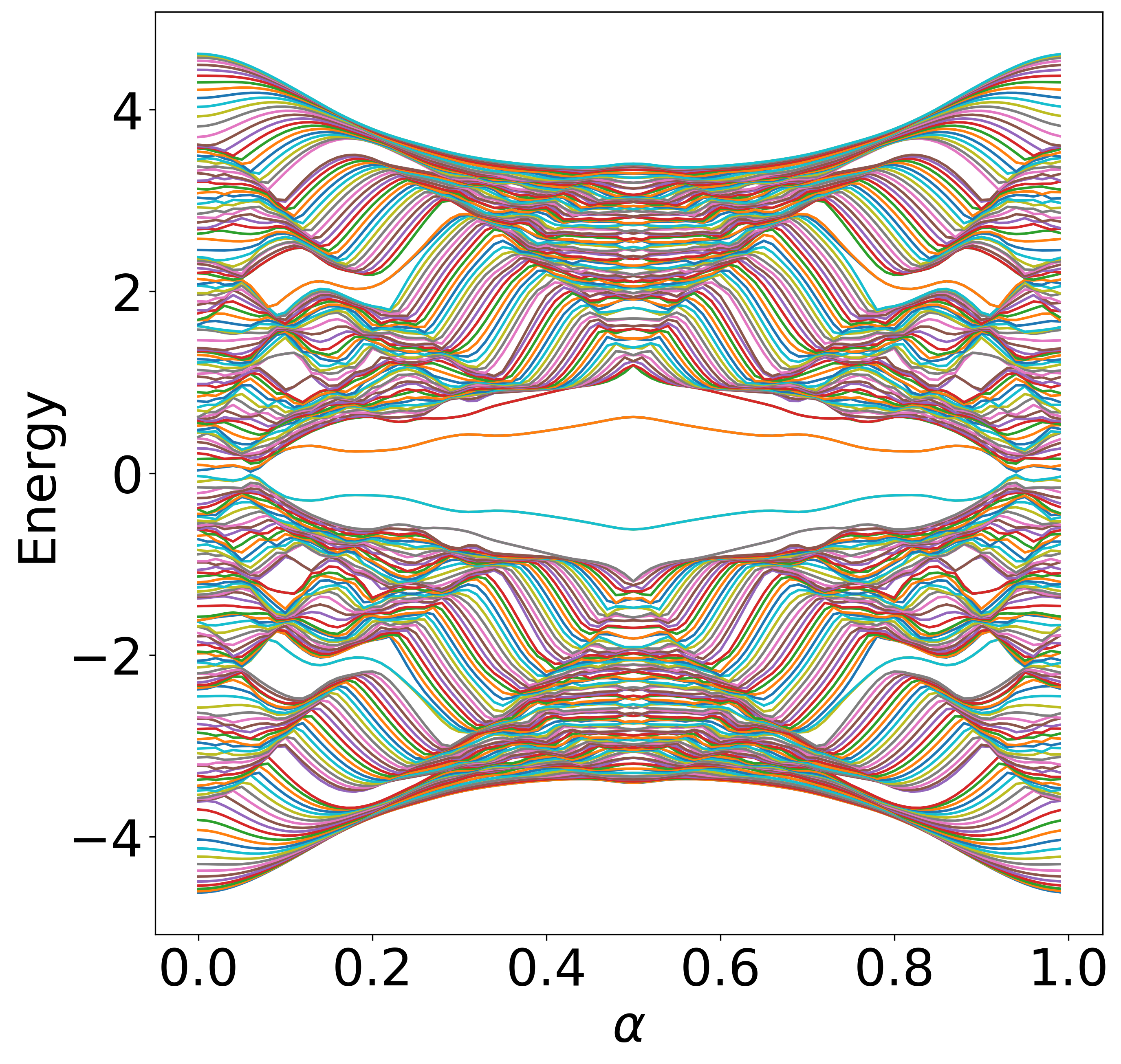}};
            \node[anchor=north west, text=black, font=\bfseries, xshift=-0.2cm, yshift=0.2cm] at (image.north west) {(b)};
        \end{tikzpicture}
    \end{subfigure}
    \vspace{-1.5\baselineskip}  
    \begin{subfigure}[t]{0.4\textwidth}
        \centering
        \begin{tikzpicture}
            \node[anchor=north west] (image) at (0,0) {\includegraphics[width=\textwidth]{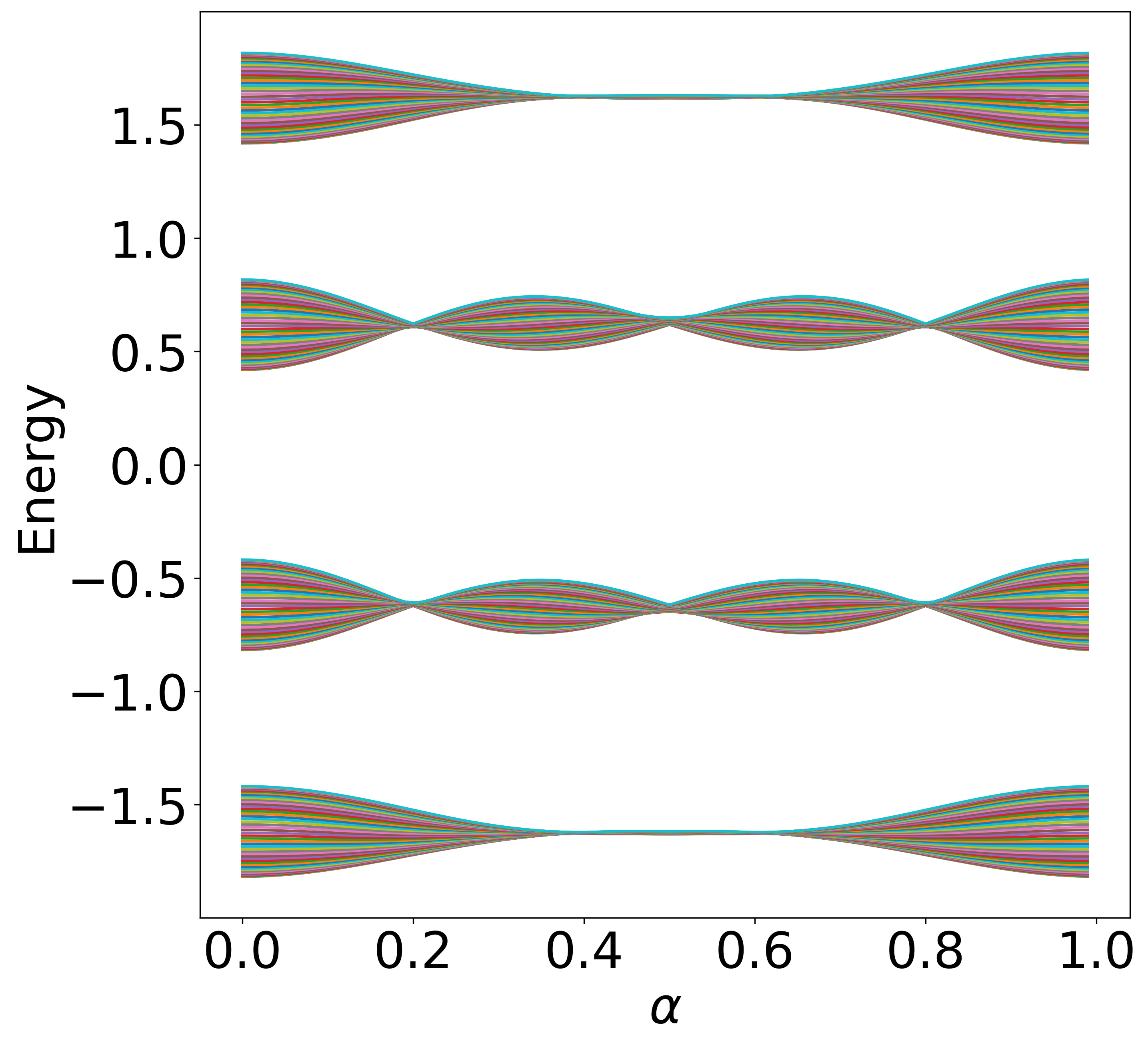}};
            \node[anchor=north west, text=black, font=\bfseries, xshift=-0.2cm, yshift=0.2cm] at (image.north west) {(c)};
        \end{tikzpicture}
    \end{subfigure}
    \hspace{0.3cm}
    \begin{subfigure}[t]{0.4\textwidth}
        \centering
        \begin{tikzpicture}
            \node[anchor=north west] (image) at (0,0) {\includegraphics[width=\textwidth]{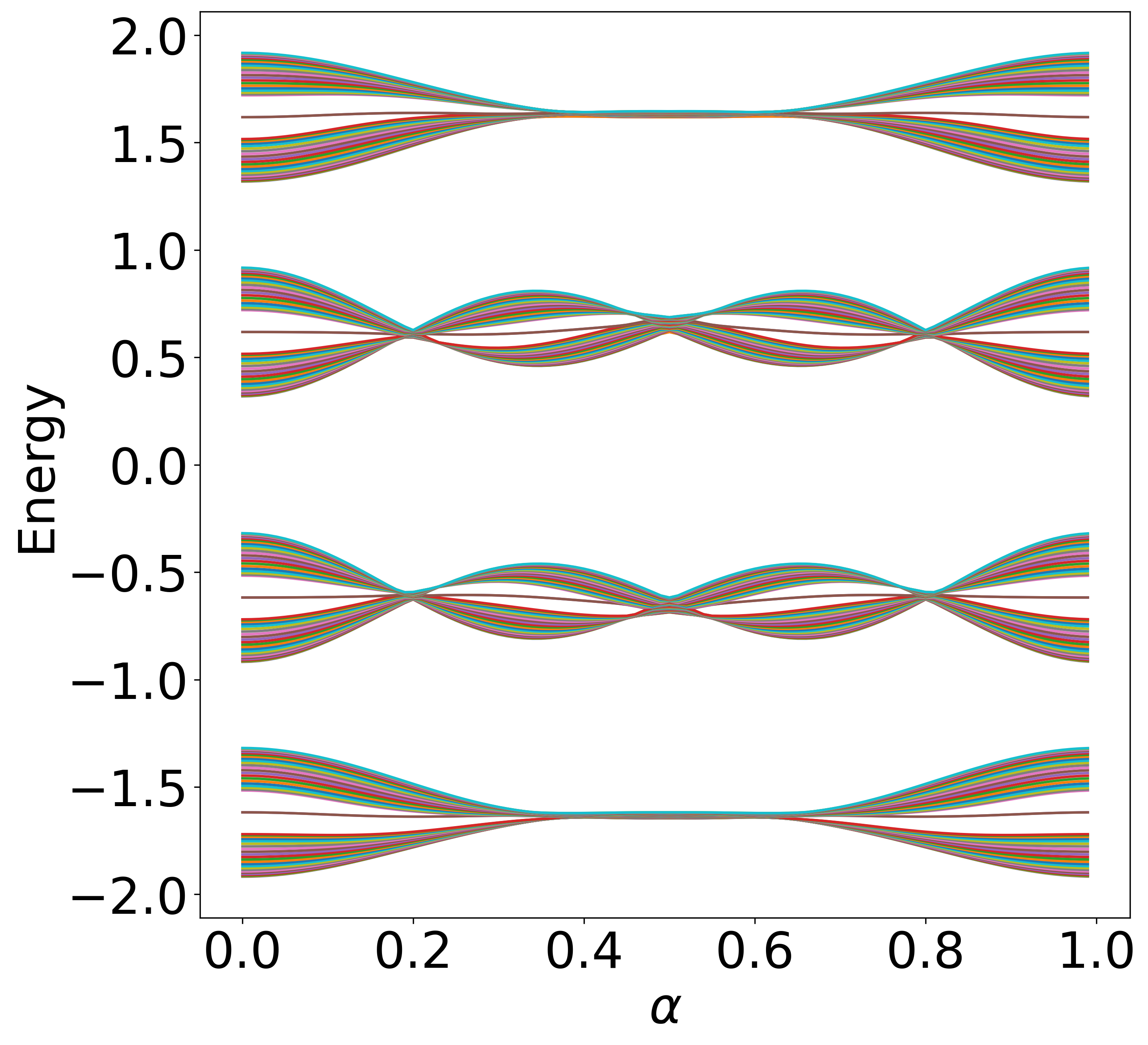}};
            \node[anchor=north west, text=black, font=\bfseries, xshift=-0.2cm, yshift=0.2cm] at (image.north west) {(d)};
        \end{tikzpicture}
    \end{subfigure}
    
    \vspace{1.5\baselineskip}  
    \begin{subfigure}[t]{0.4\textwidth}
        \centering
        \begin{tikzpicture}
            \node[anchor=north west] (image) at (0,0) {\includegraphics[width=\textwidth]{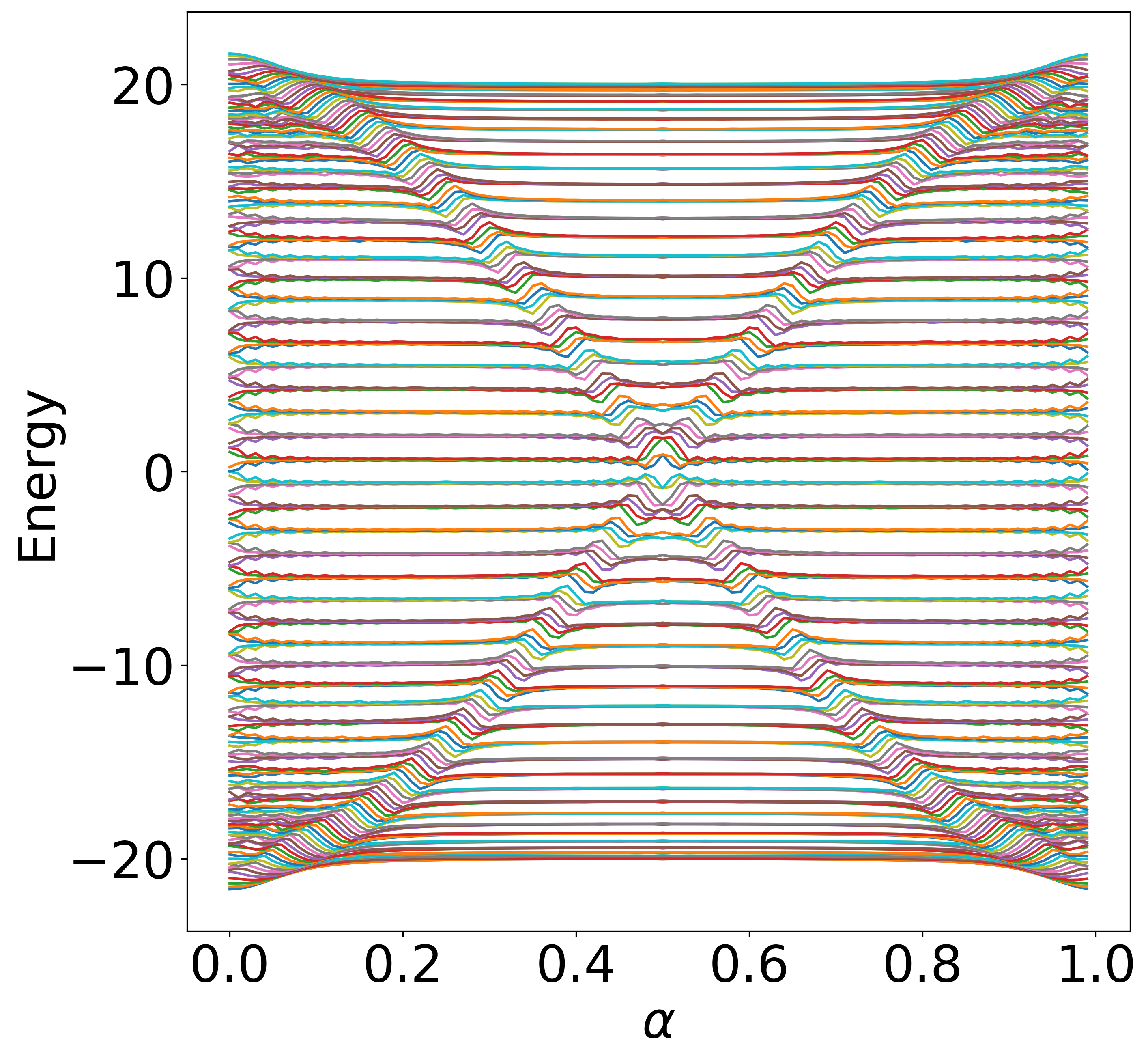}};
            \node[anchor=north west, text=black, font=\bfseries, xshift=-0.2cm, yshift=0.2cm] at (image.north west) {(e)};
        \end{tikzpicture}
    \end{subfigure}
    \hspace{0.3cm}
    \begin{subfigure}[t]{0.4\textwidth}
        \centering
        \begin{tikzpicture}
            \node[anchor=north west] (image) at (0,0) {\includegraphics[width=\textwidth]{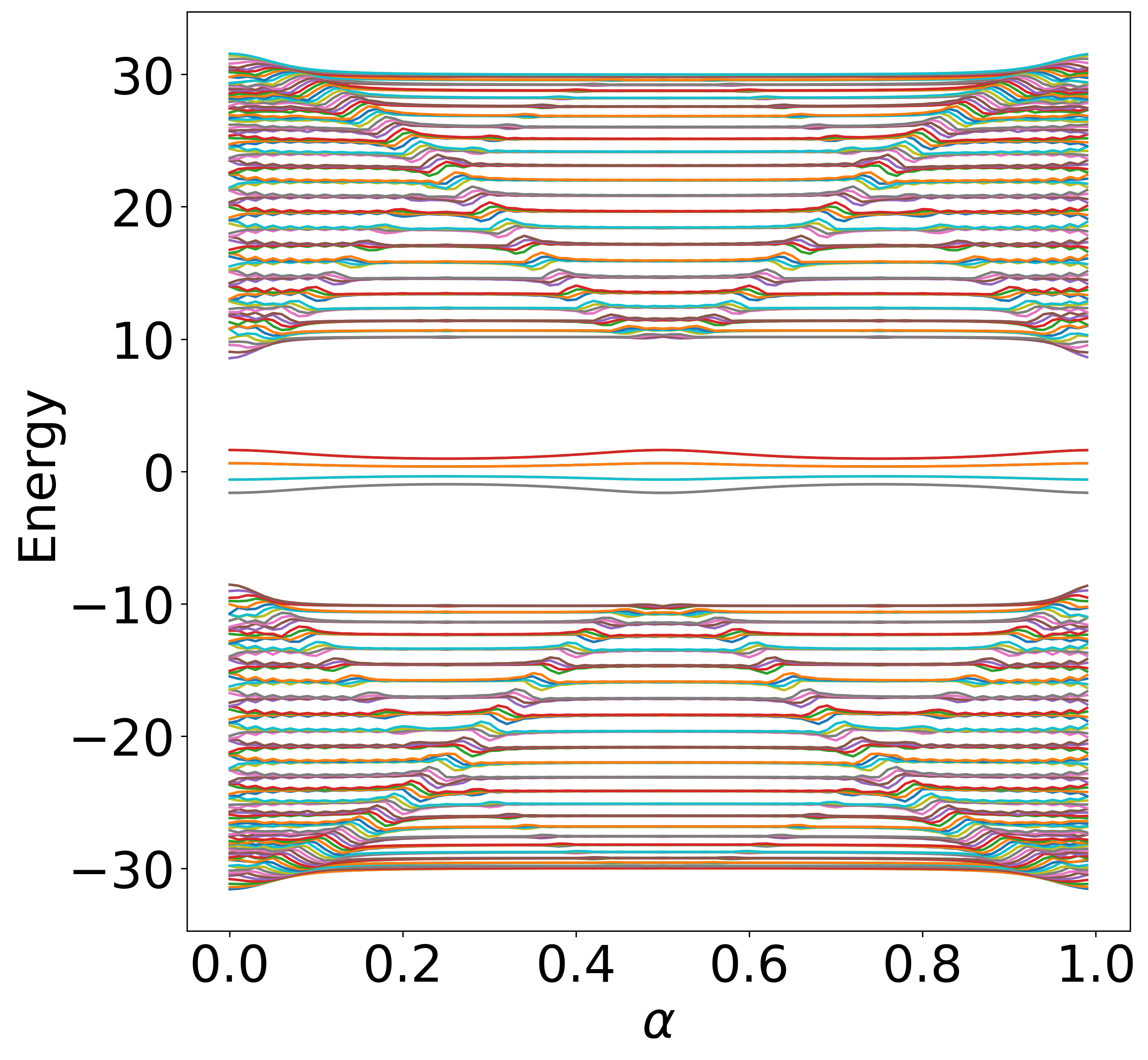}};
            \node[anchor=north west, text=black, font=\bfseries, xshift=-0.2cm, yshift=0.2cm] at (image.north west) {(f)};
        \end{tikzpicture}
    \end{subfigure}
    \caption{Hofstadter butterfly spectra of the four-leg SSH ladder. (a) Trivially dimerized phase without surface states, $t_0=1,t_2=1, t_1=2$; (b) Topologically non-trivial dimerized phase with eight two-fold degenerate surface states, $t_2=2$, $t_1=t_0=1$; (c) Trivial dimerized phase without surface states, dominant inter-chain coupling, $t_1=0.2$, $t_2=0.1$, $t_0=1$; (d) Topologically non-trivial dimerized phase, dominant inter-chain coupling, $t_1=0.1$, $t_2=0.2$, $t_0=1$;
    (e) Trivially dimerized phase without surface states, dominant intra-chain couplings, $t_1=20$, $t_2=10$, $t_0=1$; (f) Topologically non-trivial dimerized phase, dominant intra-chain couplings, $t_1=10$, $t_2=20$, $t_0=1$.}
        \label{fig:4legladder}
\end{figure}

Next, we turn to the 4-leg ladder system and examine which of the above features carry over as we begin approaching the 2D limit. The corresponding Hofstadter spectra are shown in Fig.~\ref{fig:4legladder}. In Figs. ~\ref{fig:4legladder}(a) and (b) we compare the trivially dimerized and topologically non-trivial dimerized phases for the case when all hopping parameters are of the same order of magnitude. In contrast to the two-leg ladder, we now observe multiple gapped regions in the spectrum, some of which correspond to magnetic flux induced states with non-trivial Chern numbers. While this can be checked by direct calculation of the TKNN invariant, the nontrivial topology is also reflected in the presence of topological surface states within bulk gaps due to the bulk-boundary correspondence (recall that we are using OBC here). Here, we observe that in the topologically non-trivial dimerized phase, there are surface states not only in the central gapped region (i.e., around $\alpha =0.5$) but also within other gaps. Within the central bulk gap, we find a pair of two-fold degenerate surface states, which are localized at the two horizontal ends of the ladder. More generally, if we consider say $\alpha = 0.25$, we observe in-gap surface states both at $1/4$-filling and also at $1/2$-filling. The former are chiral edge states that correspond to a non-zero Chern number $C=1$ in that gap, while the latter are protected by inversion symmetry of an effective 1D Hamiltonian~\cite{lau2015} but not by a Chern number, as that is trivial within this region of the butterfly.

For the case where the energy scales are well-separated i.e., either dominant inter-chain couplings (Figs. ~\ref{fig:4legladder}(c) and (d)) or dominant intra-chain couplings (Figs. ~\ref{fig:4legladder}(e) and (f)), the resulting Hofstadter spectra are a straightforward generalization of what is observed in the two-leg ladder. Namely, for $t_0\gg t_1,t_2$, there are now four instead of two bulk bands. In the topologically non-trivial phase, gaps open up in the center of these bands, and a two-fold degenerate surface state emerges in each of these. In the opposite limit, $t_0\ll t_1,t_2$, we observe two bulk bands, same as for the two-leg ladder, with four two-fold degenerate surface states within the central bulk gap.

\begin{figure}[h!]
    \centering
    \begin{subfigure}[t]{0.4\textwidth}
        \centering
        \begin{tikzpicture}
            \node[anchor=north west] (image) at (0,0) {\includegraphics[width=\textwidth]{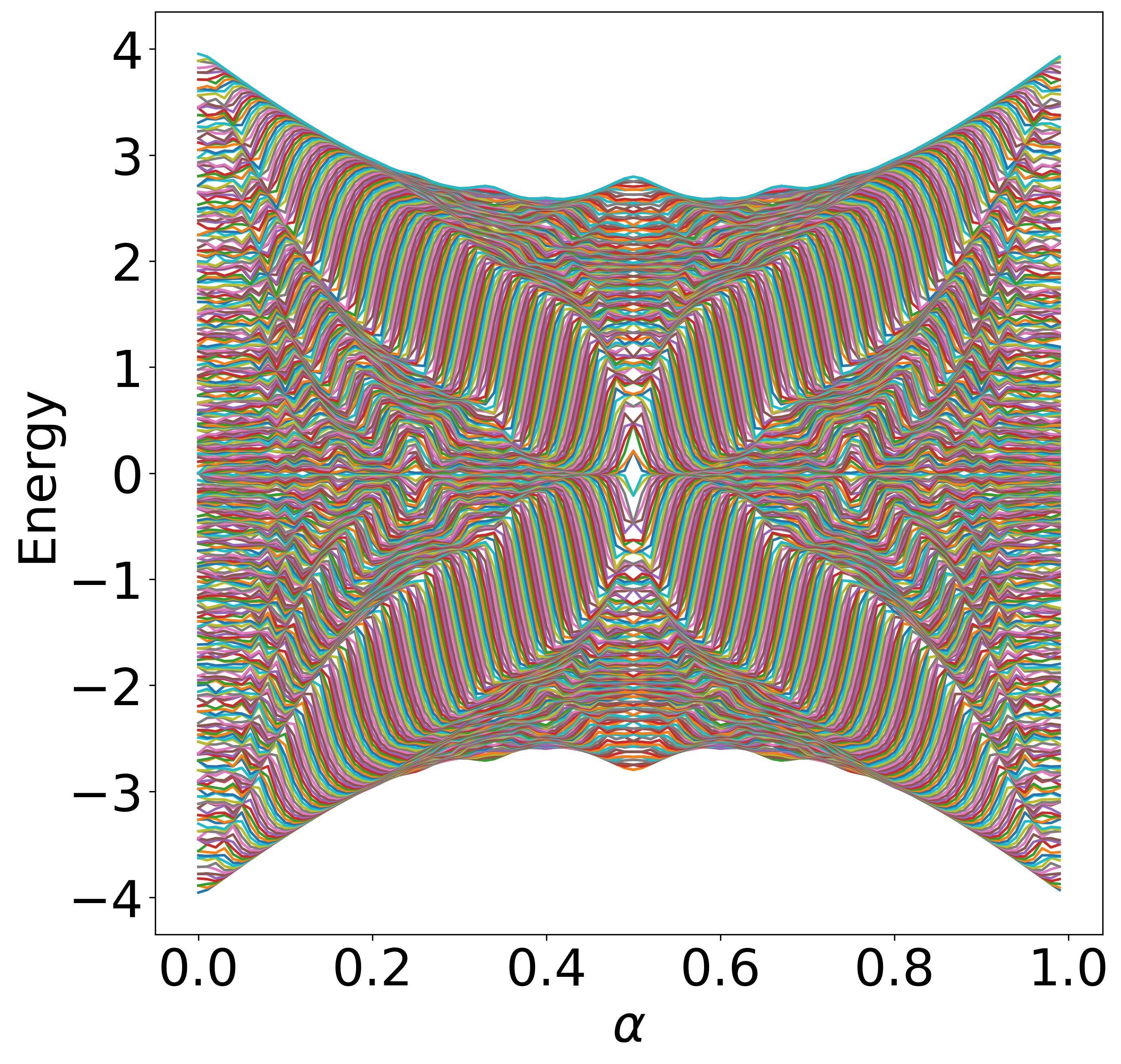}};
            \node[anchor=north west, text=black, font=\bfseries, xshift=-0.2cm, yshift=0.2cm] at (image.north west) {(a)};
        \end{tikzpicture}
    \end{subfigure}
    \hspace{0.3cm}
    \begin{subfigure}[t]{0.4\textwidth}
        \centering
        \begin{tikzpicture}
            \node[anchor=north west] (image) at (0,0) {\includegraphics[width=\textwidth]{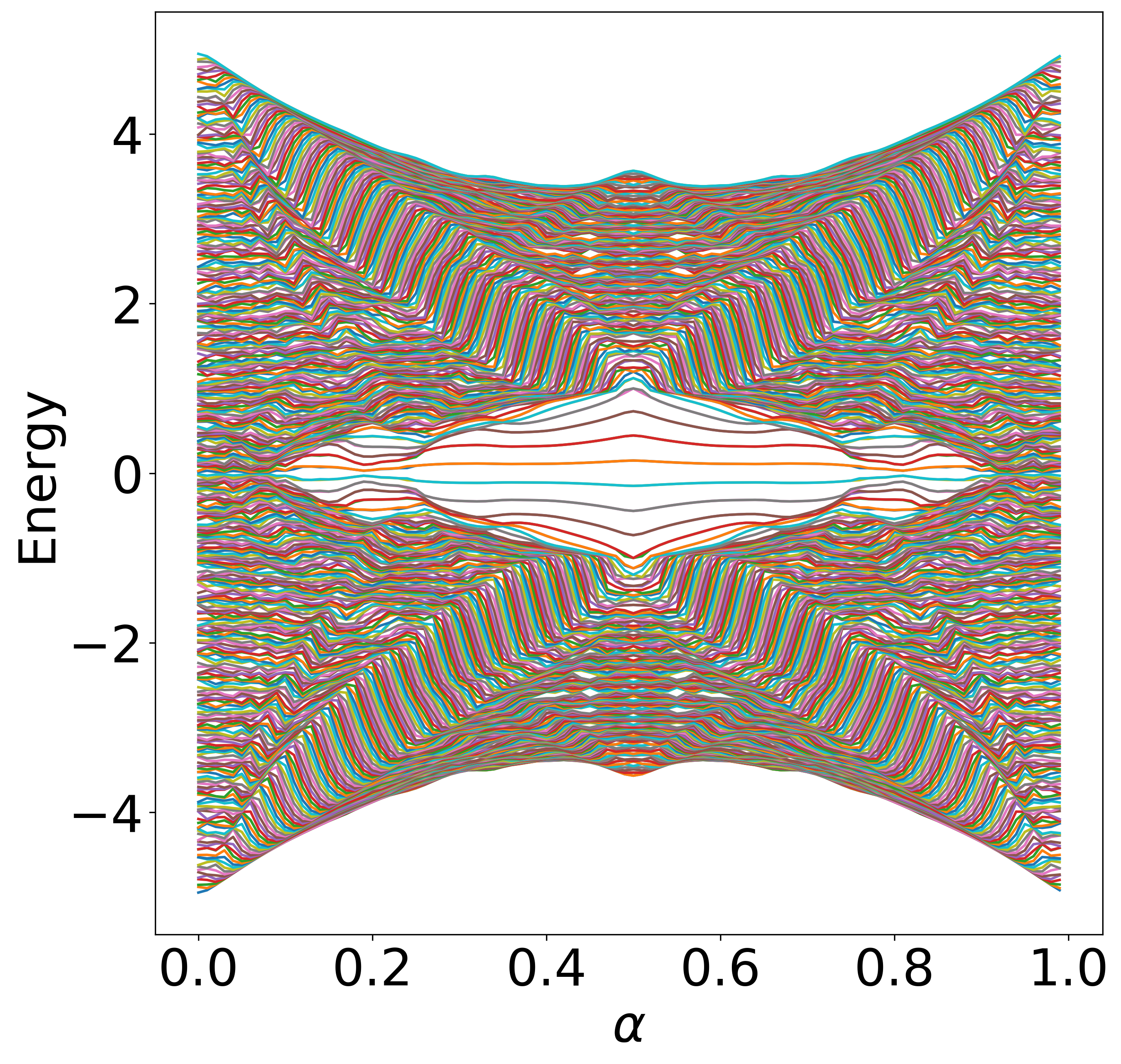}};
            \node[anchor=north west, text=black, font=\bfseries, xshift=-0.2cm, yshift=0.2cm] at (image.north west) {(b)};
        \end{tikzpicture}
    \end{subfigure}
    \vspace{-1.5\baselineskip}  
    \begin{subfigure}[t]{0.4\textwidth}
        \centering
        \begin{tikzpicture}
            \node[anchor=north west] (image) at (0,0) {\includegraphics[width=\textwidth]{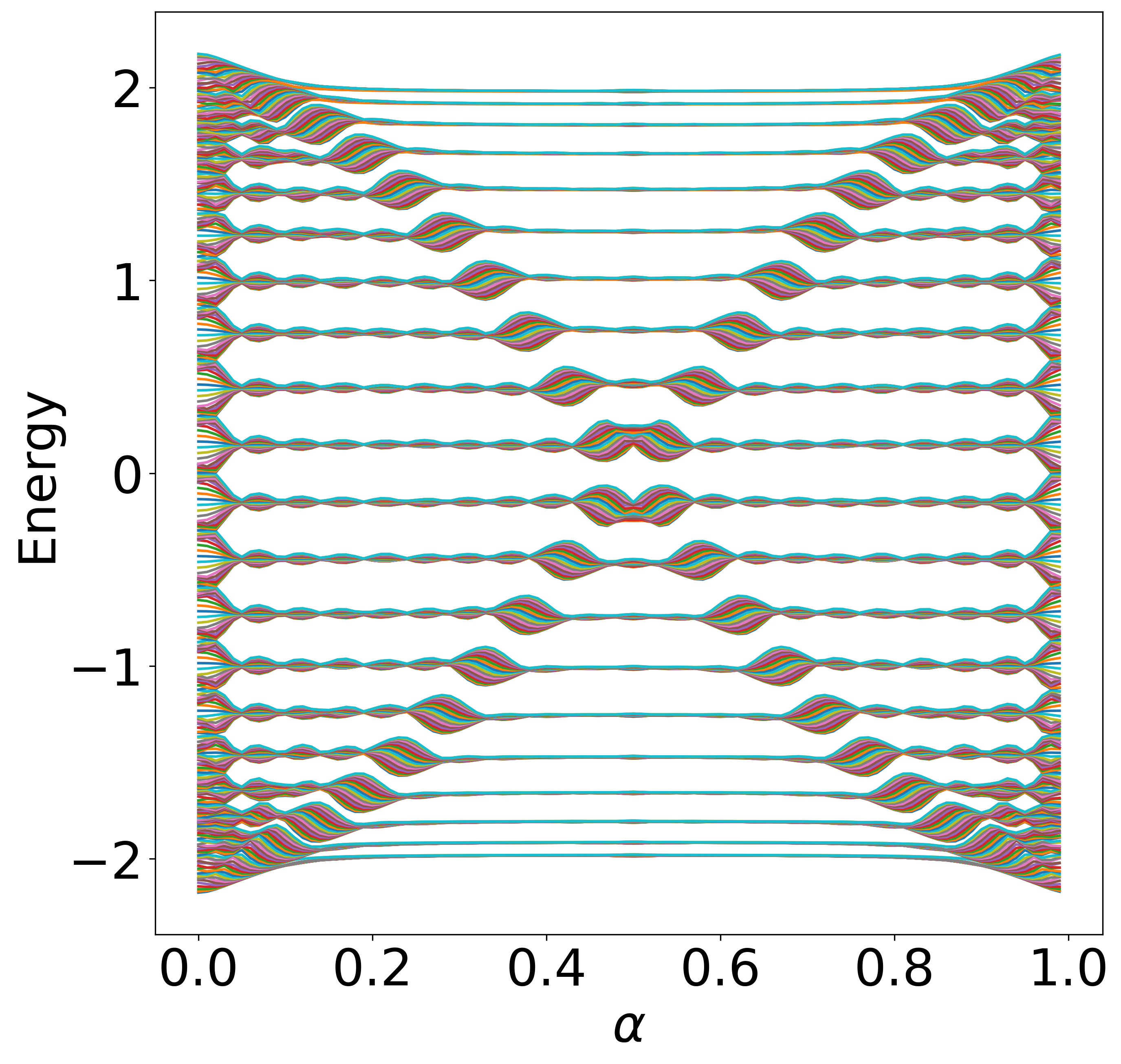}};
            \node[anchor=north west, text=black, font=\bfseries, xshift=-0.2cm, yshift=0.2cm] at (image.north west) {(c)};
        \end{tikzpicture}
    \end{subfigure}
    \hspace{0.3cm}
    \begin{subfigure}[t]{0.4\textwidth}
        \centering
        \begin{tikzpicture}
            \node[anchor=north west] (image) at (0,0) {\includegraphics[width=\textwidth]{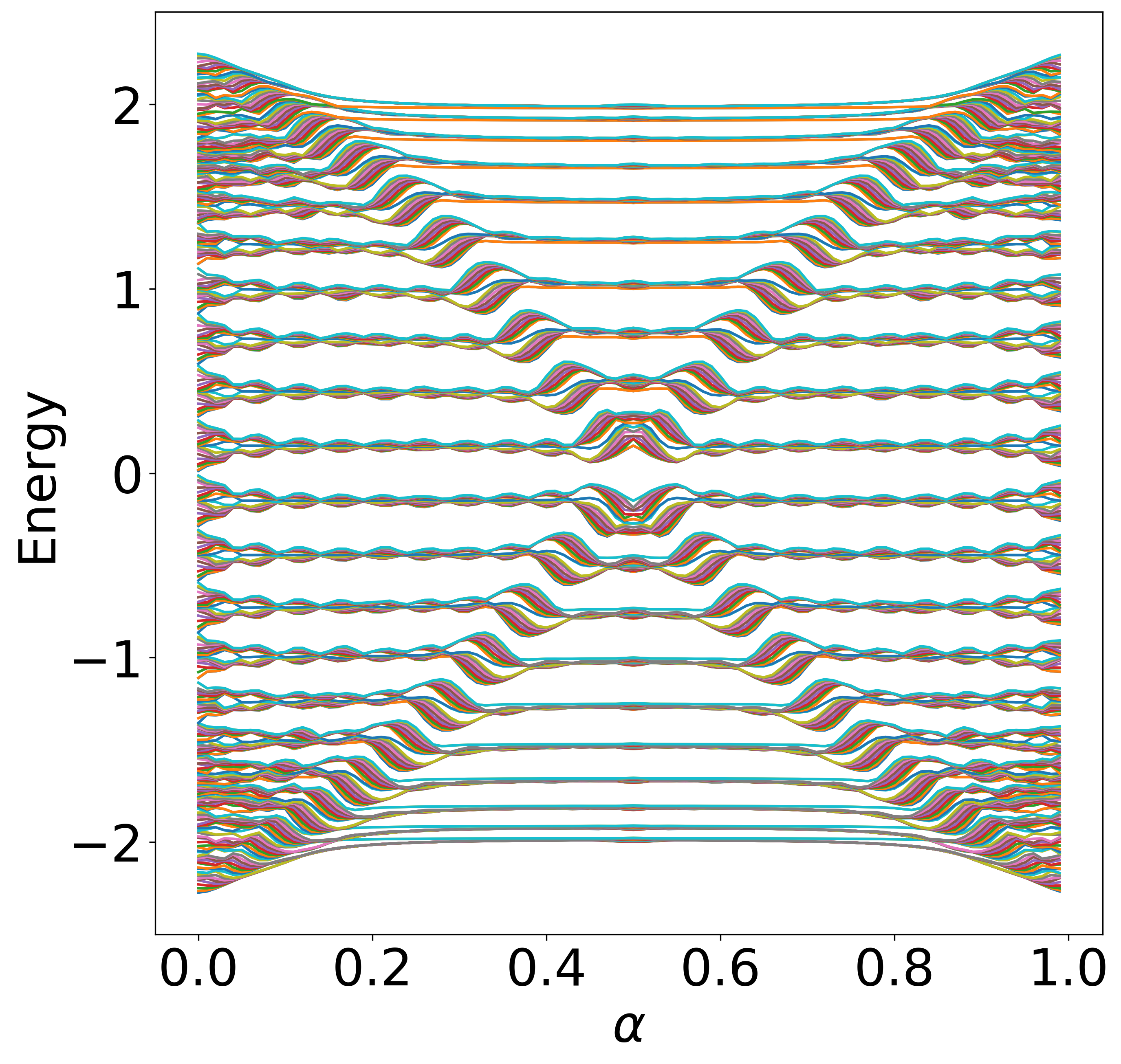}};
            \node[anchor=north west, text=black, font=\bfseries, xshift=-0.2cm, yshift=0.2cm] at (image.north west) {(d)};
        \end{tikzpicture}
    \end{subfigure}
    
    \vspace{1.5\baselineskip}  
    \begin{subfigure}[t]{0.4\textwidth}
        \centering
        \begin{tikzpicture}
            \node[anchor=north west] (image) at (0,0) {\includegraphics[width=\textwidth]{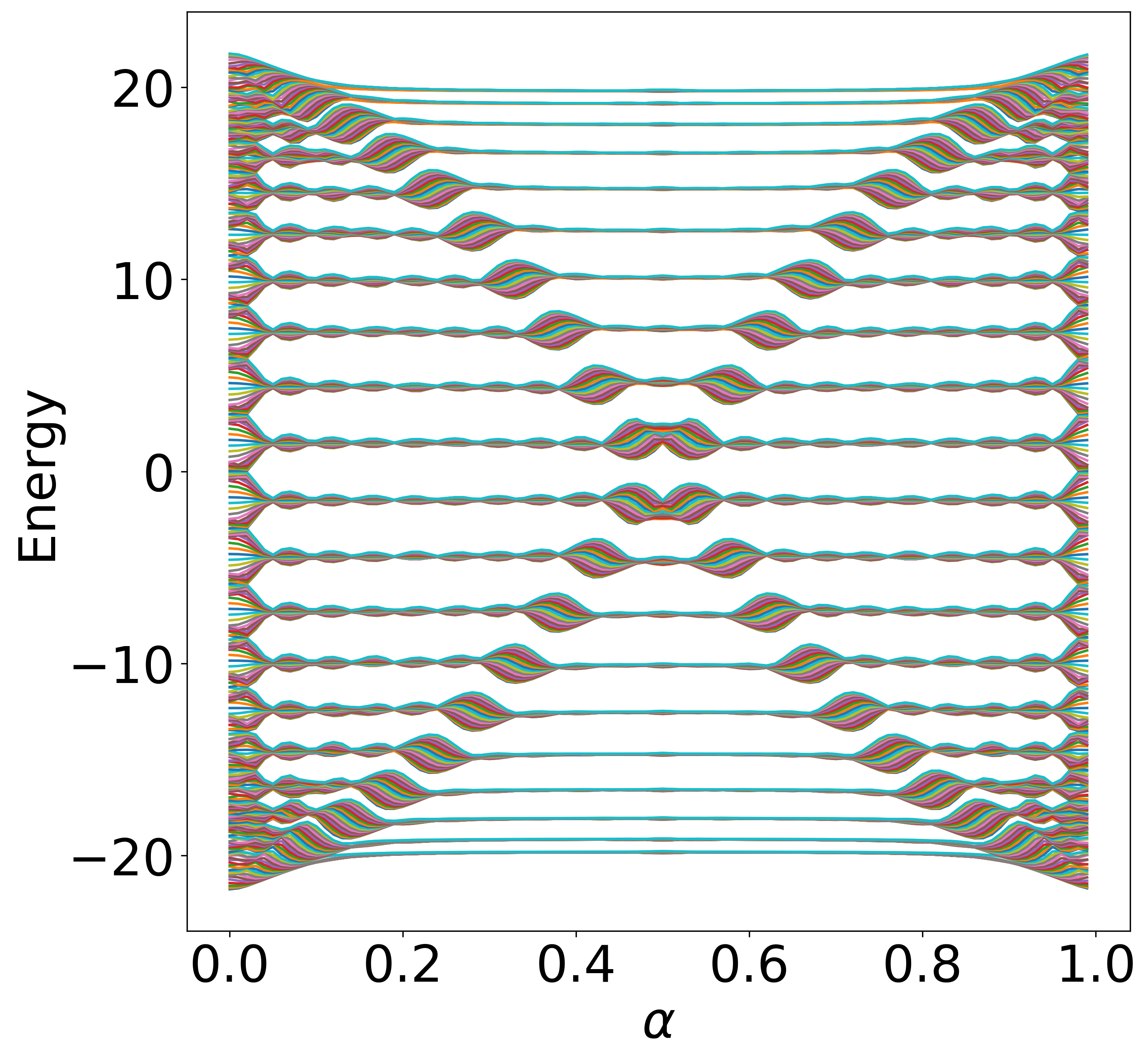}};
            \node[anchor=north west, text=black, font=\bfseries, xshift=-0.2cm, yshift=0.2cm] at (image.north west) {(e)};
        \end{tikzpicture}
    \end{subfigure}
    \hspace{0.3cm}
    \begin{subfigure}[t]{0.4\textwidth}
        \centering
        \begin{tikzpicture}
            \node[anchor=north west] (image) at (0,0) {\includegraphics[width=\textwidth]{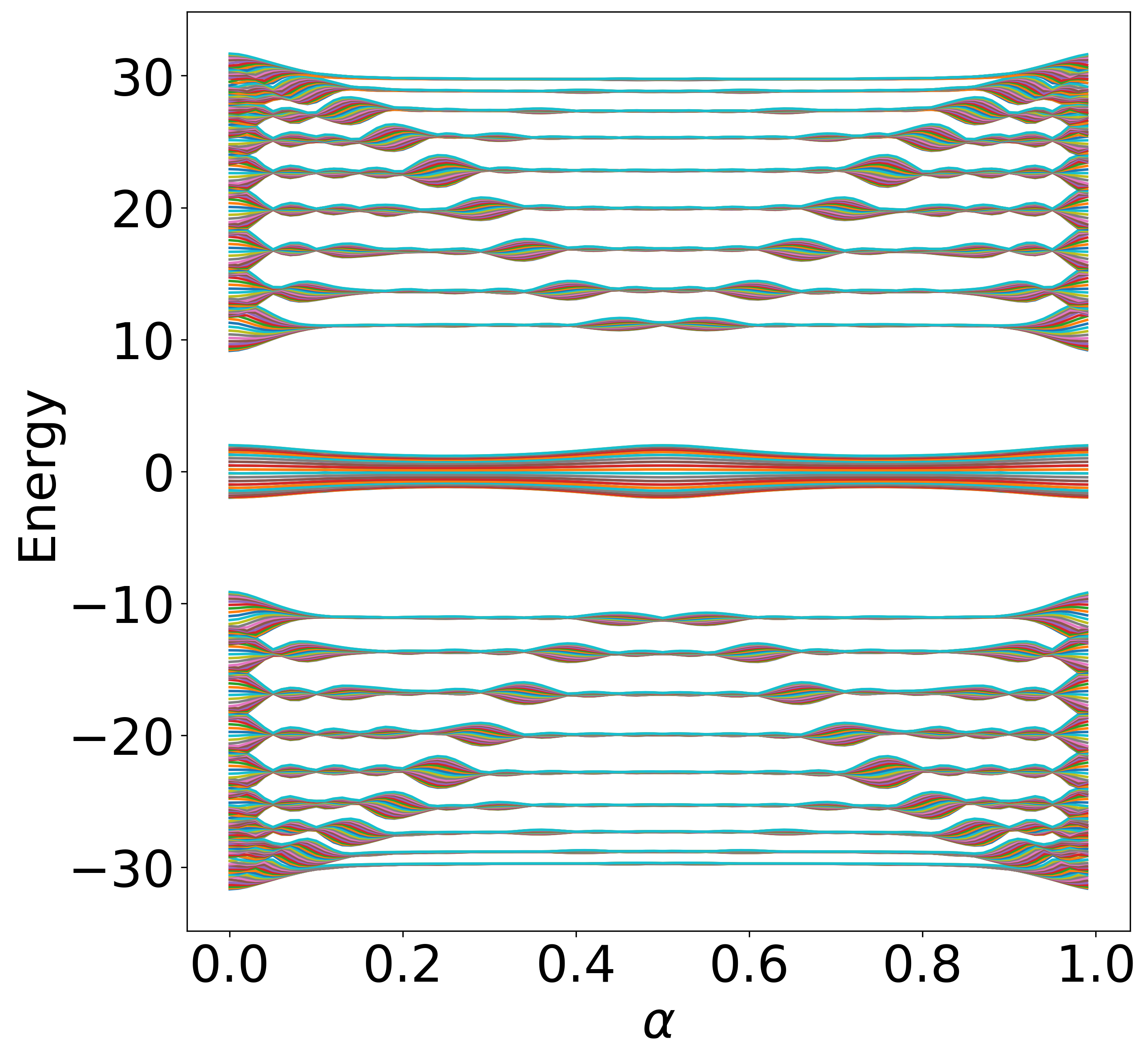}};
            \node[anchor=north west, text=black, font=\bfseries, xshift=-0.2cm, yshift=0.2cm] at (image.north west) {(f)};
        \end{tikzpicture}
    \end{subfigure}
        \caption{Hofstadter butterfly spectra of the 20-leg SSH ladder. (a) Trivially dimerized phase, $t_0=t_2=1$, $t_1=2$; (b) Topologically non-trivial dimerized phase with 10 4-fold degenerate surface states in the central bulk gao, $t_2=2$, $t_1=t_0=1$; (c) Trivially dimerized phase without surface states, dominant inter-chain coupling, $t_1=0.2$, $t_2=0.1$, $t_0=1$; (d) Topologically non-trivial dimerized phase, dominant inter-chain coupling, $t_1=0.1$, $t_2=0.2$, $t_0=1$;
    (e) Trivially dimerized phase without surface states, dominant intra-chain couplings, $t_1=20$, $t_2=10$, $t_0=1$; (f) Topologically non-trivial dimerized phase, dominant intra-chain couplings, $t_1=10$, $t_2=20$, $t_0=1$.}
            \label{fig:20legladder}
\end{figure}

We now consider the 2D case, with the numerical results for a $20 \times 20$ system with OBC shown in Fig.~\ref{fig:20legladder}. In the trivially dimerized phase, shown in Fig.~\ref{fig:20legladder}(a), we recognize the slightly altered well-known pattern of the Hofstadter spectrum in an isotropic 2D tight-binding lattice. Due to the OBC, the bulk-gap regions of the periodic system are populated by chiral surface states, which result from the non-trivial Chern number in these regions and will be further discussed below. In the topologically non-trivial dimerized phase, shown in Fig.~\ref{fig:20legladder}(b), we observe additional surface states which are not protected by a Chern number (which vanishes in this region of the butterfly) but are instead protected by inversion symmetry of an effective 1D Hamiltonian (see discussion in Ref.~\cite{lau2015}).

\begin{figure}[h!]
    \centering
    \begin{minipage}[t]{\textwidth}
        \begin{minipage}[t]{0.45\textwidth}
            \centering
            \begin{tikzpicture}
                \node[anchor=north west] (image) at (0,0) {\includegraphics[width=\textwidth]{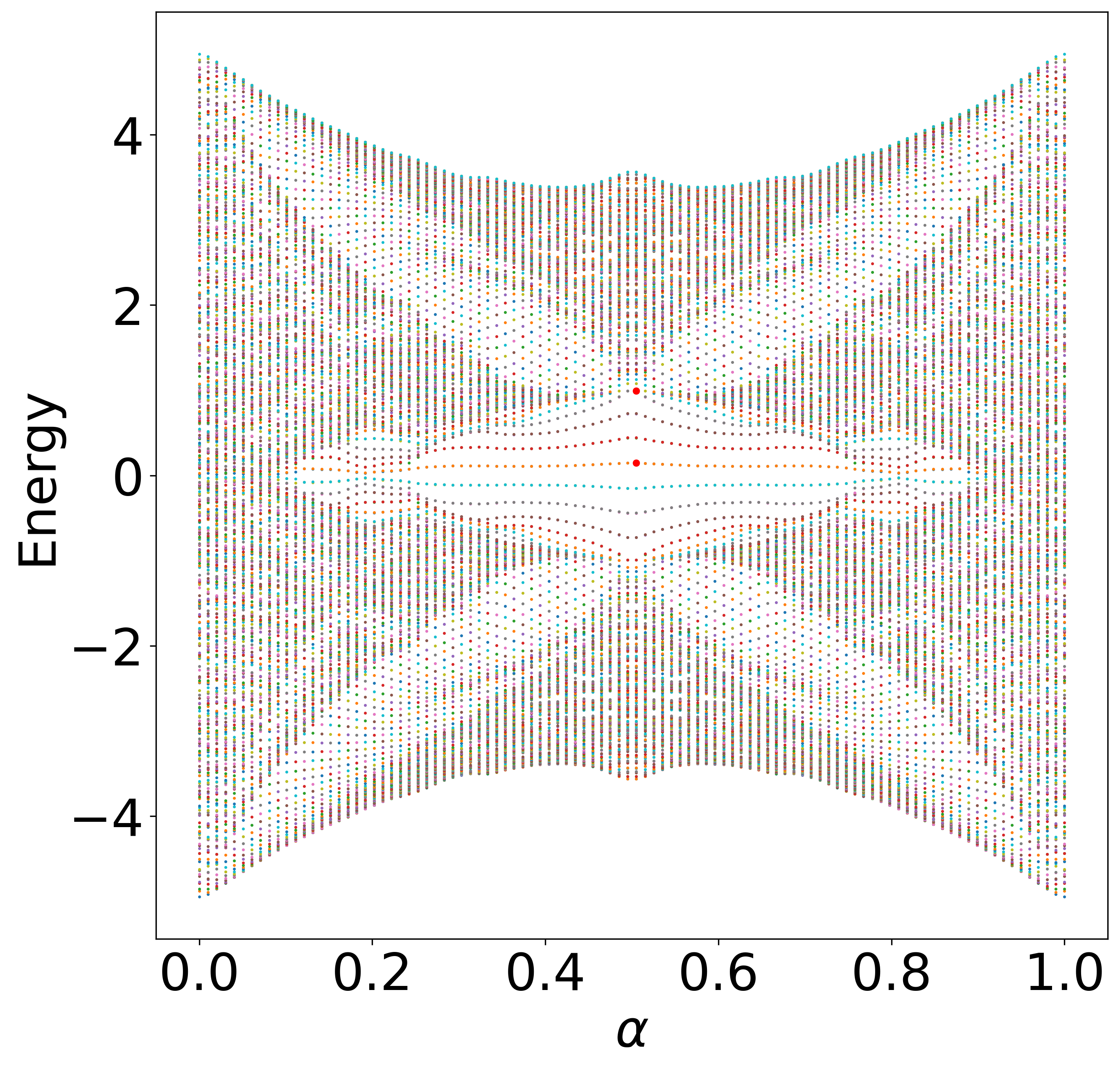}};
                \node[anchor=north west, text=black, font=\bfseries, xshift=-0.2cm, yshift=0.2cm] at (image.north west) {(a)};
            \end{tikzpicture}
            \label{fig:big}
        \end{minipage}
        \hspace{0.02\textwidth}
        \raisebox{3cm}{ 
            \begin{minipage}[t]{0.50\textwidth}
                \centering
                \begin{minipage}[t]{\linewidth}
                    \begin{minipage}[t]{0.49\textwidth}
                        \centering
                        \begin{tikzpicture}
                            \node[anchor=north west] (image) at (0,0) {\includegraphics[width=\textwidth]{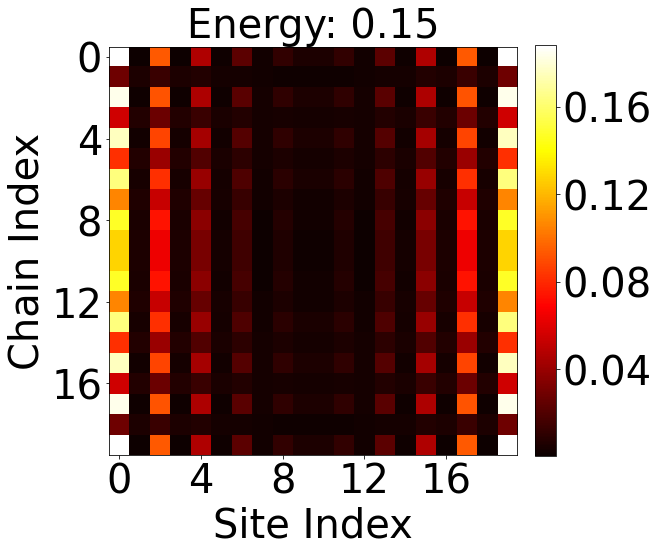}};
                            \node[anchor=north west, text=black, font=\bfseries, xshift=-0.2cm, yshift=0.2cm] at (image.north west) {(b)};
                        \end{tikzpicture}
                        \label{fig:small1}
                    \end{minipage}
                    \hfill
                    \begin{minipage}[t]{0.49\textwidth}
                        \centering
                        \begin{tikzpicture}
                            \node[anchor=north west] (image) at (0,0) {\includegraphics[width=\textwidth]{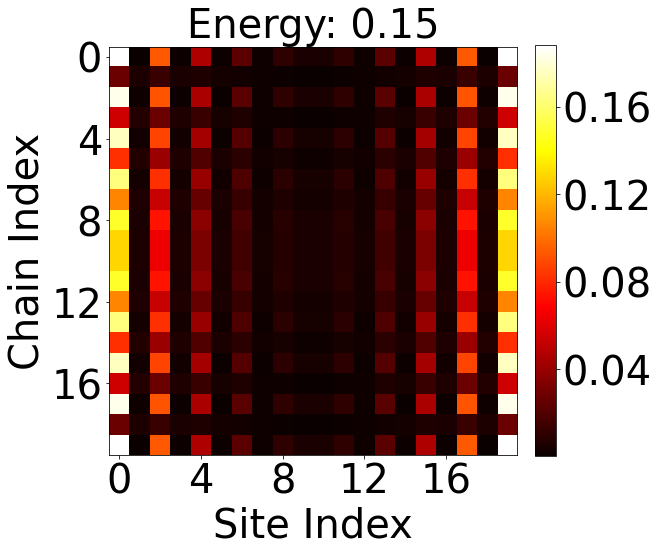}};
                            \node[anchor=north west, text=black, font=\bfseries, xshift=-0.2cm, yshift=0.2cm] at (image.north west) {(c)};
                        \end{tikzpicture}
                        \label{fig:small2}
                    \end{minipage}
                \end{minipage}
                \vspace{0.02\textwidth}
                \vspace{-1.5\baselineskip}
                \begin{minipage}[t]{\linewidth}
                    \begin{minipage}[t]{0.49\textwidth}
                        \centering
                        \begin{tikzpicture}
                            \node[anchor=north west] (image) at (0,0) {\includegraphics[width=\textwidth]{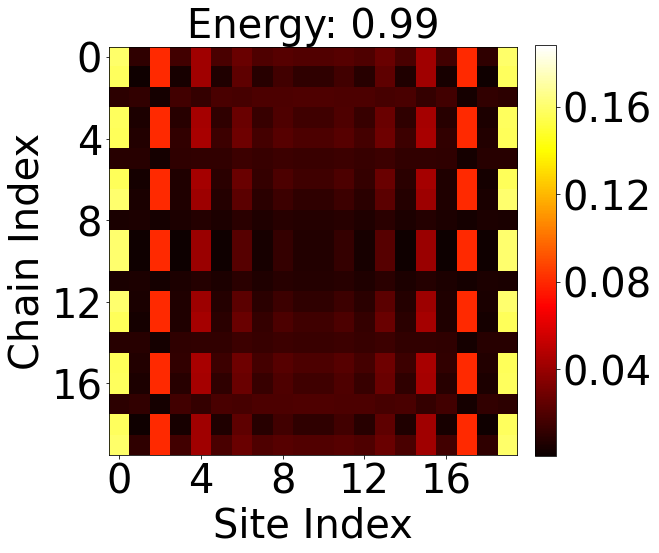}};
                            \node[anchor=north west, text=black, font=\bfseries, xshift=-0.2cm, yshift=0.2cm] at (image.north west) {(d)};
                        \end{tikzpicture}
                        \label{fig:small3}
                    \end{minipage}
                    \hfill
                    \begin{minipage}[t]{0.49\textwidth}
                        \centering
                        \begin{tikzpicture}
                            \node[anchor=north west] (image) at (0,0) {\includegraphics[width=\textwidth]{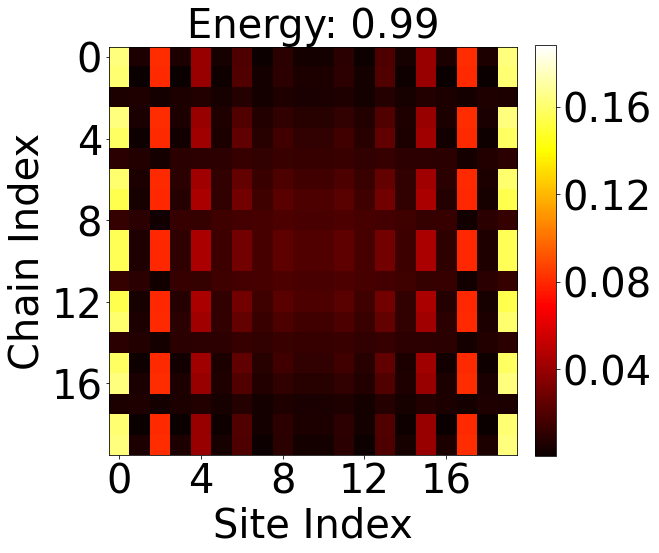}};
                            \node[anchor=north west, text=black, font=\bfseries, xshift=-0.2cm, yshift=0.2cm] at (image.north west) {(e)};
                        \end{tikzpicture}
                        \label{fig:small4}
                    \end{minipage}
                \end{minipage}
            \end{minipage}
        }
    \end{minipage}
    \caption{(a) Hofstadter  spectrum for the 20-leg SSH ladder in the topologically non-trivial dimerized phase with 10 4-fold degenerate surface states protected by inversion symmetry, $t_2=2$, $t_1=t_0=1$; (b) and (c) two of the four degenerate topological surface states at energy E=0.15; (d) and (e) two of the four degenerate topological surface states at energy E=0.99.}
    \label{fig:sshsurfacestates}
\end{figure}

Moving on to the cases of dominant inter-chain (Figs.~\ref{fig:20legladder}(c) and (d)) and intra-chain (Figs.~\ref{fig:20legladder}(e) and (f)) couplings, we observe that the spectra in the trivially dimerized regimes look quite similar in the 2D limit, although the overall energy scales are either set by $t_0$ or by $t_1$ and $t_2$. In both cases, one observes an emerging quasi-continuum. However, what is different for these two cases are the energy levels of the topological surface states in the non-trivial regimes: for dominant inter-chain couplings (Fig.~\ref{fig:20legladder}(c)(d)) they emerge within each of this 20 mini-bands, whereas for dominant intra-chain couplings ((Fig.~\ref{fig:20legladder}(f)) they form a central $ E\approx 0$ surface state band, separated from the bulk bands by a gap of order $t_2-t_1$.

We now move on to a discussion of the nature of the various surface states that can be observed in this hybrid topological system. Here, we focus on the $20 \times 20$ system, but the following observations are apply more generally. In Fig.~\ref{fig:sshsurfacestates}, we first focus on the surface states with energies within the central bulk gap, i.e., around $\alpha =0.5$ and $|E| < 1$. We have highlighted 4 of these in Fig.~\ref{fig:sshsurfacestates}(a). These disperse only slightly as a function of the applied magnetic field, and, as seen in Figs.~\ref{fig:sshsurfacestates}(b)-(e), they are localized at the two ends of the coupled ladder systems perpendicular to the direction of the magnetic vector potential. Also, note that these surface states come in degenerate pairs, with their number determined by a 1D topological invariant of an inversion symmetric effective 1D Hamiltonian~\cite{lau2015}.

\begin{figure}[h!]
    \centering
    \begin{minipage}[t]{\textwidth}
        \begin{minipage}[t]{0.45\textwidth}
            \centering
            \begin{tikzpicture}
                \node[anchor=north west] (image) at (0,0) {\includegraphics[width=\textwidth]{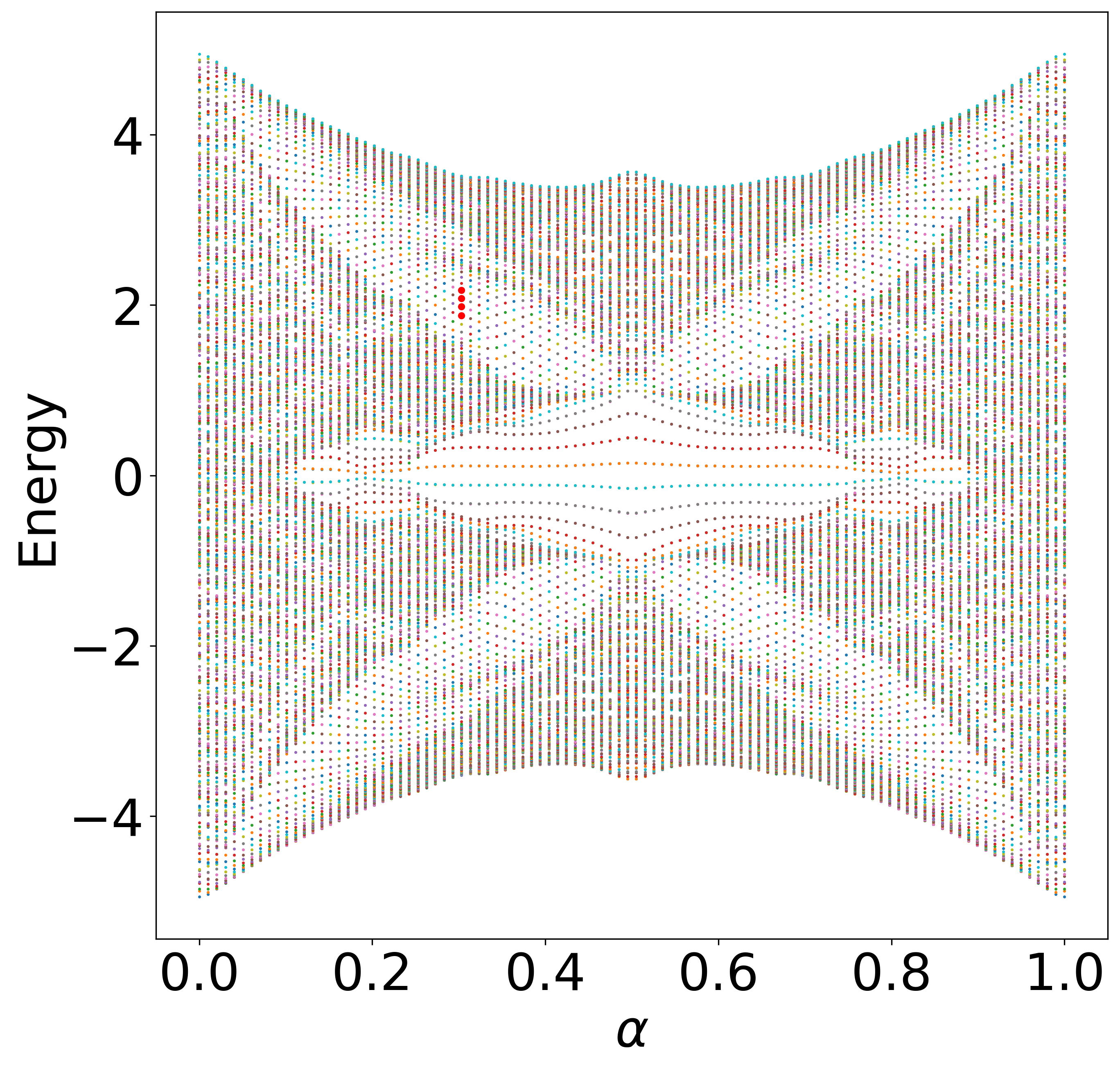}};
                \node[anchor=north west, text=black, font=\bfseries, xshift=-0.2cm, yshift=0.2cm] at (image.north west) {(a)};
            \end{tikzpicture}
            \label{fig:big}
        \end{minipage}
        \hspace{0.02\textwidth}
        \raisebox{3cm}{ 
            \begin{minipage}[t]{0.50\textwidth}
                \centering
                \begin{minipage}[t]{\linewidth}
                    \begin{minipage}[t]{0.49\textwidth}
                        \centering
                        \begin{tikzpicture}
                            \node[anchor=north west] (image) at (0,0) {\includegraphics[width=\textwidth]{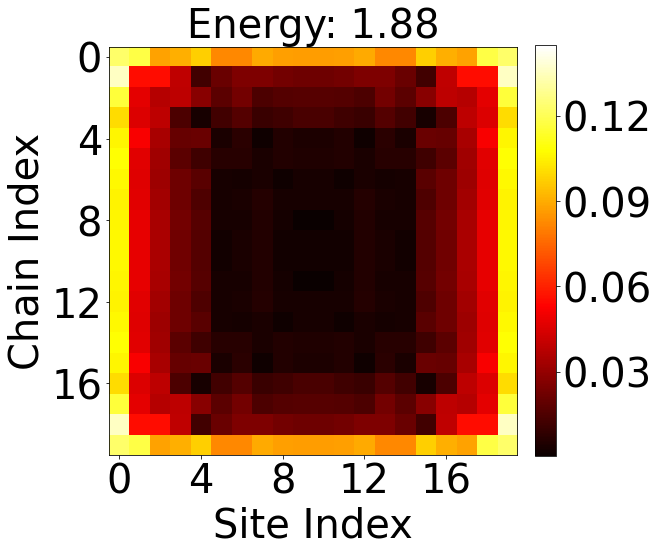}};
                            \node[anchor=north west, text=black, font=\bfseries, xshift=-0.2cm, yshift=0.2cm] at (image.north west) {(b)};
                        \end{tikzpicture}
                        \label{fig:small1}
                    \end{minipage}
                    \hfill
                    \begin{minipage}[t]{0.49\textwidth}
                        \centering
                        \begin{tikzpicture}
                            \node[anchor=north west] (image) at (0,0) {\includegraphics[width=\textwidth]{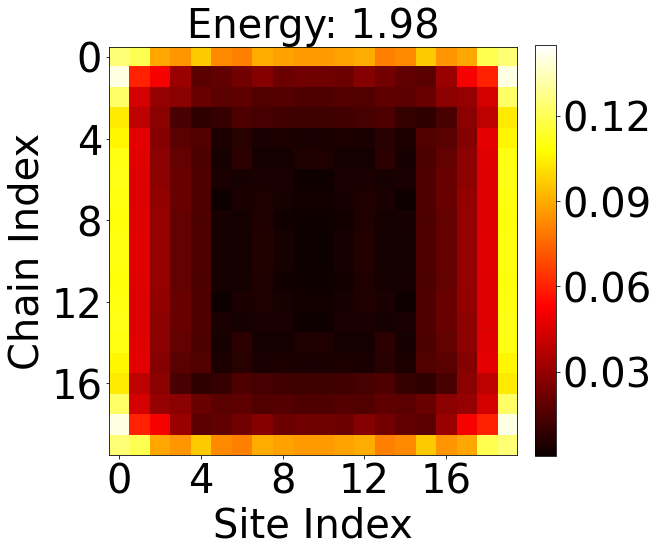}};
                            \node[anchor=north west, text=black, font=\bfseries, xshift=-0.2cm, yshift=0.2cm] at (image.north west) {(c)};
                        \end{tikzpicture}
                        \label{fig:small2}
                    \end{minipage}
                \end{minipage}
                \vspace{0.02\textwidth}
                \vspace{-1.5\baselineskip}
                \begin{minipage}[t]{\linewidth}
                    \begin{minipage}[t]{0.49\textwidth}
                        \centering
                        \begin{tikzpicture}
                            \node[anchor=north west] (image) at (0,0) {\includegraphics[width=\textwidth]{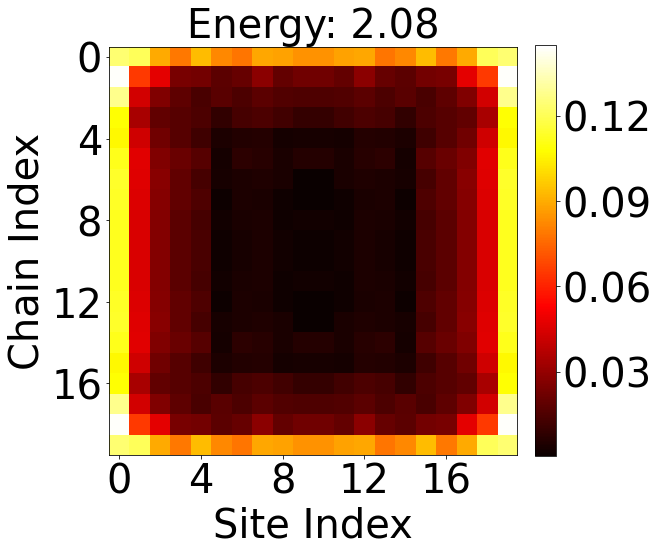}};
                            \node[anchor=north west, text=black, font=\bfseries, xshift=-0.2cm, yshift=0.2cm] at (image.north west) {(d)};
                        \end{tikzpicture}
                        \label{fig:small3}
                    \end{minipage}
                    \hfill
                    \begin{minipage}[t]{0.49\textwidth}
                        \centering
                        \begin{tikzpicture}
                            \node[anchor=north west] (image) at (0,0) {\includegraphics[width=\textwidth]{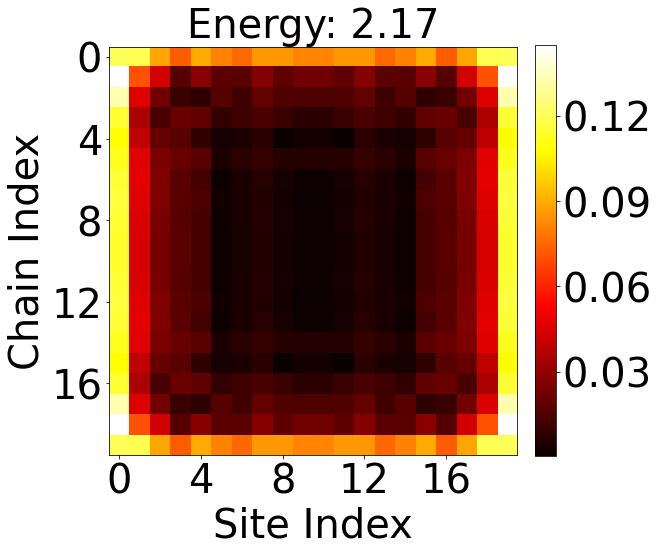}};
                            \node[anchor=north west, text=black, font=\bfseries, xshift=-0.2cm, yshift=0.2cm] at (image.north west) {(e)};
                        \end{tikzpicture}
                        \label{fig:small4}
                    \end{minipage}
                \end{minipage}
            \end{minipage}
        }
    \end{minipage}
    \caption{(a) Hofstadter  spectrum of the 20-leg SSH ladder in the topologically non-trivial dimerized phase with non-degenerate surface states due to non-trivial Chern number, $t_2=2$, $t_1=t_0=1$; (b) topological surface states at energy E=1.88; (c) E=1.98, (d) E=2.08 (e) E=2.17.}
    \label{fig:hhsurfacestates}
\end{figure}

In contrast, consider the other type of surface states, which result from a non-trivial Chern number and are shown in Fig.~\ref{fig:hhsurfacestates}. Again, we have highlighted 4 of these states, this time in the gapped region within the upper left lobe of the Hofstadter butterfly spectrum, Fig.~\ref{fig:hhsurfacestates}(a). As seen in Figs.~\ref{fig:hhsurfacestates}(b)-(e), these surface states are chiral (since they correspond to a non-trivial Chern number) and are qualitatively different from the ones shown in Fig.~\ref{fig:sshsurfacestates}. They are localized at all 4 edges of the system, and also, they do not display the spatial  modulations observed in Figs.~\ref{fig:sshsurfacestates}(b)-(e). 

\begin{figure}[h!]
    \centering
    \begin{minipage}[t]{\textwidth}
        \begin{minipage}[t]{0.45\textwidth}
            \centering
            \begin{tikzpicture}
                \node[anchor=north west] (image) at (0,0) {\includegraphics[width=\textwidth]{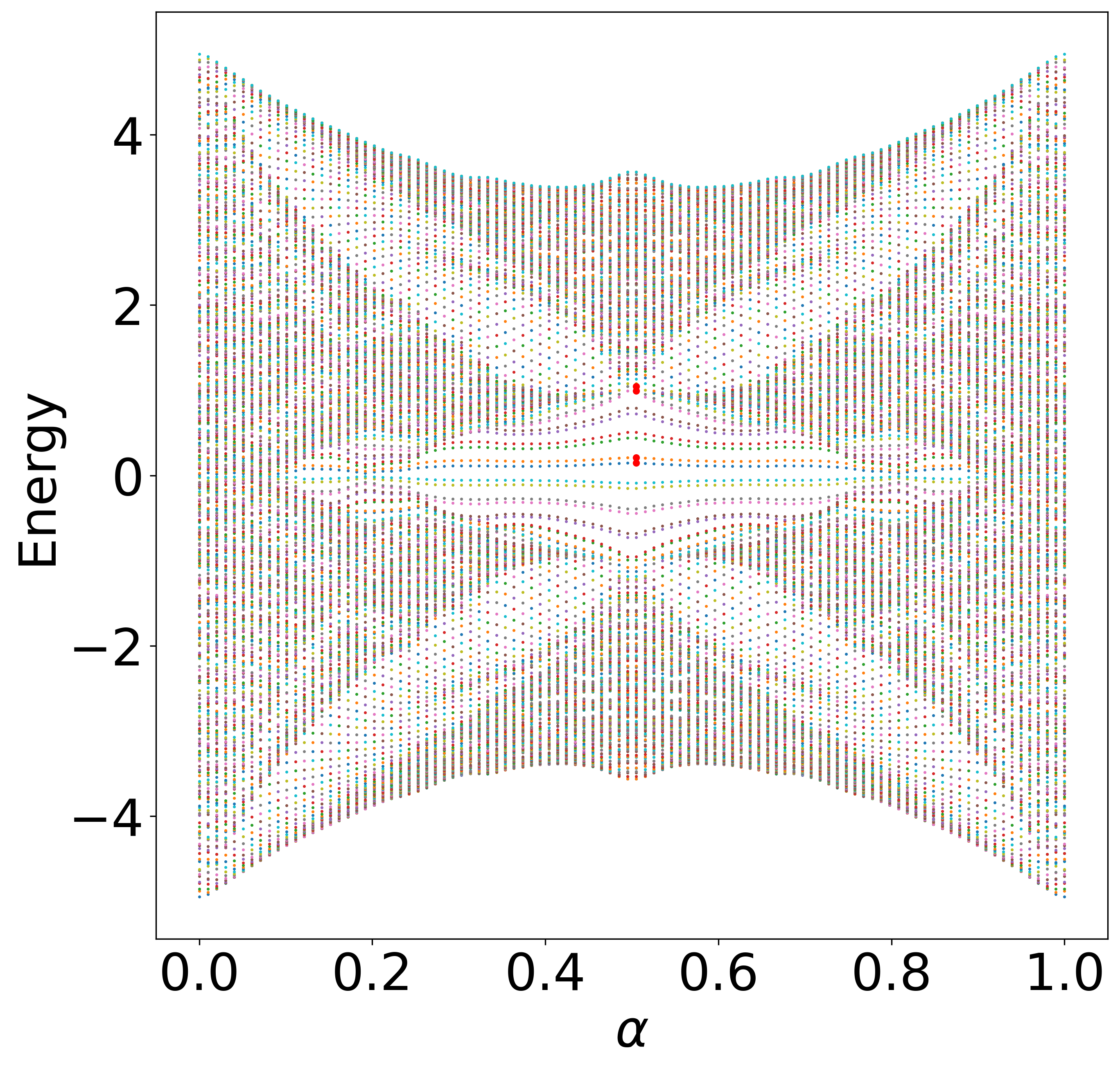}};
                \node[anchor=north west, text=black, font=\bfseries, xshift=-0.2cm, yshift=0.2cm] at (image.north west) {(a)};
            \end{tikzpicture}
            \label{fig:big}
        \end{minipage}
        \hspace{0.02\textwidth}
        \raisebox{3cm}{ 
            \begin{minipage}[t]{0.50\textwidth}
                \centering
                \begin{minipage}[t]{\linewidth}
                    \begin{minipage}[t]{0.49\textwidth}
                        \centering
                        \begin{tikzpicture}
                            \node[anchor=north west] (image) at (0,0) {\includegraphics[width=\textwidth]{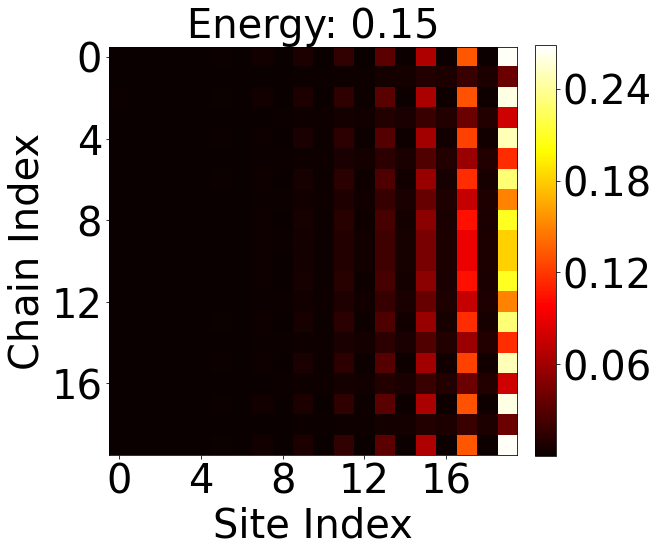}};
                            \node[anchor=north west, text=black, font=\bfseries, xshift=-0.2cm, yshift=0.2cm] at (image.north west) {(b)};
                        \end{tikzpicture}
                        \label{fig:small1}
                    \end{minipage}
                    \hfill
                    \begin{minipage}[t]{0.49\textwidth}
                        \centering
                        \begin{tikzpicture}
                            \node[anchor=north west] (image) at (0,0) {\includegraphics[width=\textwidth]{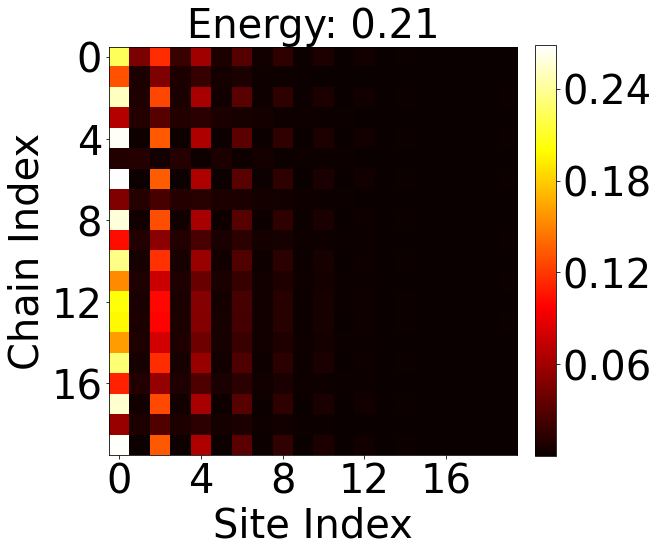}};
                            \node[anchor=north west, text=black, font=\bfseries, xshift=-0.2cm, yshift=0.2cm] at (image.north west) {(c)};
                        \end{tikzpicture}
                        \label{fig:small2}
                    \end{minipage}
                \end{minipage}
                \vspace{0.02\textwidth}
                \vspace{-1.5\baselineskip}
                \begin{minipage}[t]{\linewidth}
                    \begin{minipage}[t]{0.49\textwidth}
                        \centering
                        \begin{tikzpicture}
                            \node[anchor=north west] (image) at (0,0) {\includegraphics[width=\textwidth]{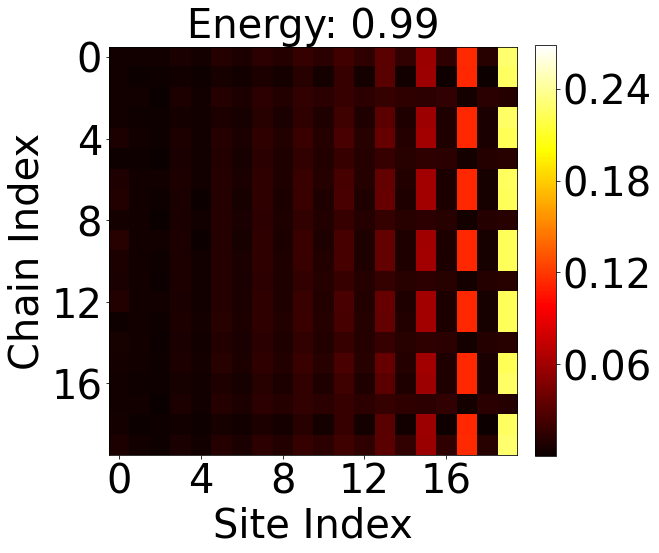}};
                            \node[anchor=north west, text=black, font=\bfseries, xshift=-0.2cm, yshift=0.2cm] at (image.north west) {(d)};
                        \end{tikzpicture}
                        \label{fig:small3}
                    \end{minipage}
                    \hfill
                    \begin{minipage}[t]{0.49\textwidth}
                        \centering
                        \begin{tikzpicture}
                            \node[anchor=north west] (image) at (0,0) {\includegraphics[width=\textwidth]{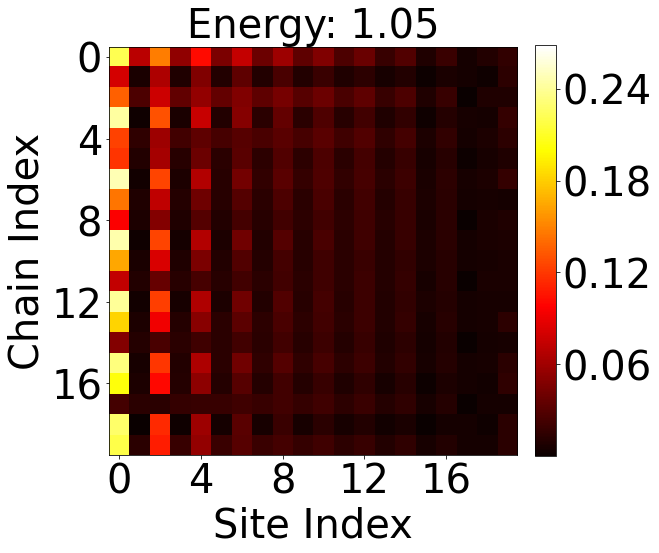}};
                            \node[anchor=north west, text=black, font=\bfseries, xshift=-0.2cm, yshift=0.2cm] at (image.north west) {(e)};
                        \end{tikzpicture}
                        \label{fig:small4}
                    \end{minipage}
                \end{minipage}
            \end{minipage}
        }
    \end{minipage}
    \caption{Effect of an on-site impurity (strength U=1) on the inversion symmetry protected surface states in the topologically non-trivial dimerized phase. The impurity is located in the lower left corner of the  20-leg SSH ladder. Parameters are chosen identical to the previous figure.}
    \label{fig:sshimpurity}
\end{figure}

Finally, we examine how these different topological surface states respond to the introduction of a local perturbation. Specifically, we consider the effects of an onsite potential $U=1$, introduced at the lower left corner of the system. This potential obviously breaks the spatial symmetries of the system, with time-reversal symmetry further broken by the magnetic field. As seen in Fig.~\ref{fig:sshimpurity}(a), this type of impurity breaks the two-fold degeneracy of the inversion symmetry protected surface states. As illustrated in Figs.~\ref{fig:sshimpurity}(b)-(e), the corresponding wave functions are now localized only on one side of the system, thus adapting to the local onsite repulsion. 

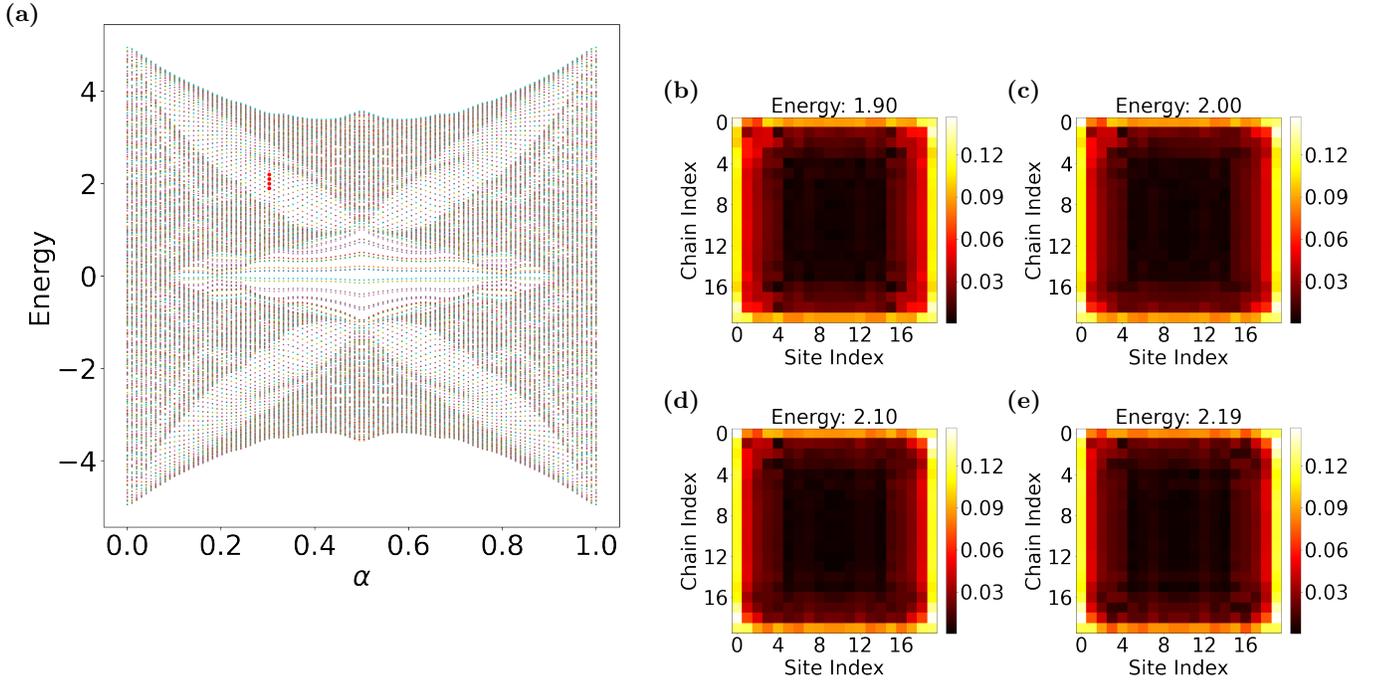
\begin{figure}[h!]
    \centering
    \begin{minipage}[t]{\textwidth}
        \begin{minipage}[t]{0.45\textwidth}
            \centering
            \begin{tikzpicture}
                \node[anchor=north west] (image) at (0,0) {\includegraphics[width=\textwidth]{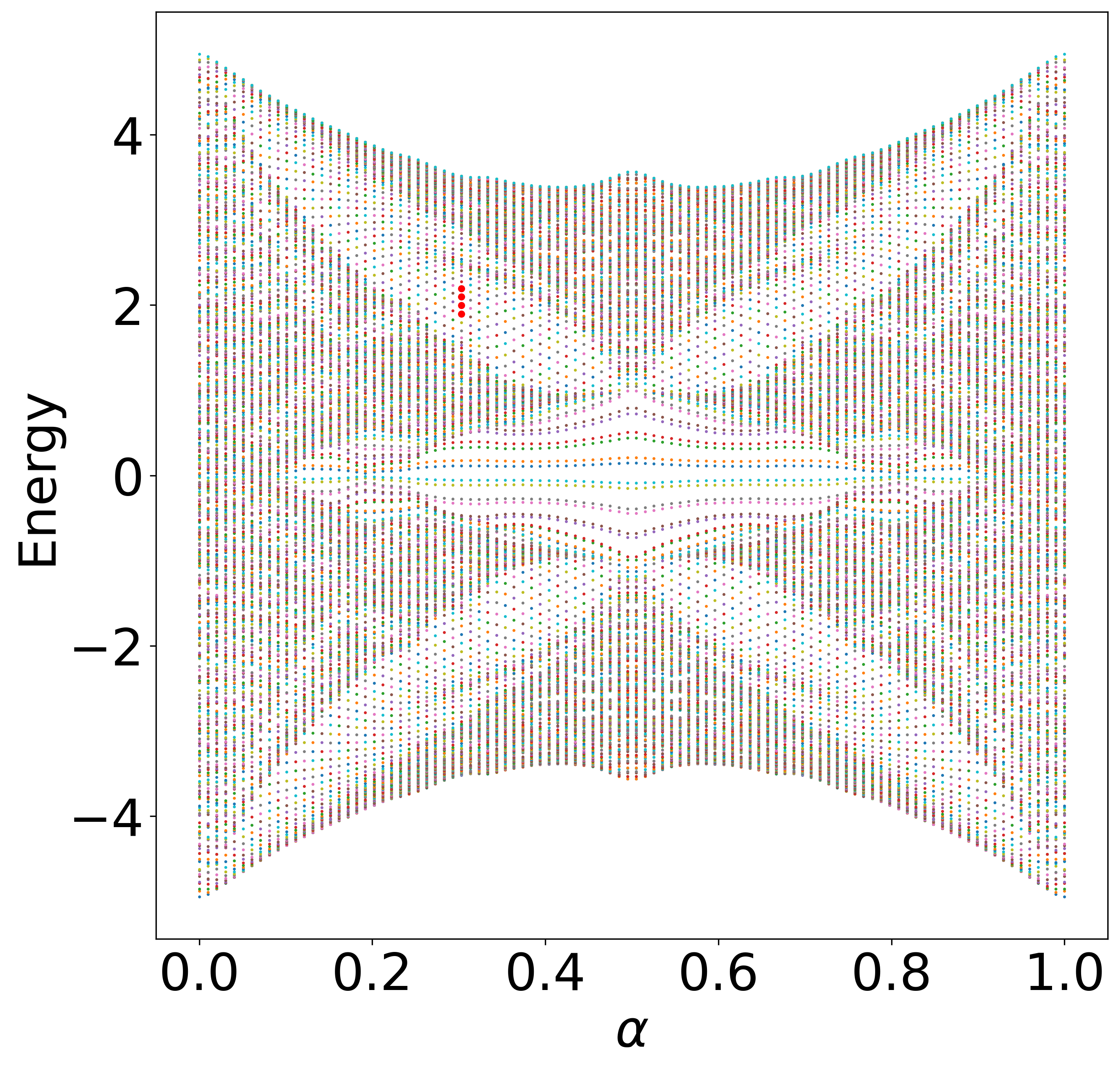}};
                \node[anchor=north west, text=black, font=\bfseries, xshift=-0.2cm, yshift=0.2cm] at (image.north west) {(a)};
            \end{tikzpicture}
            \label{fig:big}
        \end{minipage}
        \hspace{0.02\textwidth}
        \raisebox{3cm}{ 
            \begin{minipage}[t]{0.50\textwidth}
                \centering
                \begin{minipage}[t]{\linewidth}
                    \begin{minipage}[t]{0.49\textwidth}
                        \centering
                        \begin{tikzpicture}
                            \node[anchor=north west] (image) at (0,0) {\includegraphics[width=\textwidth]{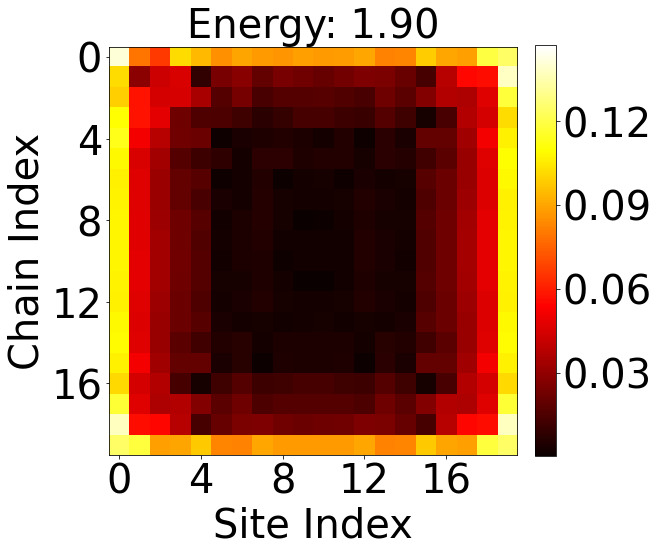}};
                            \node[anchor=north west, text=black, font=\bfseries, xshift=-0.2cm, yshift=0.2cm] at (image.north west) {(b)};
                        \end{tikzpicture}
                        \label{fig:small1}
                    \end{minipage}
                    \hfill
                    \begin{minipage}[t]{0.49\textwidth}
                        \centering
                        \begin{tikzpicture}
                            \node[anchor=north west] (image) at (0,0) {\includegraphics[width=\textwidth]{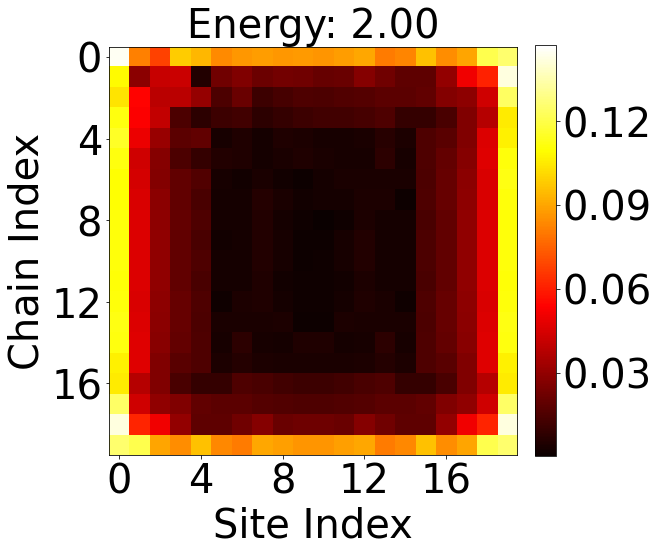}};
                            \node[anchor=north west, text=black, font=\bfseries, xshift=-0.2cm, yshift=0.2cm] at (image.north west) {(c)};
                        \end{tikzpicture}
                        \label{fig:small2}
                    \end{minipage}
                \end{minipage}
                \vspace{0.02\textwidth}
                \vspace{-1.5\baselineskip}
                \begin{minipage}[t]{\linewidth}
                    \begin{minipage}[t]{0.49\textwidth}
                        \centering
                        \begin{tikzpicture}
                            \node[anchor=north west] (image) at (0,0) {\includegraphics[width=\textwidth]{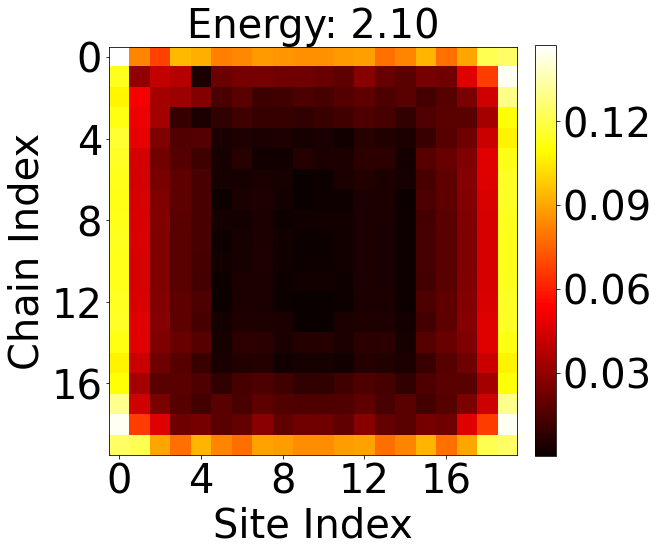}};
                            \node[anchor=north west, text=black, font=\bfseries, xshift=-0.2cm, yshift=0.2cm] at (image.north west) {(d)};
                        \end{tikzpicture}
                        \label{fig:small3}
                    \end{minipage}
                    \hfill
                    \begin{minipage}[t]{0.49\textwidth}
                        \centering
                        \begin{tikzpicture}
                            \node[anchor=north west] (image) at (0,0) {\includegraphics[width=\textwidth]{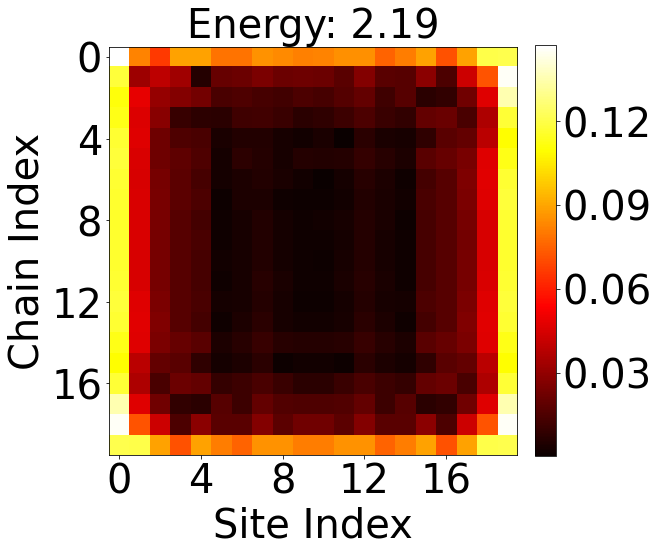}};
                            \node[anchor=north west, text=black, font=\bfseries, xshift=-0.2cm, yshift=0.2cm] at (image.north west) {(e)};
                        \end{tikzpicture}
                        \label{fig:small4}
                    \end{minipage}
                \end{minipage}
            \end{minipage}
        }
    \end{minipage}
    \caption{No effect of an on-site impurity (strength U=1) on the surface states due to the non-trivial Chern number in the topologically non-trivial dimerized phase. The impurity is located in the lower left corner of the  20-leg SSH ladder. Parameters are chosen identical to the previous figure.
    $20\times 20$ square lattice, $H[0][0]=1$ lattice $t_0=1, t_1=1, t_2=2$ at index 30}
    \label{fig:hhimpurity}
\end{figure}

In contrast, the onsite repulsive potential has no discernible effect on the chiral edge states which stem from the non-trivial Chern number, as shown in Fig.~\ref{fig:hhimpurity}. While this is clearly expected, given the nature of this hybrid topological system, it also points to potential applications that allow us to address and tune some of the prominent topological surface states, while leaving others unaltered. 

\section{Conclusions}

In this chapter, we have explored the intricate energy spectra and the behavior of bulk and surface states in topological insulators characterized by co-existing topological invariants, with a particular focus on the Hofstadter butterfly patterns that emerge. Through our analysis of coupled stacks of 1D Su-Schrieffer-Heeger (SSH) chains, we have shown that the interplay of topologically non-trivial dimerization and an external magnetic field can lead to distinct kinds of surface states.

In these hybrid topological systems, we have observed well-defined bulk energy bands which are well-separated by energy gaps. As is expected from the conventional Hofstadter model, one can calculate the Hall conductance for this system and find distinct phases which are characterized by distinct integer-valued topological invariants i.e., Chern numbers. Upon introducing boundaries into the system, these non-trivial Chern numbers result in chiral edge states which appear within the bulk gaps and are localized on the boundaries of the system. Additionally, however, the topologically non-trivial dimerization introduced by the underlying SSH model can also lead to topologically non-trivial surface states which are protected by the inversion symmetry of an effective 1D model and a corresponding 1D topological invariant (as discussed in Ref.~\cite{lau2015}). This protection ensures the resilience of these states against various perturbations, highlighting the crucial role of symmetry in maintaining their stability.

Our examination of the SSH and Hofstadter models has revealed that the nature of the symmetry-protected surface states differs significantly between these models. The SSH model, which preserves inversion symmetry, and the Hofstadter model, which breaks time-reversal symmetry and hence can have a non-trivial Chern number, each give rise to distinct types of surface states. These states are not only localized differently but also exhibit different responses to perturbations, underscoring the diverse ways in which topological protection can manifest.

More generally, we can also imagine introducing dimerization in the vertical direction, which would lead to the coexistence of 1D edge states and 0D corner states within the same system: as such, this would constitute an instance of a hybrid higher-order topological insulator in which 0D corner states, non-chiral 1D edge states, and chiral 1D edge states coexist~\cite{otaki2019,zuo2021}. This illustrates the rich tapestry of topologically protected boundary states that can emerge upon subjecting spatial symmetry protected topological states to an external magnetic field. 

In conclusion, our study highlights the fascinating physics that arises from combining strong topology (as encoded in the Chern number) and crystalline symmetries, which can lead to weak topological invariants. By investigating  relatively simple, yet paradigmatic models, we have illustrated how hybrid topology can lead to the emergence of robust surface states, further enriching our understanding of topological phases of matter. We hope that the insights gained from this work lead to further exploration of hybrid topological systems and their potential applications.



\bibliography{library}{}

\begin{thebibliography}{10}

\bibitem{kanemele}
C.~L. Kane and E.~J. Mele.
\newblock ${Z}_{2}$ topological order and the quantum spin hall effect.
\newblock {\em Phys. Rev. Lett.}, 95:146802, Sep 2005.

\bibitem{bernevigzhang}
B.~Andrei Bernevig and Shou-Cheng Zhang.
\newblock Quantum spin hall effect.
\newblock {\em Phys. Rev. Lett.}, 96:106802, Mar 2006.

\bibitem{fukanemele}
Liang Fu, C.~L. Kane, and E.~J. Mele.
\newblock Topological insulators in three dimensions.
\newblock {\em Phys. Rev. Lett.}, 98:106803, Mar 2007.

\bibitem{moorebalents}
J.~E. Moore and L.~Balents.
\newblock Topological invariants of time-reversal-invariant band structures.
\newblock {\em Phys. Rev. B}, 75:121306, Mar 2007.

\bibitem{roy2009}
Rahul Roy.
\newblock Topological phases and the quantum spin hall effect in three
  dimensions.
\newblock {\em Phys. Rev. B}, 79:195322, May 2009.

\bibitem{konig2007}
Markus {K{\"o}nig}, Steffen {Wiedmann}, Christoph {Br{\"u}ne}, Andreas {Roth},
  Hartmut {Buhmann}, Laurens~W. {Molenkamp}, Xiao-Liang {Qi}, and Shou-Cheng
  {Zhang}.
\newblock {Quantum Spin Hall Insulator State in HgTe Quantum Wells}.
\newblock {\em Science}, 318(5851):766, Nov 2007.

\bibitem{hsieh2008}
David Hsieh, Dong Qian, Lewis Wray, YuQi Xia, Yew~San Hor, Robert~Joseph Cava,
  and M~Zahid Hasan.
\newblock A topological dirac insulator in a quantum spin hall phase.
\newblock {\em Nature}, 452(7190):970, 2008.

\bibitem{readgreen}
N.~Read and Dmitry Green.
\newblock Paired states of fermions in two dimensions with breaking of parity
  and time-reversal symmetries and the fractional quantum hall effect.
\newblock {\em Phys. Rev. B}, 61:10267--10297, Apr 2000.

\bibitem{ivanov}
D.~A. Ivanov.
\newblock Non-abelian statistics of half-quantum vortices in p-wave
  superconductors.
\newblock {\em Phys. Rev. Lett.}, 86:268--271, Jan 2001.

\bibitem{stoneroy}
Michael Stone and Rahul Roy.
\newblock Edge modes, edge currents, and gauge invariance in ${p}_{x}{+ip}_{y}$
  superfluids and superconductors.
\newblock {\em Phys. Rev. B}, 69:184511, May 2004.

\bibitem{zhang2018}
Peng Zhang, Koichiro Yaji, Takahiro Hashimoto, Yuichi Ota, Takeshi Kondo, Kozo
  Okazaki, Zhijun Wang, Jinsheng Wen, G.~D. Gu, Hong Ding, and Shik Shin.
\newblock Observation of topological superconductivity on the surface of an
  iron-based superconductor.
\newblock {\em Science}, 360(6385):182--186, 2018.

\bibitem{ryu2010}
Shinsei Ryu, Andreas~P Schnyder, Akira Furusaki, and Andreas W~W Ludwig.
\newblock Topological insulators and superconductors: tenfold way and
  dimensional hierarchy.
\newblock {\em New Journal of Physics}, 12(6):065010, jun 2010.

\bibitem{kitaev2009}
Alexei Kitaev.
\newblock Periodic table for topological insulators and superconductors.
\newblock {\em AIP Conference Proceedings}, 1134(1):22--30, 2009.

\bibitem{chiu2016rmp}
Ching-Kai Chiu, Jeffrey C.~Y. Teo, Andreas~P. Schnyder, and Shinsei Ryu.
\newblock Classification of topological quantum matter with symmetries.
\newblock {\em Rev. Mod. Phys.}, 88:035005, Aug 2016.

\bibitem{AZclass}
Alexander Altland and Martin~R. Zirnbauer.
\newblock Nonstandard symmetry classes in mesoscopic normal-superconducting
  hybrid structures.
\newblock {\em Phys. Rev. B}, 55:1142--1161, Jan 1997.

\bibitem{hasankanermp}
M.~Z. Hasan and C.~L. Kane.
\newblock Colloquium: Topological insulators.
\newblock {\em Rev. Mod. Phys.}, 82:3045--3067, Nov 2010.

\bibitem{qizhangrmp}
Xiao-Liang Qi and Shou-Cheng Zhang.
\newblock Topological insulators and superconductors.
\newblock {\em Rev. Mod. Phys.}, 83:1057--1110, Oct 2011.

\bibitem{hasanmoore}
M.~Zahid {Hasan} and Joel~E. {Moore}.
\newblock {Three-Dimensional Topological Insulators}.
\newblock {\em Annual Review of Condensed Matter Physics}, 2:55--78, March
  2011.

\bibitem{fuTCI}
Liang Fu.
\newblock Topological crystalline insulators.
\newblock {\em Phys. Rev. Lett.}, 106:106802, Mar 2011.

\bibitem{hsiehTCI}
Timothy~H. {Hsieh}, Hsin {Lin}, Junwei {Liu}, Wenhui {Duan}, Arun {Bansil}, and
  Liang {Fu}.
\newblock {Topological crystalline insulators in the SnTe material class}.
\newblock {\em Nature Communications}, 3:982, Jul 2012.

\bibitem{okada2013}
Yoshinori Okada, Maksym Serbyn, Hsin Lin, Daniel Walkup, Wenwen Zhou, Chetan
  Dhital, Madhab Neupane, Suyang Xu, Yung~Jui Wang, R.~Sankar, Fangcheng Chou,
  Arun Bansil, M.~Zahid Hasan, Stephen~D. Wilson, Liang Fu, and Vidya Madhavan.
\newblock Observation of dirac node formation and mass acquisition in a
  topological crystalline insulator.
\newblock {\em Science}, 341(6153):1496--1499, 2013.

\bibitem{sessi2016}
Paolo Sessi, Domenico Di~Sante, Andrzej Szczerbakow, Florian Glott, Stefan
  Wilfert, Henrik Schmidt, Thomas Bathon, Piotr Dziawa, Martin Greiter, Titus
  Neupert, Giorgio Sangiovanni, Tomasz Story, Ronny Thomale, and Matthias Bode.
\newblock Robust spin-polarized midgap states at step edges of topological
  crystalline insulators.
\newblock {\em Science}, 354(6317):1269--1273, 2016.

\bibitem{ma2017}
Junzhang Ma, Changjiang Yi, Baiqing Lv, ZhiJun Wang, Simin Nie, Le~Wang,
  Lingyuan Kong, Yaobo Huang, Pierre Richard, Peng Zhang, Koichiro Yaji, Kenta
  Kuroda, Shik Shin, Hongming Weng, Bogdan~Andrei Bernevig, Youguo Shi, Tian
  Qian, and Hong Ding.
\newblock Experimental evidence of hourglass fermion in the candidate
  nonsymmorphic topological insulator khgsb.
\newblock {\em Science Advances}, 3(5), 2017.

\bibitem{schindlerHOTI}
Frank Schindler, Ashley~M. Cook, Maia~G. Vergniory, Zhijun Wang, Stuart S.~P.
  Parkin, B.~Andrei Bernevig, and Titus Neupert.
\newblock Higher-order topological insulators.
\newblock {\em Science Advances}, 4(6), 2018.

\bibitem{benalcazar2017}
Wladimir~A. Benalcazar, B.~Andrei Bernevig, and Taylor~L. Hughes.
\newblock Quantized electric multipole insulators.
\newblock {\em Science}, 357(6346):61--66, 2017.

\bibitem{langbehnHOTI}
Josias Langbehn, Yang Peng, Luka Trifunovic, Felix von Oppen, and Piet~W.
  Brouwer.
\newblock Reflection-symmetric second-order topological insulators and
  superconductors.
\newblock {\em Phys. Rev. Lett.}, 119:246401, Dec 2017.

\bibitem{songHOTI}
Zhida Song, Zhong Fang, and Chen Fang.
\newblock $(d\ensuremath{-}2)$-dimensional edge states of rotation symmetry
  protected topological states.
\newblock {\em Phys. Rev. Lett.}, 119:246402, Dec 2017.

\bibitem{khalaf2018}
Eslam Khalaf, Hoi~Chun Po, Ashvin Vishwanath, and Haruki Watanabe.
\newblock Symmetry indicators and anomalous surface states of topological
  crystalline insulators.
\newblock {\em Phys. Rev. X}, 8:031070, Sep 2018.

\bibitem{khalaf2021botp}
Eslam Khalaf, Wladimir~A. Benalcazar, Taylor~L. Hughes, and Raquel Queiroz.
\newblock Boundary-obstructed topological phases.
\newblock {\em Phys. Rev. Res.}, 3:013239, Mar 2021.

\bibitem{chen2013}
Xie Chen, Zheng-Cheng Gu, Zheng-Xin Liu, and Xiao-Gang Wen.
\newblock Symmetry protected topological orders and the group cohomology of
  their symmetry group.
\newblock {\em Phys. Rev. B}, 87:155114, Apr 2013.

\bibitem{guwen2014}
Zheng-Cheng Gu and Xiao-Gang Wen.
\newblock Symmetry-protected topological orders for interacting fermions:
  Fermionic topological nonlinear $\ensuremath{\sigma}$ models and a special
  group supercohomology theory.
\newblock {\em Phys. Rev. B}, 90:115141, Sep 2014.

\bibitem{song2017}
Hao Song, Sheng-Jie Huang, Liang Fu, and Michael Hermele.
\newblock Topological phases protected by point group symmetry.
\newblock {\em Phys. Rev. X}, 7:011020, Feb 2017.

\bibitem{buildingblock}
Sheng-Jie Huang, Hao Song, Yi-Ping Huang, and Michael Hermele.
\newblock Building crystalline topological phases from lower-dimensional
  states.
\newblock {\em Phys. Rev. B}, 96:205106, Nov 2017.

\bibitem{elsethorngren}
Ryan Thorngren and Dominic~V. Else.
\newblock Gauging spatial symmetries and the classification of topological
  crystalline phases.
\newblock {\em Phys. Rev. X}, 8:011040, Mar 2018.

\bibitem{else2021}
Dominic~V. Else, Sheng-Jie Huang, Abhinav Prem, and Andrey Gromov.
\newblock Quantum many-body topology of quasicrystals.
\newblock {\em Phys. Rev. X}, 11:041051, Dec 2021.

\bibitem{senthilSPT}
T.~Senthil.
\newblock Symmetry-protected topological phases of quantum matter.
\newblock {\em Annual Review of Condensed Matter Physics}, 6(1):299--324, 2015.

\bibitem{otaki2019}
Yuria Otaki and Takahiro Fukui.
\newblock Higher-order topological insulators in a magnetic field.
\newblock {\em Phys. Rev. B}, 100:245108, Dec 2019.

\bibitem{jonah2020}
Jonah Herzog-Arbeitman, Zhi-Da Song, Nicolas Regnault, and B.~Andrei Bernevig.
\newblock Hofstadter topology: Noncrystalline topological materials at high
  flux.
\newblock {\em Phys. Rev. Lett.}, 125:236804, Dec 2020.

\bibitem{zuo2021}
Zheng-Wei {Zuo}, Wladimir~A. {Benalcazar}, Yunzhe {Liu}, and Chao-Xing {Liu}.
\newblock {Topological phases of the dimerized Hofstadter butterfly}.
\newblock {\em Journal of Physics D Applied Physics}, 54(41):414004, October
  2021.

\bibitem{hybrid1}
Md~Shafayat {Hossain}, Frank {Schindler}, Rajibul {Islam}, Zahir {Muhammad},
  Yu-Xiao {Jiang}, Zi-Jia {Cheng}, Qi~{Zhang}, Tao {Hou}, Hongyu {Chen}, Maksim
  {Litskevich}, Brian {Casas}, Jia-Xin {Yin}, Tyler~A. {Cochran}, Mohammad
  {Yahyavi}, Xian~P. {Yang}, Luis {Balicas}, Guoqing {Chang}, Weisheng {Zhao},
  Titus {Neupert}, and M.~Zahid {Hasan}.
\newblock {A hybrid topological quantum state in an elemental solid}.
\newblock {\em Nature}, 628(8008):527--533, April 2024.

\bibitem{hybrid2}
Sheng-Jie {Huang}, Kyungwha {Park}, and Yi-Ting {Hsu}.
\newblock {Hybrid-order topological superconductivity in a topological metal
  1T'-MoTe$_{2}$}.
\newblock {\em npj Quantum Materials}, 9:21, January 2024.

\bibitem{kim2022}
Sun-Woo {Kim}, Sunam {Jeon}, Moon~Jip {Park}, and Youngkuk {Kim}.
\newblock {Replica higher-order topology of Hofstadter butterflies in twisted
  bilayer graphene}.
\newblock {\em npj Computational Mathematics}, 9:152, January 2023.

\bibitem{jonah2023}
Jonah Herzog-Arbeitman, Zhi-Da Song, Luis Elcoro, and B.~Andrei Bernevig.
\newblock Hofstadter topology with real space invariants and reentrant
  projective symmetries.
\newblock {\em Phys. Rev. Lett.}, 130:236601, Jun 2023.

\bibitem{ssh1979}
W.~P. Su, J.~R. Schrieffer, and A.~J. Heeger.
\newblock Solitons in polyacetylene.
\newblock {\em Phys. Rev. Lett.}, 42:1698--1701, Jun 1979.

\bibitem{hofstadter1976}
Douglas~R. Hofstadter.
\newblock Energy levels and wave functions of bloch electrons in rational and
  irrational magnetic fields.
\newblock {\em Phys. Rev. B}, 14:2239--2249, Sep 1976.

\bibitem{asboth2016short}
J{\'a}nos~K Asb{\'o}th, L{\'a}szl{\'o} Oroszl{\'a}ny, and Andr{\'a}s P{\'a}lyi.
\newblock A short course on topological insulators.
\newblock {\em Lecture notes in physics}, 919:166, 2016.

\bibitem{liu2017novel}
Feng Liu and Katsunori Wakabayashi.
\newblock Novel topological phase with a zero berry curvature.
\newblock {\em Phys. Rev. Lett.}, 118:076803, Feb 2017.

\bibitem{macdonaldrev}
A.~H. {MacDonald}.
\newblock {Introduction to the Physics of the Quantum Hall Regime}.
\newblock {\em arXiv e-prints}, pages cond--mat/9410047, October 1994.

\bibitem{zak}
J.~Zak.
\newblock Magnetic translation group.
\newblock {\em Phys. Rev.}, 134:A1602--A1606, Jun 1964.

\bibitem{harper}
P.~G. {Harper}.
\newblock {Single Band Motion of Conduction Electrons in a Uniform Magnetic
  Field}.
\newblock {\em Proceedings of the Physical Society A}, 68(10):874--878, October
  1955.

\bibitem{wenzee}
X.~G. {Wen} and A.~{Zee}.
\newblock {Winding number, family index theorem, and electron hopping in a
  magnetic field}.
\newblock {\em Nuclear Physics B}, 316(3):641--662, April 1989.

\bibitem{TKNN}
D.~J. Thouless, M.~Kohmoto, M.~P. Nightingale, and M.~den Nijs.
\newblock Quantized hall conductance in a two-dimensional periodic potential.
\newblock {\em Phys. Rev. Lett.}, 49:405--408, Aug 1982.

\bibitem{Laughlin}
R.~B. Laughlin.
\newblock Quantized hall conductivity in two dimensions.
\newblock {\em Phys. Rev. B}, 23:5632--5633, May 1981.

\bibitem{Halperin}
B.~I. Halperin.
\newblock Quantized hall conductance, current-carrying edge states, and the
  existence of extended states in a two-dimensional disordered potential.
\newblock {\em Phys. Rev. B}, 25:2185--2190, Feb 1982.

\bibitem{Hatsugai}
Yasuhiro Hatsugai.
\newblock Edge states in the integer quantum hall effect and the riemann
  surface of the bloch function.
\newblock {\em Phys. Rev. B}, 48:11851--11862, Oct 1993.

\bibitem{lau2015}
Alexander Lau, Carmine Ortix, and Jeroen van~den Brink.
\newblock Topological edge states with zero hall conductivity in a dimerized
  hofstadter model.
\newblock {\em Phys. Rev. Lett.}, 115:216805, Nov 2015.

\bibitem{he2022}
Ai-Lei He, Xiuyun Zhang, and Yongjun Liu.
\newblock Topological states in a dimerized square-octagon lattice with
  staggered magnetic fluxes.
\newblock {\em Phys. Rev. B}, 106:125147, Sep 2022.

\end{thebibliography}
\bibliographystyle{unsrt}


\end{document}